\documentclass[aps,pra,preprint]{revtex4}
\usepackage{amsmath}
\usepackage{epsfig,amssymb,amsfonts}
\usepackage{graphicx}
\usepackage{natbib}
\usepackage{hyperref}

\begin{document}

\title{QUANTUM MECHANICS AS A REVERSIBLE DIFFUSION THEORY:AN EXTENDED PROBABILITY APPROACH}

\author{Charalampos Antonakos}

\email{bantonakos200@gmail.com}

\today

\begin{abstract}
This paper proposes an interpretation of quantum mechanics, relying on the time-symmetric stochastic dynamics of quantum particles and on extended probability theory. Our main purpose is to demonstrate that the wave function and its complex conjugate can be interpreted as complex probability distributions in two complex diffusion equations related to non-real forward and backward in time stochastic motions respectively. We say non-real because Schroedinger forward and backward diffusions describe both reversible (real trajectories) and irreversible trajectories (non-real trajectories). The reversible trajectories are the only real trajectories and are given by the intersection of those forward and backward processes. It turns out that if we translate this intersection using set-theoretic language, we are led to a reversible diffusion described by Born rule probabilities. This proposal is useful also for explaining more about the role of complex numbers in quantum mechanics that produces this so-called "wave-like" nature of quantum reality. Our perspective also challenges the notion of physical superposition and aims at a derivation of superposition principle not based on the linearity of Schroedinger's equation but relying on pure probability theory. Moreover, it is suggested that, embracing the idea of stochastic processes in quantum theory, explains the reasons for the appearance of classical behavior in large objects, in contrast to the quantum behavior of small ones. In other words, we claim that a combination of a probabilistic and no-ontic view (neither epistemic though) of the wave function with a stochastic hidden-variables approach, may provide some insight into the quantum physical reality and potentially establish the groundwork for a novel interpretation of quantum mechanics.

\end{abstract}
\maketitle

\section{Introduction}

Until now, the Copenhagen interpretation is considered for many as the most satisfying framework that explains a variety of quantum phenomena. However, the status of the wave function and the quantum formalism (i.e. Schroedinger equation, Born rule, superposition, complex-valuedeness etc.) remain unexplained both mathematically and physically. Moreover, there is no mechanism for the so-called "collapse" of the wave function as well as for what actually exists before measurement (the reality problem).

The scope of this paper is to demonstrate that QM is a reversible diffusion theory, whose description necessitates the use of complex probability theory. In this picture, Schroedinger’s equation emerges from extension of sets outside the real sample space and Born rule is derived from the intersection of forward and backward ensemble paths which are described by the forward and backward Schroedinger equations/complex diffusion equations (Chapter III,IV and Appendix G). The extension of sets is needed to ensure statistical reversibility of trajectories (Appendices A, B and D).
Moreover, linear superposition emerges from a probabilistic (or set) description of the inability of the quantum particle to occupy more than one states simultaneously (see Appendices F and J).  In general, one of the scopes of this paper is to present a hyper-realist theory for all observables. In that sense, the state of the particle is definite in a contextual way before measurement. The analysis is sufficiently done for energy states in Appendices F, J and L, while the extension to spin (in Appendix L) and thus to other observables is a bit more difficult and requires further research. However, an attempt is made. In the same Appendix, reality problem is solved completely and it is also mentioned that the intersection process which leads to the Born rule, could also be understood as an intersection of forward and backward individual (and not ensemble) paths due to reversibility at the microscopic level  which is said to exist in classical diffusion too (where irreversibility emerges only at the statistical/macroscopic level). An SDE analysis is done to support that idea (it is important to read) and the view that QM is a reversible Markovian theory. Also, measurement problem is addressed in Appendix P.

I have to note that a possible connection of quantum phenomena with time-reversible diffusion processes had already been pointed out by Furth \cite{peliti2023r}. \textit{Furth also derived an uncertainty relation between velocity and position for classical diffusion, which is completely analogous to quantum uncertainty relation} and I want to show that this should be taken seriously into account and not just as a mathematical analogy between quantum theory and classical diffusion. Thus, Appendices S, T and V are extremely important too.

Additionally, the status of the wave function is neither ontological (like in Bohmian mechanics or Many-Worlds) nor epistemic (like in RQM \cite{rovelli1996relational}  or QBism \cite{caves2007subjective}), but instead it has non-subjective probabilistic character (complex probability distribution), which means that it is closer having a law-like/nomological role \cite{goldstein2013reality}.

\section{INTERPRETATIONS OF THE WAVE FUNCTION}

For decades, researchers have endeavored to comprehend quantum mechanics and its intricate phenomena (i.e. superposition). In 1959, Everett proposed a theory known as the MWI or Many Worlds Interpretation \cite{dewitt2015many}. A notable feature of the MWI is that all the possible outcomes of an experiment are existent in different worlds. That's a way to explain superposition and avoid the wave function collapse. Additionally, there have been efforts to derive the Born rule within this framework (\cite{carroll2014many}, \cite{wallace2010prove}, \cite{saunders2024finite}), although none of them has gained so far widespread acceptance in the scientific community.

Other theories attempt to provide a clearer understanding of the wave function's nature. For instance, Bohmian mechanics, introduced by Bohm \cite{bohm1952suggested} and further developed by others \cite{durr2009bohmian}, posits that the wave function is a guiding field living in the 3N-dimensional configuration space, where N is the number of particles that are guided deterministically by the wave function. According to this theory, there is a guidance equation which in the one-particle case is given by:
\begin{equation} \label{eq:1}
\overrightarrow{\eta}(r,t)=\frac{\hbar}{m_p}Im\left(\frac{\overrightarrow{\nabla} \Psi(r,t)}{\Psi(r,t)}\right)
\end{equation}
where $m_\rho$ is the mass of the particle, $\hbar$ the Planck's constant and $\Psi(r,t)$ the wave function of the system. Furthermore, $\overrightarrow{\eta}(r,t)$ is called the current velocity.
It is derived from the quantum continuity equation
$$\frac{\partial \rho(r,t)} {\partial t}=-\overrightarrow{\nabla} \cdot \left(\rho(r,t) \overrightarrow{\eta}(r,t)\right)$$
which ensures the conservation of probability. This equation can be regarded as an indirect method for deriving the Born rule $\rho(r,t)=\Psi^*(r,t)\Psi(r,t)$ as referred by Deotto and Ghirardi \cite{deotto1998bohmian} and given that the so-called Quantum Equilibrium Hypothesis is valid (recently, in the context of BM there have also existed numerical approaches for the establishment of the Born rule \cite{tzemos2023unstable}, \cite{tzemos2022born}). The wave function guidance is attributed to the presence of a quantum force, which is mathematically described by the equation:
$$\overrightarrow{F}_{Q}(r,t)=\frac{\hbar^2}{2m_p} \overrightarrow{\nabla}\left(\frac{\nabla^2\sqrt{\rho(r,t)}}{\sqrt{\rho(r,t)}}\right)=- \overrightarrow{\nabla} Q(r,t)$$
where $Q(r,t)=-\frac{\hbar^2}{2m_\rho}\frac{\nabla^2\sqrt{\rho(r,t)}}{\sqrt{\rho(r,t)}}$ is defined as the quantum potential. The latter is a significant additional term to the Hamilton-Jacobi equation, which is the following:
$$\frac{\partial S(r,t)}{\partial t}+\frac{(\overrightarrow{\nabla}S(r,t))^2}{2m_p}+V(r,t)+Q(r,t)=0$$
It is often stated that in the the classical limit $\hbar \rightarrow 0$, we get that $Q \rightarrow 0$ and we obtain the classical Hamilton-Jacobi equation, but that statement is problematic because the probability density is almost never independent of Planck's constant and thus at the limit $\hbar \rightarrow0$, we can not be sure that we will not be led to an indeterminate form.

In Bohmian mechanics, a significant issue arises with the stationary states, where the particle has zero current velocity meaning that the particle is motionless. This phenomenon, however, is not physically plausible. This problem can be solved by introducing Nelson's stochastic mechanics \cite{nelson1966derivation}-this theory has been further discussed by other researchers (\cite{beyer2021stochastic}, \cite{wallstrom1989derivation}, \cite{davidson1979generalization}, \cite{bacciagaluppi2005conceptual})-that also adds a new velocity, which is called the osmotic velocity and is given by the equation:
$$\overrightarrow{u}(r,t)=\frac{\hbar}{m_p}Re\left(\frac{\overrightarrow{\nabla} \Psi(r,t)}{\Psi(r,t)}\right)$$
which, via the transformation $\Psi=Re^{iS}$, leads us to obtain:
\begin{equation} \label{eq:2}
\overrightarrow{u}(r,t)=\beta^2 \frac{\overrightarrow{\nabla} \rho(r,t)}{\rho(r,t)}
\end{equation}
where $\beta^2=\hbar/2m_p$ is considered as the diffusion coefficient of the quantum stochastic process. The result of introducing this velocity is the derivation of the average kinetic energy of the particle, which was also presented by Mita \cite{mita2021schrodinger} and which is the following:
$$E_K=-\frac{\hbar^2}{2m_{\rho}}\int_{\mathbb{R}^3}\Psi^*(r,t)\overrightarrow{\nabla}^2\Psi(r,t)d^3r=\frac{1}{2}m_{\rho}\int_{\mathbb{R}^3}\rho(r,t)\left((\overrightarrow{\eta}(r,t))^2+(\overrightarrow{u}(r,t))^2\right)d^3r$$

However, our approach is closest to  Bohm and Vigier's stochastic approach \cite{bohm1989non} and does not include osmotic velocities in the dynamics. In fact, it's the evolution of the ensemble of particles and not of each individual particle (which behaves in a stochastic way) that is given by Bohmian guidance equation (Appendix D) and plays the key role in the so-called statistical reversibility of quantum diffusion (the possibility of individual path reversibility however is discussed in the Appendices).

The only common element of our approach with Nelson's is the ontological existence of a quantum background field. The nature of this background field is briefly presented in Appendix K and R and the mean "force field" exerted by it is the quantum force. This force though does not support causal relations between two entangled particles (see Appendix R and V).
\section{DERIVATION OF THE FIRST TYPE OF THE BORN RULE}

 Our starting axiom is that QM is a reversible non-equilibrium stochastic theory (reasons are analyzed in Appendices C, D). Prior to the derivation, we will express the two Schrödinger equations (for simplicity, we have chosen $V(x,t)=0$ and our quantities $x$ and $t$ to be dimensionless). The forward evolution takes place at times $t \in [0,t_0]$ and the backward evolution takes place again in the same time interval, but this time $t'=-t+t_0$ (more rigorous justification for why those must be satisfied given the reversible diffusion axiom can be found in Appendices L, S and V and from our complex probability theory developed in the Appendices):

$$\frac{\partial \Psi(r,t\in [0,t_0])}{\partial t}=\frac{i}{2}\nabla^2\Psi(r,t\in [0,t_0])$$

$$\frac{\partial \Psi^*(r,t'\in [0,t_0])}{\partial t}=\frac{i}{2}\nabla^2\Psi^*(r,t'\in [0,t_0])$$

From those two equations we can interpret the wave function and its complex conjugate as complex probability distributions, since these two quantities take the place of real probability densities, considering the fact that they are solutions to two diffusion equations. We will consider the first equation as the complex diffusion equation of the quasi-particle moving forward in time ($0 \rightarrow t_0)$ and the second as the complex diffusion equation of the quasi-particle moving backwards in time ($t_0 \rightarrow 0$) as seen in Appendix C. We say "quasi" because the trajectories derived from these equations are not actual physical trajectories but rather mathematical constructs that when intersected, give us the physical time-reversible trajectories of the quantum particle (see Appendices A, O). And all this process is done in order to construct a reversible diffusion dynamics (see Appendices B, C). 

Let's pause and note that there are theories that have referred to the time-symmetric nature of quantum mechanics, although not within the context of Brownian motion. One of them is the "Transactional Interpretation of Quantum Mechanics" \cite{cramer1986transactional}. In this interpretation, the wave function  represents a physical wave propagating forward in time. However, it is important to note that the "Transactional Interpretation" seeks to address the measurement problem by applying the notion of Retrocausality which is a concept irrelevant to what we're gonna do here.
(information regarding time-symmetry in quantum mechanics and time-invariance of the Schroedinger equation can be found in more detail in Appendices C and D).

Now, we will try to derive the Born rule $\rho(x,t_c)=|\Psi(x,t_c)|^2$. Since we have assumed that the wave function and its complex conjugate have properties of probability distributions, then it's wise to suppose that the probabilities derived by those distributions according to the following equations:

$$\textit{Prob}\left(T\rightarrow\{ x_1 \in F,  t_c \}\right)=\int_{L_1}^{L_2}\Psi\left(x_1,  t_c\right)dx_1$$
and 
$$Prob\left(A\rightarrow\{ x_2 \in F,t_c \}\right)=\int_{L_1}^{L_2}\Psi^*\left(x_2,t_c\right)dx_2$$
will have properties of probability. In the above equations, we must state that $T\rightarrow\{ x_1 \in F,  t_c \}$ is the probability set that describes the forward-moving particle as being located at a specific position $x_1 \in F=[L_1,L_2]$ at a time $t_c$. It's obvious that $ A \rightarrow \{x_2 \in F,t_c \} $ is the set that describes the backward-moving particle.

We have to enforce a condition that is essential for accurately reflecting the time-reversible nature of quantum dynamics (see Appendices C and D as well as Furth's parallelism with classical Brownian motion), which implies that a trajectory can be reversed, maintaining the same path but with opposite velocity. To be more specific, let's consider the motion of a particle at the time interval $t \in [0,t_0]$, with the motion starting at $t=0$ (forward motion). If the dynamics is time-reversible, then the process of reversing time-which means reversing the time ordering of events-is completely allowed. So, for a time-symmetric particle for which we assign the set $B\rightarrow\{ x \in F,t_c \}$, we can make the following proposition (the reader must see Appendix K too): $\forall$ trajectory, $\forall$ $ t_c \in [0,t_0]$ we have that $x_{\textit{backward}}(t'=t_c)=x_{\textit{forward}}(t=t_c)$, where $t'=-t+t_0$. Thus, we write: $B\rightarrow\{ x \in F,t_c \}=T\rightarrow\{ x \in F,t_c \} \cap A\rightarrow\{ x \in F,t_c \}$, which holds $\forall$ $F \subseteq \mathbb{R}$, $\forall$ $x \in F$. This gives:
$$Prob\left(B\rightarrow\{ x \in F,t_c \}\right)=Prob\left(T\rightarrow\{ x \in F,t_c \} \cap A\rightarrow\{ x \in F,t_c \}\right)=$$
$$Prob\left(T\rightarrow\{ x_1 \in F,t_c \} \cap A\rightarrow\{ x_2 \in F,t_c \}|x_1 \equiv x_2  \right)=\left[\int_{L_1}^{L_2}\int_{L_1}^{L_2}\Psi\left(x_1,t_c\right)\Psi^*\left(x_2,t_c\right)dx_1dx_2\right]_{x_1\equiv x_2}=$$
$$\int_{L_1}^{L_2}\int_{L_1}^{L_2}\Psi\left(x_1,t_c\right)\Psi^*\left(x_2,t_c\right)\delta(x_1-x_2)dx_1dx_2=\int_{L_1}^{L_2}\mid \Psi\left(x,t_c\right) \mid^2 dx$$
A more rigorous proof is provided in Appendix G, where the correct result is derived and which includes also the normalization constant (Appendix K is extremely useful too).

The time-reversible quantum motion will emerge from the intersection of forward and backward ensemble paths, some of which are not real and lead to complex-valuedeness of probabilities (see why wave function is complex in Appendix A and B).
In Appendix B, in fact, we further demonstrate that two distinct events, which either consistently occur together or don't occur at all ("entangled" events), can be made "artificially" independent by using extension beyond the real sample space. That leads to the existence of non-real sets and complex probabilities. This analysis is also applicable in quantum mechanics, wherein every quantum trajectory has always both (but never only one) directions in time, as demonstrated in the equation $B\rightarrow\{ x \in F,t_c \}=T\rightarrow\{ x \in F,t_c \} \cap A\rightarrow\{ x \in F,t_c \}$ above. A more complete description of how non-real stochastic motions and non-real probabilities in general \textit{must} emerge in QM, in order for it to be a reversible diffusion theory (this was Nelson's starting axiom too), is made in Appendices A, B, E and O.

We must mention though, that it's true that the time-reversal invariance of the Schroedinger equation does not, of course, imply the existence of time-reversible trajectories. However, the Born rule, as we saw, provides an indication of such fact. Also, in Appendices C and D there is a clear justification for the existence of reversible motion. In Appendix D, we also discuss the issue that this intersection process describes the ensemble of trajectories and not each individual trajectory (see also Appendix S). However, in Appendices L, Q and V we discuss how the intersection process we use to derive Born rule may emerge from individual path reversibility (SDEs are also involved in L and Q, but not in V).

\section{DERIVATION OF THE SECOND TYPE OF THE BORN RULE}

In this chapter, we derive the second formulation of the Born rule for energy states, that says that for a wave function $\Psi\left(x_1,t_c\right)=\sum_{n=1}^{\infty}c_n\Psi_n\left(x_1,t_c\right)$ , the probability of detecting the particle in the $n-$th state is equal to $|c_n|^2$. By integrating both sides of the above equation in the domain $F=[L_1,L_2]$, we obtain:
 \begin{equation} \label{eq:7}
\int_{L_1}^{L_2}\Psi\left(x_1,t_c\right)dx_1=\sum_{n=1}^{\infty}\int_{L_1}^{L_2}c_n\Psi_n\left(x_{1},t_c\right)dx_{1}
\end{equation}

Additionally, considering that the forward stochastic particle can occupy whichever energy state out of the infinite energy states, We can write:
$$T\rightarrow\{ x_1 \in F,t_c \}=\bigcup_{n=1}^{\infty}T_n\rightarrow\{ x_{1} \in F,t_c \} \Rightarrow$$
$$\Rightarrow Prob\left(T\rightarrow\{ x_1 \in F,t_c \}\right)=Prob\left(\bigcup_{n=1}^{\infty}T_n\rightarrow\{ x_{1} \in F,t_c \}\right)$$

The set $T\rightarrow\{ x_1 \in F,t_c \}$ describes the forward moving particle being located in the region $[L_1,L_2]$, while $T_n\rightarrow\{ x_{1} \in F,t_c \}$ describes the  same particle being located in the same region specifically in the n-th state. Now, since 
\begin{equation} \label{eq:8}
T_i\rightarrow\{ x_{1} \in F,t_c \} \cap T_j\rightarrow\{ x_{1} \in F,t_c \}=\varnothing
\end{equation}
for $i \neq j$ and thus 
$$Prob\left(T_i\rightarrow\{ x_{1} \in F,t_c \} \cap T_j\rightarrow\{ x_{1} \in F,t_c \}\right)=0$$
we get
\begin{equation} \label{eq:9}
Prob\left(T\rightarrow\{ x_1 \in F,t_c \}\right)=\sum_{n=1}^{\infty}Prob\left(T_n\rightarrow\{ x_{1} \in F,t_c \}\right)
\end{equation}
Taking into account the above equations for the forward-moving particle, we obtain
$$Prob\left(T_n\rightarrow\{ x_{1} \in F,t_c \}\right)=c_n\int_{L_1}^{L_2}\Psi_n\left(x_{1},t_c\right)dx_{1}$$
while for the backward moving particle, in the same way, we obtain:
$$Prob\left(A_n\rightarrow\{ x_{2} \in F,t_c \}\right)=c_n^*\int_{L_1}^{L_2}\Psi_n^*\left(x_{2},t_c\right)dx_{2}$$ 

Now, in order to determine the probability of finding the quantum particle at the $n$-th state within the interval $F$ as we did before, we calculate the probability of the intersection of the corresponding sets, yielding:
$$Prob\left(T_n\rightarrow\{ x_{1} \in F,t_c \} \cap A_n\rightarrow\{ x_{2} \in F,t_c \}\right)=Prob\left(T_n\rightarrow\{ x_{1} \in F,t_c \}\right)Prob\left(A_n\rightarrow\{ x_{2} \in F,t_c \}\right)$$
Applying the identical methodology as previously and acknowledging the necessity of $x_{1} \equiv x_{2}$, we find that 
$$Prob\left(T_n\rightarrow\{ x \in F,t_c \} \cap A_n\rightarrow\{ x \in F,t_c \}\right)= \mid c_n \mid^2  \int_{L_1}^{L_2} \mid \Psi_n(x,t_c) \mid^2dx $$

But, we know that the probability of finding a particle at a specific state (denoted as the $n$-th state in our case) is independent of the particle's position within that state. Consequently, in that case, $F$ corresponds to $\mathbb{R}$, indicating that the particle can occupy any position and thus $L_1 \rightarrow -\infty$ and $L_2 \rightarrow \infty $. In this manner, we obtain
$$ Prob\left(T_n\rightarrow\{ x \in \mathbb{R}, t_c \} \cap A_n\rightarrow\{ x \in \mathbb{R},t_c \}\right)=\mid c_n \mid^2$$
or
$$Prob\left(B_n\rightarrow\{ x \in \mathbb{R}, t_c \}\right)=\mid c_n \mid^2$$

To justify Eq. (4), it is essential to highlight that the two Schrödinger equations correspond to the forward and backward-moving stochastic quasi-particles. These quasi-particles, despite their abstract mathematical nature, exhibit attributes akin to physical particles. For instance, we discuss their trajectories and employ probabilities to describe these paths. According to this perspective, since these mathematical entities exhibit properties of physical particles—participating in diffusion equations (complex though), as previously discussed—it is reasonable to infer that they cannot exist in two distinct states simultaneously, as no physical particle exhibits such behavior. In this way, by rejecting the physical superposition of states for quasi-particles, we were able to derive the well-known mathematical superposition of the wave function in quantum mechanics. This derivation was achieved not by relying on the linearity of the Schrödinger equation, which naturally leads to linear superposition, but by considering purely physical probabilistic properties. Without this approach, the derivation would not have been possible. Also, it is evident that if quasi-particles cannot exist in a superposition of states, then the same principle must apply to quantum particles (see Chapter VI) and thus wave function superposition becomes ultimately tied to the absence of physical superposition of states for the quantum particle. This topic will be explored mathematically in greater detail in Chapter VI, while a more formal proof for the inverse statement will be provided in Appendices F and J. There, we see that linear superposition emerges just from the absence of physical state-superposition for the quantum particle.

\section{COMPLEX PROBABILITIES AND THE DOUBLE-SLIT EXPERIMENT}

The idea of taking into consideration sets that don't describe reality as non-real instead of empty (see Appendix A and B) also applies effectively in the double-slit experiment, where we observe interference. In other words, nature seems to "care" about those sets and this is experimentally confirmed. More specifically, given that $x_1 \equiv x_2$, we have that:
$$Prob_{T1A1}=  Prob\left(T_1\rightarrow\{ x \in F,t_c \} \cap A_1\rightarrow\{ x \in F,t_c \}\right)=\int_{L_1}^{L_2}\mid \Psi_1\left(x,t_c\right) \mid^2 dx \in \mathbb{R}$$
$$Prob_{T2A2}=Prob\left(T_2\rightarrow\{ x \in F ,t_c \} \cap A_2\rightarrow\{ x \in F,t_c \}\right)=\int_{L_1}^{L_2}\mid \Psi_2\left(x,t_c\right) \mid^2 dx \in \mathbb{R}$$
$$Prob_{T1A2}=Prob\left(T_1\rightarrow\{ x\in F,t_c \} \cap A_2\rightarrow\{ x \in F,t_c \}\right)=\int_{L_1}^{L_2}\Psi_1\left(x,t\right)\Psi_2^*\left(x,t_c\right) dx \in \mathbb{C}$$
$$Prob_{T2A1}=Prob\left(T_2\rightarrow\{ x \in F,t_c \} \cap A_1\rightarrow\{ x \in F,t_c \}\right)=\int_{L_1}^{L_2}\Psi_2\left(x,t_c\right)\Psi_1^*\left(x,t_c\right) dx \in \mathbb{C}$$

Real probabilities represent quantum particles exhibiting real physical processes. That's why $Prob_{T1A1}$, $Prob_{T2A2} \in \mathbb{R}$. Because each one of them refers to particles passing from one single slit exhibiting time-symmetric characteristics, and that, is accomplished when considering both forward and backward-moving particles passing from the same slit. So, this is also why $Prob_{T1A2}$, $Prob_{T2A1} \in \mathbb{C}$. Because it is obvious that these probabilities describe processes in which the forward-moving particle passes from a different slit from the backward-moving particle. They represent purely mathematical abstracts and tools detached from real physical processes and more closely tied to irreversible behavior. This assertion similarly applies to the probabilities associated with the wave function and its complex conjugate-as previously noted-that are also complex (see Appendices A, B, L, O).

Of course, someone may wonder why $T_n\rightarrow\{x \in F,t_c\} \cap A_m \rightarrow \{x\in F,t_c\}\neq \varnothing$, so that $Prob_{TnAm} \neq 0$ while $T_n\rightarrow\{x \in F,t_c\} \cap T_m\rightarrow\{x \in F,t_c\} =\varnothing$ for $n \neq m$. That's because, from a mathematical set perspective, T and A sets describe two independent random experiments, the one evolving forward and the other backwards in time. As a result, the intersection of those two sets in the product sample space (see also product space discussion on Appendix V) will always be a non-empty set. To be more precise, if $\Gamma,\Delta$ are related to distinct random experiments then $\Gamma \cap \Delta \neq \varnothing$  unless $\Gamma=\varnothing$ or $\Delta=\varnothing$. And that, in combination with what we said in the previous paragraph explains how complex "interference" terms emerge (see also Appendix P).

\section{discussions and conclusions}

The presence of complex functions in quantum mechanics can be attributed to the principle of time-reversibility inherent in quantum systems, necessitating the use of complex probability theory for their description. So, in such case, "wave-particle" duality is demystified since it just emerges from the description of a time-symmetric stochastic process. And, of course, this discussion can be performed only when we talk about quantum Brownian motion and not about classical Brownian motion where it is clear that, there, the processes are characterized by time-irreversibility, as explained in Appendix C, and thus, there is no need for complex-valued distributions. Also, there is a discussion by Guerra \cite{guerra1981structural} about the \textit{microscopic} reversibility of quantum stochastic processes in opposition to the\textit{ macroscopic} irreversibility of classical diffusion processes where many particles are involved (look at Appendix K). Also, we accomplish to derive both $|\psi|^2$ and $|c_n|^2$ and demystify superposition, More specifically, given that we accept the diffusion-like character of the forward and backward stochastic particles (once we accept the stochastic and not wave-like nature of forward and backward Schroedinger equations) it is logical to posit the assumption that physical superposition of states is impossible for those particles namely:
 $T_n \rightarrow  \{ x \in \mathbb{R}  ,t_c \} \cap T_m\rightarrow  \{ x \in \mathbb{R}  ,t_c \}= \varnothing$ and $A_n \rightarrow  \{ x \in \mathbb{R}  ,t_c \} \cap A_m\rightarrow  \{ x \in \mathbb{R}  ,t_c \}= \varnothing $. 
But the consequence of this exotic description is to obtain:
$$T_n \rightarrow  \{ x \in \mathbb{R}  ,t_c \} \cap T_m \rightarrow  \{ x \in \mathbb{R}  ,t_c \}\cap A_n \rightarrow  \{ x \in \mathbb{R}  ,t_c \} \cap A_m \rightarrow  \{ x \in \mathbb{R}  ,t_c \}= \varnothing \cap \varnothing=\varnothing$$
For two sets $\Gamma,\Delta $, we know that $\Gamma \cap \Delta=\Delta \cap \Gamma$, and as a result, we get
$$\left\{T_n \rightarrow  \{ x \in \mathbb{R}  ,t_c \} \cap A_n\rightarrow  \{ x \in \mathbb{R}  ,t_c \}\right\}\cap \left \{T_m \rightarrow  \{ x \in \mathbb{R}  ,t_c \} \cap A_m \rightarrow  \{ x \in \mathbb{R}  ,t_c \}\right\}=\varnothing$$
which leads to 
$$B_n \rightarrow  \{ x \in \mathbb{R}  ,t_c \} \cap B_m \rightarrow  \{ x \in \mathbb{R}  ,t_c \}=\varnothing$$

Translating the aforementioned expression into words, it indicates that a quantum particle cannot simultaneously exist in two distinct states. Consequently,accepting the stochastic nature of "quasi-particles" leads naturally to the superposition principle and results in the absence of physical superposition for the quantum particle. However, we could go the other way around and show how linear superposition emerges from the impossibility of (physical) superposition for the quantum particle as shown in Appendices F and J.

Subsequently, we derive the expression for the probability distribution in the double-slit experiment, reinforcing the concept that complex probabilities are not intended to describe a real particle  behavior but an irreversible (purely mathematical) behavior. This concept was initially introduced when discussing the wave function and its complex conjugate.

Our view diverges from the traditional epistemic views of the wave function and it states the following:"we should not account agents/observers at all, but instead consider the linear superposition of states as a probabilistic expression for the existence of the possible quantum states of the particle with not subjective but objective/frequentist complex probabilities. In this case, of course, there is no meaning in talking about the collapse of the wave function at all. And why should we? Do we use collapses when discussing conventional probabilities?" In this way, we preserve the non-ontological character of the wave function, rejecting at the same time its subjective view (that QBists support) and the participation of an agent. However, a derivation of an "effective non-Bayesian collapse" is provided in Appendix P.

Now, how can we be so sure about the existence of this sub-quantum field? Because due to the fluctuations of this field in QM, we necessitate the consideration of average values for position and momentum (see uncertainty principle discussion and its relation to diffusion in Appendix R). Furthermore, talking about average values of energy in a quantum system simply means that during the whole process, the particle does not have a constant energy value. What is responsible for the non-conservation of instantaneous mechanical energy of the particle? The continuous exchanges of energy between particle and field (the opposite view is presented in Appendix K). Another indication for the existence of the sub-quantum field is that it can explain why small particles exhibit quantum mechanical behavior, while large objects tend to behave classically. \textit{In fact, the smaller the mass, the easier for the particle to be guided (or affected) by the random fluctuations, resulting in a stochastic, and thus quantum behavior, while in the classical limit, we have $\lim_{m _\rho\to \infty}sup\{\Delta x^{(q)}\Delta u^{(q)}\}=\lim_{m_{\rho} \to \infty}\beta^2=0$, which implies the existence of well-defined non-stochastic trajectories (we remind that $\beta^2=\frac{\hbar}{2m_\rho}$ is the Nelson's diffusion coefficient).
}

 When it comes to quantum "interference", it occurs as a consequence of the Born rule in combination with the wave function superposition. But since both features can be explained by our reversible diffusion (or complex probability) theory, the "interference" is just a mathematical consequence of this theory. And the asset of our approach is that the occurence of "interference" effects does not imply physical superposition (see also Appendices F,J) or state indefiniteness (Appendix L), in opposition to Quantum Logic \cite{redei2013quantum} where there is no clear picture about the actual state of the particle. More analysis regarding the mathematical structure behind "interference" was done in Chapter V (see Appendix P also for a more set-theoretic analysis).

To sum up, the probabilistic view of the wave function may eliminate all the ontological views associated with it, but in fact, it strongly presents to us the physical reality. This physical reality has nothing to do with the wave function, but with the quantum background field that is much more fundamental and is responsible for the time-symmetric stochastic dynamics. The existence of the wave function in a complex diffusion equation-the Schroedinger equation-that we have to solve first in order to find the physically real probability distribution via the Born rule, should not reveal to us an ontological reality of the wave function, but in fact, the distinct type of the quantum stochastic processes that arise because of the sub-quantum field.

\appendix

\section{}

Let's consider a hyper-sample space within quantum mechanics denoted as $\Omega_h^{(q)}$, such that $Prob(\Omega_h^{\left(q\right)})=1$. In our probability measure theory we can have (if we want) $\Omega^{(q)} \subset \Omega_h^{\left(q\right)} $, but still $\mu\left(\Omega^{(q)}\right)=\mu(\Omega_h^{(q)})=1$ since our theory is not a classical positive measure theory (see more details in \cite{folland1999real} and Appendix H) for which we know that if $\Gamma \subset \Delta$ then that demands that $\mu(\Gamma)<\mu(\Delta)$. Thus, we can say:
$$T \rightarrow  \{ x \in \mathbb{R}  ,t_c \} \cup A \rightarrow  \{ x \in \mathbb{R}  ,t_c \} \subseteq \Omega_h^{(q)}$$
(see Appendix H). In other words, this hyper-sample space will contain both sets that correspond to forward and backward motion. We can say that:
$$T \rightarrow  \{ x \in F  ,t_c \}=\left(T \rightarrow  \{ x \in F  ,t_c \} \cap A \rightarrow  \{ x \in F  ,t_c \}\right)\cup\left(T \rightarrow  \{ x \in F  ,t_c \} \cap \left(A \rightarrow  \{ x \in  F ,t_c \}\right)^C\right) $$
$$A \rightarrow  \{ x \in F  ,t_c \}=\left(T \rightarrow  \{ x \in F  ,t_c \} \cap A \rightarrow  \{ x \in F  ,t_c \}\right)\cup\left(A \rightarrow  \{ x \in F  ,t_c \} \cap \left(T \rightarrow  \{ x \in F  ,t_c \}\right)^C\right) $$
where $ (A \rightarrow  \{ x \in F  ,t_c \})^C$ is the complementary set of $A \rightarrow  \{ x \in F  ,t_c \}$ and $ (T \rightarrow  \{ x \in F  ,t_c \})^C$ is the complementary set of $T \rightarrow  \{ x \in F  ,t_c \}$.
As a result, due to the fact that $T \rightarrow  \{ x \in F  ,t_c \} \cap A \rightarrow  \{ x \in F  ,t_c \}=S^{(q)}$ , we obtain the following equation:
$$Prob\left(T \rightarrow  \{ x \in F ,t_c \} \right)=Prob(S^{(q)})+Prob\left(T \rightarrow  \{ x \in F  ,t_c \} \cap \left(A \rightarrow  \{ x \in F  ,t_c \}\right)^C\right) \Rightarrow$$
\begin{equation} 
\Rightarrow Prob\left(T \rightarrow  \{ x \in F  ,t_c \} \right)=s+Prob\left(T \rightarrow  \{ x \in F  ,t_c \} \cap \left(A \rightarrow  \{ x \in F  ,t_c \}\right)^C\right)
\label{eq:B.1}
\end{equation}
where $s\in \mathbb{R}$ since $S^{(q)}\subset \Omega^{(q)}$. In addition, we also obtain:
$$Prob\left(A \rightarrow  \{ x \in F  ,t_c \} \right)=Prob(S^{(q)} )+Prob\left(A \rightarrow  \{ x \in F  ,t_c \} \cap \left(T \rightarrow  \{ x \in F  ,t_c \}\right)^C\right) \Rightarrow$$
\begin{equation} \label{eq:14}
\Rightarrow Prob\left(A \rightarrow  \{ x \in F  ,t_c \} \right)=s+Prob\left(A \rightarrow  \{ x \in F  ,t_c \} \cap \left(T \rightarrow  \{ x \in F  ,t_c \}\right)^C\right)
\end{equation}
We previously stated that the particle does not have a preferred direction in time, and thus, we can not have $Prob\left(T \rightarrow  \{ x \in F  ,t_c \} \cap \left(A \rightarrow  \{ x \in F  ,t_c \}\right)^C\right)$, $Prob\left(A \rightarrow  \{ x \in F  ,t_c \} \cap \left(T \rightarrow  \{ x \in F  ,t_c \}\right)^C\right)>0$. 

Now, there exists another scenario in which the above equations hold. And this is the obvious fact that $T \rightarrow  \{ x \in F  ,t_c \} \cap \left(A \rightarrow  \{ x \in F  ,t_c \}\right)^C = A \rightarrow  \{ x \in F  ,t_c \} \cap \left(T \rightarrow  \{ x \in F  ,t_c \}\right)^C=\varnothing$. But this can not be true since we would get $T \rightarrow  \{ x \in F  ,t_c \} = A \rightarrow  \{ x \in F  ,t_c \}=S^{(q)}$ and that can not hold for reasons that are analyzed in Appendix B.
In fact, $\Omega_h^{(q)}$ is our hyper-sample space and thus, the only empty set we can have in this context is  $\left(\Omega_h^{(q)}\right)^C= \varnothing$.

Consequently, since those second terms in the Eqs. (A1), (A2) can not obtain any real values, they must obtain complex values, and, as a result, we must get that $Prob\left(T \rightarrow  \{ x \in F  ,t_c \} \right)$, $Prob\left(A \rightarrow  \{ x \in F  ,t_c \} \right) \in \mathbb{C} $ also. 

Therefore, we conclude that if we are referring to the set $T \rightarrow  \{ x \in F  ,t_c \} \cap \left( A \rightarrow  \{ x \in F  ,t_c \}\right)^C$, we have to mention that it describes a particle exhibiting solely time-asymmetric motion, making the set $T \rightarrow  \{ x \in F  ,t_c \}$ and the probabilities associated with it non-real. That is the reason why the wave function plays the role of a complex probability distribution. The same analysis must be done for $A \rightarrow  \{ x \in F  ,t_c \}$ and the complex conjugate of the wave function. In general, we consider the aforementioned empty sets (according to conventional probability theory) as non-real instead of empty, and, as we said, the reason why we must do that is analyzed in Appendix B and more rigorously in O.

\section{}

Before reaching at the main point of this Appendix, we will repeat that our axiom for deriving QM is the fact that the quantum particle is subject to a reversible diffusion process. We say diffusion due to the uncertainty principle, which doesn't allow us to assign a specific velocity field of the particle at each point in space-time (see Appendix C, while Appendices L and Q are helpful too), as well as due to the probabilistic structure of the theory, which allows us to compute only the average values of physical quantities (i.e. position, energy etc.). Moreover, we say reversible for reasons that are analyzed in Appendices C and D.

So, why didn't we accept that $T \rightarrow  \{ x \in F  ,t_c \}=A \rightarrow  \{ x \in F  ,t_c \}$ in Appendix A? Because in that case, we would have that $Prob(T\rightarrow \{x \in F,t_c\})=\int_{F}\rho^{T}(x,t_c)dx$, where $\rho^{T}(x,t_c) \in \mathbb{R}$. The next thing we would naturally say in order to obtain stochasticity is that this quantum probability distribution will obey to a Fokker-Planck-type equation:
$$\frac{\partial \rho^{T}(x,t)}{\partial t}=-\overrightarrow{\nabla}\cdot (\rho^T(x,t)\overrightarrow b(x,t)) +D\overrightarrow{\nabla}^2\rho^{T}(x,t)$$
where $ \rho^{T}(x,t_c)$, $D \in \mathbb{R}$ and $\overrightarrow{b}(x,t)$ is the so-called drift velocity, which always carries the information about the existence of a deterministic external force field (however we will focus first on the case where such force field is absent in quantum diffusion and we will discuss on Appendix T how external forces enter into our picture). However, such an equation is by definition time-irreversible, meaning that reversing time on it will give $\rho^{A}(x,t_c) \neq \rho^{T}(x,t_c) $ which contradicts with our set equality. So, the other choice we have is to treat those sets as independent events that extend outside the real (product) sample space in order to avoid such contradictions, which eventually leads, in the no-external source case, to $\rho^{T}(x,t_c)$, $D \in \mathbb{C}$ and thus to Schroedinger's equation.

Let us use a toy example. Let $Z_1$, $Z_2$ be inseparable events (or "entangled" sets) from distinct experiments such that $Prob(Z_1)=z_1$ and $Prob(Z_2)=z_2$ in the product sample space. Such sets appear in our QM case. Clearly, conventional probability theory enables us to say that $Z_1=Z_2=Z_1 \cap Z_2=S$ and thus $Prob(Z_1)=Prob(Z_2)=z_1=z_2=s \in\mathbb{R}$. But, mathematically speaking, there is another way of description of reality ($S$) and this is by "artificially" treating them as independent despite the fact that \textit{in reality} they always co-occur. Such extension is valid mathematically because: We can consider $Prob(Z_1 \cap Z_2)=z_1z_2 =s\in \mathbb{R}$ since $Z_1 \cap Z_2=S \subset \Omega$ (space of real events), $Prob(Z_1 \cap Z_2^C)=\delta_1 \in \mathbb{C}$ and $Prob(Z_2 \cap Z_1^C)=\delta_2 \in \mathbb{C}$ since $Z_1 \cap Z_2^C$ and $Z_2 \cap Z_1^C$ non-real events. Now, for the pair of the two equations $z_1=z_1z_2+\delta_1$ and $z_2=z_1z_2+\delta_2$ to be valid - since $Prob(Z_1)=Prob((Z_1 \cap Z_2) \cup (Z_1 \cap Z_2^C))$ and 
$Prob(Z_2)=Prob((Z_2 \cap Z_1) \cup (Z_2 \cap Z_1^C))$- as well as for $s\in \mathbb{R}$, we must have $z_1=z$ and $z_2=z^*$, as well as $\delta_1=\delta_2^*$. In other words, we produced $S$ by "artificially" making the two events $Z_1$, $Z_2$ independent and the same logic applies in our quantum sets $T \rightarrow  \{ x \in F  ,t_c \} $ and $A \rightarrow  \{ x \in F  ,t_c \}$, by making them describe two independent motions after giving "birth" to the non-real sets $T \rightarrow  \{ x \in F  ,t_c \} \cap \left(A \rightarrow  \{ x \in F  ,t_c \}\right)^C$ and $A \rightarrow  \{ x \in F  ,t_c \} \cap \left(T \rightarrow  \{ x \in F  ,t_c \}\right)^C$. For a more rigorous analysis on the independence of those motions and complex nature of probabilities in QM see urgently Appendix O (for the origin of Schroedinger equation, see also L, S, V).

\textit{ Maybe, such extension, for describing classical world should not hold. However, it is definitely necessary for describing quantum diffusion processes, since non-extension of sets would unavoidably lead the quantum probability distribution to obey to an irreversible diffusion-type equation, while we want a reversible diffusion to take place (see Appendices below). And of course, predictions like wave-like nature, diffusion-type form of Schroedinger's equation, Born rule, interference and mathematical superposition arise naturally in such case.
}

\section{}

In driftless classical Brownian motion, the evolution of the probability distribution of particles suspended in a fluid with diffusion coefficient $D$ is governed by the equation: 

$$\frac{\partial \rho^{(cl)}(r,t \in [0,t_0])}{\partial t}=D\nabla^2\rho^{(cl)}(r,t \in [0,t_0])$$
Upon applying the transformation $t \rightarrow -t+t_0$, we can easily see that:
$$\frac{\partial \rho^{(cl)}(r,t'\in [0,t_0])}{\partial t} \neq D\nabla^2\rho^{(cl)}(r,t'\in[0,t_0])$$

From the above relations, we can observe that the diffusion equation lacks time-reversal invariance, as the reversed probability distribution doesn't satisfy the original diffusion equation.

On the other hand, in the Schroedinger equation, if we apply the transformation $t' \rightarrow -t+t_0$ and then take the complex conjugate of each side of the equation, we obtain the following equation:
$$\frac{\partial \Psi^*(r,t'\in [0,t_0])}{\partial t}=\frac{i\hbar}{2m_p}\nabla^2\Psi^*(r,t'\in [0,t_0])$$
which has the same form as the original. Thus, if $\Psi$ is related to forward diffusion, then $\Psi^*$ is related to diffusion evolving backwards in time.

However, the time-reversibility of quantum dynamics is evident not from the time invariance of the Schroedinger equation, but when applying the anti-linear operator $T$ that changes $i$ to $-i$ and $t$ to $-t$ (while also causes $\Psi(r,t)  \rightarrow \Psi^*(r,-t)$) to the operators of position and momentum. In this case, according to Ardakani \cite{ardakani2018time}, we get the following equations:
$$T \hat{p} T^{-1}=-\hat{p}$$
$$T \hat{r} T^{-1}=\hat{r}$$

These transformations, according to those who "believe" in quantum trajectories, could indicate that the time reversal of motion preserves the position of quantum particles but inverts the sign of their velocities (at least at a statistical level). This result is also obvious in classical Newton's equation, where time-reversibility is also a fundamental property, or in E/M systems, where the motion of charged particles also get reversed (including the charges that participate in the electric current, if it is involved in the system) under time reversal.

Reversibility is well-understood and it's logical to hypothesize its existence since quantum motion is a micro-level process (see Appendices D and K). But why should we have stochasticity in the first place (in Appendix B, we take stochasticity for granted by beginning from a diffusion equation, from which later Schroedinger equation emerges from set extension). And the answer to that is not due to the probabilistic structure of QM but due to the uncertainty principle, since in Furth's paper it is explicitly mentioned that $sup\{\Delta x^{(q)}\Delta u^{(q)}\}=\beta^2=\frac{\hbar}{2m_\rho}$ is completely analogous to $sup\{\Delta x^{(cl)}\Delta u^{(cl)}\}=D$ in classical diffusion.

\section{}

We must make a clarification and remark that time-reversibility doesn't appear mathematically at individual trajectories but appears in the "average trajectories" that are in fact the Bohmian trajectories. \textit{And that can be easily recognized if we look at the current velocity which is actually the local mean velocity of the process - that's how it is also defined in Bohm and Vigier's stochastic interpretation \cite{bohm1989non}, since uncertainty principle prevents us from assigning exact velocities at each space-time point}- and which appears in the following quantum Newton's equation:
$$\frac{d \overrightarrow{\eta}(r,t)}{d t}=-\frac{1}{m_p}\overrightarrow{\nabla}\left(V(r,t)+Q(r,t)\right)=\frac{ \overrightarrow{F}_{tot}(r,t)}{m_p}$$
This velocity is reversed if we perform the transformation $t'=-t+t_0$ and the same holds for the Newtonian classical velocity (of course under the assumption that $V(r,t)=V(r)$ in both classical and quantum case). More specifically, for two "mean (or Bohmian) trajectories", the one evolving from $0$ to $t_0$ and the other from $t_0$ to 0, we obtain:
$$\overrightarrow{\eta}^{0 \rightarrow t_0}\left(r,t\in [0,t_0] \right)=\frac{\hbar}{m_p}Im\left(\frac{\overrightarrow{\nabla}\Psi\left(r,t\in [0,t_0] \right)}{\Psi\left(r,t\in [0,t_0] \right)}\right)$$
and
$$\overrightarrow{\eta}^{t_0 \rightarrow 0}\left(r,t'\in [0,t_0] \right)=\frac{\hbar}{m_p}Im\left(\frac{\overrightarrow{\nabla}\Psi^*\left(r,t'\in [0,t_0] \right)}{\Psi^*\left(r,t'\in [0,t_0] \right)}\right)$$
from which we conclude that
$$\overrightarrow{\eta}^{t_0 \rightarrow 0}\left(r,t'=t_c \right)=-\overrightarrow{\eta}^{0 \rightarrow t_0}\left(r,t=t_c \right)$$
In Appendix E, without making reference to wave functions, it is shown that the probability distribution, in order to evolve symmetrically in time, must have the form:
$$\rho(r,t_c)=\rho^{T}(r,t=t_c)\rho^{A}(r,t'=t_c) = \Psi(r,t=t_c)\Psi^*(r,t'=t_c) \rightarrow |\Psi\left(r,t_c\right)|^2 $$

Those results imply to us that, if we reverse time, the ensemble of all possible quantum trajectories, on average, remains the same, but reversed. It should also be mentioned that the reversal of the current velocity and invariance of the probability distribution under time-reversal have also been stated in the context of stochastic mechanics by Guerra \cite{guerra1981structural}. Furthermore, there is also discussion by Callender \cite{callender2024insights} about the reversal of the current velocity in the context of Bohmian mechanics that confirms the existence of time symmetry in quantum systems. An analogous perspective was also proposed by Gao \cite{gao2022understanding}, who instead refers to the reversal of the probability current and time-invariance of probability distribution. It is suggested that those - or in other words the time-invariance of the continuity equation - are in fact responsible for the appearance of time-symmetry in QM and the complex conjugation of the wave function under time-reversal.
In the case that we have described, of course, the set $B\rightarrow\{ x \in F,t_c \}$ describes a particle whose each possible average trajectory is time-reversible but not each individual trajectory is reversible. And, of course, it makes sense to care about the "average behavior" of the particle, because the form and evolution of the probability distribution are determined by it (Bohmian trajectories have been experimentally observed as average trajectories \cite{kocsis2011observing}). \textit{On the other hand, someone could suggest that individual and not the ensemble path reversibility is expressed by the intersection process of Chapter III and this ensures reversibility at the micro-level (see more information on Appendices L and V) } and consequently at the ensemble level too.

Moreover, time-reversibility is also obvious due to the time-invariance of the equations of motion, i.e. the quantum Hamilton-Jacobi equation, the quantum Newton's equation or the quantum continuity equation (of course again in the case $V(r,t)=V(r)$). All the Bohmian equations which are the fundamental equations of motion in our theory (despite its stochastic character) are time-symmetric.

\section{}

 As we can see, the Born probability rule reminds us of the expression for the joint probability distribution, which is $\rho(x_1,t_c; x_2,t_c)=\rho^{T}(x_1, t_c)\rho^{A}(x_2,t_c)$ where $x_1$ independent from $x_2$. On the other hand, a strict mathematical condition $x_{backward}(t'=t_c)=x_{forward}(t=t_c)$ -as in QM- would of course demand that $T^{(q)} \rightarrow \{x \in F , t_c\}=A^{(q)} \rightarrow  \{x \in F, t_c\}$ in conventional probability theory. For that reason, we can not obtain a joint probability distribution, leading us to eventually get that $\rho(x, t_c)=\rho^{T}(x,t_c)=\rho^{A}(x,t_c)$.

So, in order to obtain a joint probability distribution that has two directions in time for a case that describes distinct entangled events (case where $x_{backward}(t'=t_c)=x_{forward}(t=t_c)$)  we must treat the forward and backward processes as independent which naturally gives rise to $T^{(q)} \rightarrow  \{ x \in F  ,t_c \} \neq  A^{(q)} \rightarrow  \{ x \in F  ,t_c \}$, while also to $T^{(q)} \rightarrow  \{ x \in F  ,t_c \}\cap A^{(q)} \rightarrow  \{ x \in F  ,t_c \}= S \neq \varnothing$ where $S\subset \Omega^{(q)}$ , As a consequence, we can now apply the multiplication rule, but our probability distributions will be complex due to the extension we made via inclusion of time-irreversible/non-real trajectories (see Appendices A, B) to ensure independence and non-equality between forward and backward sets. So, this leads to $\rho(x,t_c)=\rho^{T}(x_1= x,t_c)\rho^{A}(x_2 = x,t_c)$, where $\rho^{T}(x_1 = x,t_c)=\Psi(x,t_c)$ and $\rho^{A}(x_2= x,t_c)=\Psi^*(x,t_c)$, given that $\rho(x,t_c) \in \mathbb{R}$ or $T^{(q)} \rightarrow  \{ x \in F  ,t_c \} \cap  A^{(q)} \rightarrow  \{ x \in F  ,t_c \}=S \subset \Omega^{(q)}$, where $\Omega^{(q)}$ the space of real probabilities.

In more detail, in the classic case, where we have two independent motions-$x_1$ independent from $x_2$-we have that $T_1 \rightarrow \{x_1 \in F,t_c\}\cap T_2 \rightarrow \{x_2 \in F_2,t_c\}$, $T_1 \rightarrow \{x_1 \in F_1,t_c\}\cap (T_2 \rightarrow \{x_2 \in F_2,t_c\})^C$, $T_2 \rightarrow \{x_2 \in F_2,t_c\}\cap (T_1 \rightarrow \{x_1 \in F_1,t_c\})^C \neq \varnothing $, and in addition, $Prob(T_1 \rightarrow \{x_1 \in F_1,t_c\}\cap T_2 \rightarrow \{x_2 \in F_2,t_c\})=\int_{F_2}\int_{F_1}\rho_1(x_1,t_c)\rho_2(x_2,t_c)dx_1dx_2 \in \mathbb{R}$, which gives us that $\rho_{\text{joint}}(x_1,t_c ;x_2,t_c)=\rho_1(x_1,t_c)\rho_2(x_2,t_c)$. Moreover, $Prob(T_1 \rightarrow \{x_1 \in F_1,t_c\}\cap (T_2 \rightarrow \{x_2 \in F_2,t_c\})^C)=\int_{F_1}\rho_1(x_1,t_c)dx_1\left(1-\int_{F_2}\rho_2(x_2,t_c)dx_2\right)\in \mathbb{R}$ and $Prob(T_2 \rightarrow \{x_2 \in F_2,t_c\}\cap (T_1 \rightarrow \{x_1 \in F_1,t_c\})^C)=\int_{F_2}\rho_2(x_2,t_c)dx_2\left(1-\int_{F_1}\rho_1(x_1,t_c)dx_1\right)\in \mathbb{R}$. As a consequence, in this case that we study, we have $\rho_1(x_1,t_c)$, $\rho_2(x_2,t_c)\in \mathbb{R}$ (obvious).

In the quantum mechanical case, where we have (statistical) reversibility of trajectories, we must enforce the condition $x_{backward}(t'=t_c)=x_{forward}(t=t_c)$ (see Appendices C and D). As a consequence, we would have according to conventional probability theory that $T^{(q)} \rightarrow  \{ x \in F  ,t_c \} \cap \left(A^{(q)} \rightarrow  \{ x \in F  ,t_c \}\right)^C$, $A^{(q)} \rightarrow  \{ x \in F  ,t_c \} \cap \left(T^{(q)} \rightarrow  \{ x \in F  ,t_c \}\right)^C= \varnothing$. However, for reasons that are analyzed in Appendix B, we will have that $T^{(q)} \rightarrow  \{ x \in F  ,t_c \} \cap \left(A^{(q)} \rightarrow  \{ x \in F  ,t_c \}\right)^C$, $A^{(q)} \rightarrow  \{ x \in F  ,t_c \} \cap \left(T^{(q)} \rightarrow  \{ x \in F  ,t_c \}\right)^C \neq \varnothing$. However, those sets will describe non-real events. Therefore, $Prob(T^{(q)} \rightarrow  \{ x \in F  ,t_c \} \cap \left(A^{(q)} \rightarrow  \{ x \in F  ,t_c \}\right)^C)=\delta_1^{(q)} \in \mathbb{C}$ and $Prob(A^{(q)} \rightarrow  \{ x \in F  ,t_c \} \cap \left(T^{(q)} \rightarrow  \{ x \in F  ,t_c \}\right)^C)=\delta_2^{(q)} \in \mathbb{C}$. Moreover, $T^{(q)} \rightarrow  \{ x \in F  ,t_c \}\cap A^{(q)} \rightarrow  \{ x \in F  ,t_c \}=\left(T^{(q)} \rightarrow  \{ x_1 \in F  ,t_c \}\cap A^{(q)} \rightarrow  \{ x_2 \in F  ,t_c \}|x_1 \equiv x_2\right)$. Therefore, $Prob(T^{(q)} \rightarrow  \{ x \in F  ,t_c \}\cap A^{(q)} \rightarrow  \{ x \in F  ,t_c \})=Prob(T^{(q)} \rightarrow  \{ x_1 \in F  ,t_c \}\cap A^{(q)} \rightarrow  \{ x_2 \in F  ,t_c \}|x_1 \equiv x_2)=\left[\int_{F}\int_{F}\rho^T(x_1,t_c)\rho^A(x_2,t_c)dx_1dx_2\right]_{x_1 \equiv x_2}=\int_{F}\int_{F}\rho^T(x_1,t_c)\rho^A(x_2,t_c)\delta(x_1-x_2)dx_1dx_2=\int_{F}\rho^T(x,t_c)\rho^A(x,t_c)dx$. From that, we obtain that $\rho_{\text{joint}}^{\text{quantum}}(x,t_c)=\rho^T(x,t_c)\rho^A(x,t_c)$. The use of $\delta$-function is justified in Appendix G.

On the other hand, the equations 
$T^{(q)} \rightarrow  \{ x \in F  ,t_c \}=\left(T^{(q)} \rightarrow  \{ x \in F  ,t_c \} \cap A^{(q)} \rightarrow  \{ x \in F  ,t_c \}\right)\cup\left(T^{(q)} \rightarrow  \{ x \in F  ,t_c \} \cap \left(A^{(q)} \rightarrow  \{ x \in  F ,t_c \}\right)^C\right)$ and $A^{(q)} \rightarrow  \{ x \in F  ,t_c \}=\left(T^{(q)} \rightarrow  \{ x \in F  ,t_c \} \cap A^{(q)} \rightarrow  \{ x \in F  ,t_c \}\right)\cup\left(A^{(q)} \rightarrow  \{ x \in F  ,t_c \} \cap \left(T^{(q)} \rightarrow  \{ x \in  F ,t_c \}\right)^C\right)$ lead to $$\int_{F}\rho^{T}(x,t_c)dx=\int_{F}\rho^{T}(x,t_c)\rho^{A}(x,t_c)dx+\delta_1^{(q)}$$ and 
$$\int_{F}\rho^{A}(x,t_c)dx=\int_{F}\rho^{T}(x,t_c)\rho^{A}(x,t_c)dx+\delta_2^{(q)}$$

Analogously to Appendix B, the equations above can be valid for $\delta_1^{(q)}=(\delta_2^{(q)})^*$ and $\rho^T=(\rho^A)^*$ where $\rho^T,\rho^A \in \mathbb{C}$ due to the existence of the terms $\delta_1^{(q)}$ and $\delta_2^{(q)}$ (in fact $\rho^T=\Psi$ and $\rho^A=\Psi^*$). In other words, the reversibility of quantum trajectories enforces the consolidation of variables $x_1$ and $x_2$ into one variable $x$ in the quantum joint probability distribution expression and also is responsible for the complex valuedeness of $\rho^T$ and $\rho^A$ due to the "entanglement" of sets, which enables the extension of the sets outside of the real sample space (see Appendix M). For a more rigorous analysis see Appendices G, O.

\section{}

In this Appendix, we'll start from the fact that our quantum particle can not be in a physical superposition of energy states and derive from that the superposition principle without assuming Schroedinger equation (the reader should also read Appendix J for a more complete derivation). More specifically, $\forall $ $F\subseteq \mathbb{R}$ we have:
$$B_n \rightarrow  \{ x \in F  ,t_c \} \cap B_m \rightarrow  \{ x \in F  ,t_c \}=\varnothing \Rightarrow$$
$$\left\{T_n \rightarrow  \{ x \in F ,t_c \} \cap A_n\rightarrow  \{ x \in F  ,t_c \}\right\}\cap \left \{T_m \rightarrow  \{ x \in F  ,t_c \} \cap A_m \rightarrow  \{ x \in F  ,t_c \}\right\}=\varnothing \Rightarrow$$
$$\left\{T_n \rightarrow  \{ x \in F  ,t_c \} \cap T_m\rightarrow  \{ x \in F  ,t_c \}\right\}\cap \left \{A_n \rightarrow  \{ x \in F ,t_c \} \cap A_m \rightarrow  \{ x \in F  ,t_c \}\right\}=\varnothing \Rightarrow$$
$$T_{n,m} \rightarrow  \{ x \in F  ,t_c \}\cap A_{n,m} \rightarrow  \{ x \in F  ,t_c \}=\varnothing$$

Since $T_{n,m} \rightarrow  \{ x \in F  ,t_c \}$, $A_{n,m} \rightarrow  \{ x \in F  ,t_c \}$ describe independent events, that leads to $T_{n,m} \rightarrow  \{ x \in F  ,t_c \}=\varnothing$ or $A_{n,m} \rightarrow  \{ x \in F ,t_c \}=\varnothing$. It is obvious that both cases lead us to the superposition principle of the wave function. For the one case, $Prob(T \rightarrow  \{ x \in F  ,t_c \})=\sum_{n=1}^{\infty}Prob\left(T_n\rightarrow\{ x_{} \in F,t_c \}\right)=\sum_{n=1}^{\infty}Prob\left(T\rightarrow\{ x_{} \in F,t_c \}\cap T\rightarrow \textit{\{n-th state}\}\right)=\sum_{n=1}^{\infty}Prob\left(T\rightarrow\{ x_{} \in F,t_c \}|T\rightarrow \textit{\{n-th state}\}\right)Prob(T \rightarrow \{\textit{n-th state}\})$, which leads to 
$\int_{F}\Psi(x,t_c)dx=\sum_{n=1}^{\infty}c_n\int_{F}\Psi_n(x,t_c)dx=\int_{F}\sum_{n=1}^{\infty}c_n\Psi_n(x,t_c)dx$. And since that relation holds $\forall$ $ F \subseteq \mathbb{R}$ and since $\Psi$, $\Psi_n$ are continuous in $\mathbb{R}$, we obtain $\Psi(x,t_c)=\sum_{n=1}^{\infty}c_n\Psi_n(x,t_c)$.

\section{}

In this Appendix, we'll present a more rigorous derivation of the Born rule without relying on the delta function. More specifically, suppose we are given a non-normalized solution $\Psi(x,t_c)$ we derive the Born rule together with normalization in the following way:
$$Prob\left(B\rightarrow\{ x \in F,t_c \}\right)=Prob\left(T\rightarrow\{ x_1 \in F,t_c \} \cap A\rightarrow\{ x_2 \in F,t_c \}|x_1 \equiv x_2  \right)=$$
$$=\lim_{\epsilon \to 0} Prob\left(T\rightarrow\{ x_1 \in F,t_c \} \cap A\rightarrow\{ x_2 \in F,t_c \}||x_1-x_2|\leq \epsilon  \right)$$
$$=\lim_{\epsilon \to 0}\frac{Prob\left(T\rightarrow\{ x_1 \in F,t_c \} \cap A\rightarrow\{ x_2 \in F,t_c \}\cap |x_1-x_2|\leq\epsilon  \right)}{Prob(|x_1-x_2|\leq \epsilon)} $$
but since for the denominator there is no restriction for $x_1$, $x_2$ ($x_1$, $x_2$ $\in \mathbb{R)}$, the event implies that the forward \textit{and} backward moving  particles can be at whatever point in spacetime:
$$\lim_{\epsilon \to 0}\frac{Prob\left(T\rightarrow\{ x_1 \in F,t_c \} \cap A\rightarrow\{ x_2 \in F,t_c \}\cap |x_1-x_2|\leq\epsilon  \right)}{Prob\left(T\rightarrow\{ x_1 \in \mathbb{R},t_c \} \cap A\rightarrow\{ x_2 \in \mathbb{R},t_c \}\cap |x_1-x_2|\leq\epsilon  \right)}\equiv$$
$$\equiv \lim_{\epsilon \to 0}\frac{Prob\left(T\rightarrow\{ x_1 \in [x_2-\epsilon,x_2+\epsilon],t_c \} \cap A\rightarrow\{ x_2 \in F,t_c \}\ \right)}{Prob\left(T\rightarrow\{ x_1 \in [x_2-\epsilon,x_2+\epsilon],t_c \} \cap A\rightarrow\{ x_2 \in \mathbb{R},t_c \}  \right)}=$$
$$=\lim_{\epsilon \to 0}\frac{\int_{F}\int_{x_2-\epsilon}^{x_2+\epsilon}\Psi(x_1,t_c)\Psi^*\left(x_2,t_c\right)dx_1dx_2}{\int_{\mathbb{R}}\int_{x_2-\epsilon}^{x_2+\epsilon}\Psi(x_1,t_c)\Psi^*\left(x_2,t_c\right)dx_1dx_2}$$
$$=\frac{\lim_{\epsilon \to 0}\frac{1}{2\epsilon}\int_{F}\int_{x_2-\epsilon}^{x_2+\epsilon}\Psi(x_1,t_c)\Psi^*\left(x_2,t_c\right)dx_1dx_2}{\lim_{\epsilon \to 0}\frac{1}{2 \epsilon}\int_{\mathbb{R}}\int_{x_2-\epsilon}^{x_2+\epsilon}\Psi(x_1,t_c)\Psi^*\left(x_2,t_c\right)dx_1dx_2}$$
$$=\frac{\int_{F}|\Psi(x,t_c)|^2dx}{\int_{\mathbb{R}}|\Psi(x,t_c)|^2dx}$$
which arises due to the fact that $\lim_{\epsilon \to 0}\frac{1}{2\epsilon}\int_{x-\epsilon}^{x+\epsilon}f(y)dy=f(x)$ and corresponds to the Born rule (that derivation also aims to justify the use of delta function to enforce $x_1 \equiv x_2$ inside the integral). And, of course, the reasons behind the complex valuedeness of our probability distributions are analyzed in detail in Appendix E. In Appendix J, we will more rigorously explain using the same methodology why the condition $Prob(T_{n,m} \rightarrow  \{ x \in F  ,t_c \}\cap A_{n,m} \rightarrow  \{ x \in F  ,t_c \})=0$ of the previous Appendix leads to $Prob(T_{n,m} \rightarrow  \{ x \in F  ,t_c \})=0$ or  $Prob(A_{n,m} \rightarrow  \{ x \in F  ,t_c \})=0$. In Appendix O, we explain further the philosophy behind this whole limiting procedure.

\section{}
In order to have $\mu(\Omega_h^{(q)})=1$ (normalizability) simultaneously with $\Omega^{(q)}$, $T^{(q)} \rightarrow  \{ x \in \mathbb{R}  ,t_c \} \cap \left(A^{(q)} \rightarrow  \{ x \in \mathbb{R}  ,t_c \}\right)^c$, $A^{(q)} \rightarrow  \{ x \in \mathbb{R}  ,t_c \} \cap \left(T^{(q)} \rightarrow  \{ x \in \mathbb{R}  ,t_c \}\right)^c$ $\subset \Omega_h^{(q)}$, we define a non-real set $J=J_1 \cup J_2\neq \varnothing$, with $J_1 \cap J_2= \varnothing$, such that $\Omega^{(q)}\cup J=\Omega_h^{(q)}$ and $\Omega^{(q)}\cap J=\varnothing$. In such case, we will have $\mu(\Omega^{(q)})+\mu(J)=\mu(\Omega_h^{(q)})\Rightarrow\mu(J)=0\Rightarrow \mu(J_1)+\mu(J_2)=0$, that can be achieved if $\mu(J_1)$, $\mu(J_2)\in \mathbb{C}$, or in other words if we allow for complex measures to exist. Nevertheless, someone may say that the hyper-sample space should have necessarily measure equal to 1 since the probabilities of non-real events are complex and thus don't exceed 1 anyway and therefore there is no need for normalization. In such case, we should just say that $\Omega_h^{(q)}=T^{(q)} \rightarrow  \{ x \in \mathbb{R}  ,t_c \} \cup A^{(q)} \rightarrow  \{ x \in \mathbb{R}  ,t_c \}$ where $\mu(\Omega_h^{(q)})\neq 1$. However, I personally support the view that this hyperspace does have measure 1 (that is the actual sample space of Schroedinger forward and backward trajectories while the real sample space is emergent as a subset of it as shown in Appendices G, L, V) and this is actually addressed in beginning of Appendix T as well as more rigorously in Appendix V.

\section{}

When it comes to non-locality, our theory obviously must be non-local since it embraces hidden-variables \cite{bell1966problem}. All the appropriate analysis about the source of entanglement between quantum particles has already been done by Bohm in his original paper in Chapter 6 where he introduced the quantum potential in a system of $N$ particles with equal mass
$$Q(r_1,r_2 \dots r_N,t)=-\frac{\hbar^2}{2m_p}\sum_{i=1}^{N} \frac{\overrightarrow{\nabla_i}^2\sqrt{\rho(r_1,r_2 \dots r_N,t)}}{\sqrt{\rho(r_1,r_2 \dots r_N,t)}} $$
as a generator of quantum correlations between many particles. Indeed, in a system, there may not exist an interaction potential $V(r_1,r_2 \dots r_N,t)$, but still there is a non-local influence between entangled particles. It's true that those types of correlations have their roots in the past (when the particles had initially interacted with each other) and continue to exist even after we cease that interaction. Quantum potential may thus "carry" that information in many-particle systems. At least, this is what it does in single-particle systems where the evolution of free particle wave packets is totally determined by the classical preparation of the initial state before $t=0$. More specifically, in \cite{antonakos2024aharonov} there is an analysis by Antonakos (the author) and Terzis about the fact that quantum forces, under which the states evolve at all times $t>0$ in free space, seem to have the same form but opposite direction to the classical external force fields that we used in order to prepare our initial time-independent states. In other words, the quantum force has "memory" of the past interactions of our particle with previous force fields. Similarly here, once the particles interact, they preserve the information of their interaction - as a result they produce the so-called entanglement - and they do so via the quantum force, which from its nature is a non-local entity since the quantum force exerted on one particle depends on the position of all the other particles. Caution: we must remind the reader that the quantum force in our approach represents a mean force field exerted by the background field, not a force produced by a wave.

In this paper I provided, Airy and Gaussian free-particle wave packets are studied in Chapter 3 (Airy wave packets have non-dispersive properties and accelerate in free space \cite{balazs1979nonspreading}, but nevertheless quantum force and its past classical source have the answer for this "paradoxical" behavior). For a more intuitive explanation of entanglement see Appendix N.

\section{}

According to Appendix F, we could easily derive that $T_{n,m} \rightarrow  \{ x \in F  ,t_c \}=\varnothing$ or $A_{n,m} \rightarrow  \{ x \in F  ,t_c \}=\varnothing$ due to independence of forward and backward evolving random experiments. However, if we want a more detailed analysis, we could say that:
$$ Prob(T_{n,m} \rightarrow  \{ x \in F  ,t_c \}\cap A_{n,m} \rightarrow  \{ x \in F  ,t_c \} )=0\Rightarrow$$
$$\lim_{\epsilon \to 0} Prob\left(T_{n,m}\rightarrow\{ x_1 \in F,t_c \} \cap A_{n,m}\rightarrow\{ x_2 \in F,t_c \}||x_1-x_2|\leq \epsilon  \right)=0\Rightarrow$$
$$\lim_{\epsilon \to 0}\frac{Prob\left(T_{n,m}\rightarrow\{ x_1 \in F,t_c \} \cap A_{n,m}\rightarrow\{ x_2 \in F,t_c \}\cap |x_1-x_2|\leq\epsilon  \right)}{Prob(|x_1-x_2|\leq \epsilon)}=0\Rightarrow$$
$$\lim_{\epsilon \to 0}\frac{Prob\left(T_{n,m}\rightarrow\{ x_1 \in F,t_c \} \cap A_{n,m}\rightarrow\{ x_2 \in F,t_c \}\cap |x_1-x_2|\leq\epsilon  \right)}{Prob\left(T_{}\rightarrow\{ x_1 \in \mathbb{R},t_c \} \cap A_{}\rightarrow\{ x_2 \in \mathbb{R},t_c \}\cap |x_1-x_2|\leq\epsilon  \right)}=0 \Rightarrow$$
$$\lim_{\epsilon \to 0}\frac{\int_{F}\int_{x_2-\epsilon}^{x_2+\epsilon}\rho^T_{n,m}(x_1,t_c)\rho^A_{n,m}\left(x_2,t_c\right)dx_1dx_2}{\int_{\mathbb{R}}\int_{x_2-\epsilon}^{x_2+\epsilon}\Psi(x_1,t_c)\Psi^*_{}\left(x_2,t_c\right)dx_1dx_2}=0 \Rightarrow$$
$$\frac{\int_{F}|\rho^T_{n,m}(x,t_c)|^2dx}{\int_{\mathbb{R}}|\Psi(x,t_c)|^2dx}=0\Rightarrow$$
$$\rho^T_{n,m}(x,t_c)=0$$
$\forall F \subseteq \mathbb{R}$, which leads to 
$$Prob(T_{n,m} \rightarrow  \{ x \in F  ,t_c \})=\int_{F}\rho^T_{n,m}(x,t_c)dx=0$$
$\forall F \subseteq \mathbb{R}$ (in the same way, we can derive $Prob(A_{n,m} \rightarrow  \{ x \in F  ,t_c \})=0$). And from that, as we see in Appendix F, we easily derive the superposition principle. I must mention that the reason why $\rho^T_{n,m}(x,t_c)=(\rho^A_{n,m}(x,t_c))^*$ is justified again using trajectory (statistical) reversibility argument as was done in Appendices A, B, E.

\section{}
What we will note in this Appendix is that this sub-quantum field that is responsible for time-symmetric stochastic dynamics is not an environment with a very large number of massive molecules, where information from collisions are distributed across the medium. Also, some may say that in contrast to Nelson's theory, here we do not necessarily need to abandon dissipation to ensure reversibility (see next Appendix). Dissipation is about kinetic energy, which constantly converts into quantum potential energy (which is a constantly fluctuating entity) in a way that respects total local energy conservation and thus doesn't break reversibility. In that case, quantum potential is responsible for all quantum properties including entanglement generation and time-symmetric stochastic dynamics. For me though, instantaneous total energy is a fluctuating entity, but average total energy is not and is constant in time ensuring statistical reversibility (see Nelson) and all local stochastic deviations from local Bohm's energy are related mathematically to the quantum potential.

It must also be clarified that in Chapter III, our non-real sets correspond to forward and backward evolving motion, and thus we can write that $T \rightarrow \{ x \in F, t_c \} \equiv T \rightarrow \{ x \in F, t=t_c \}$ and $A \rightarrow \{ x \in F, t_c \} \equiv A \rightarrow \{ x \in F, t'=t_c \}$. As a result, we should write that $B \rightarrow \{ x \in F, t_c \} = T \rightarrow \{ x \in F, t=t_c \} \cap A \rightarrow \{ x \in F, t'=t_c \}$ -- the set $B \rightarrow \{ x \in F, t_c \}$ refers to the particle that moves in both directions in time - which leads consequently to $\rho(x,t_c)=\Psi(x,t=t_c)\Psi^*(x,t'=t_c)\rightarrow |\Psi(x,t_c)|^2$.
The set equation that describes the time-symmetric quantum particle comes from the proposition: $\forall$ $x\in F$, $\forall$ $t_c\in [0,t_0]$, we have that: 
\textit{\{particle is located at space-time point $(x,t_c)${ exhibiting time-symmetric motion}\} = \{particle is located at $(x,t_c)$ moving forward in time\} $\wedge$ \{particle is located at $(x,t_c)$ moving backwards in time\}}.

\textit{In this way, we obtain a quantum probability distribution $\rho(x,t_c)$ (as shown in Appendix G), which is given by the Born rule and its evolution has in fact both time directions. Those two results agree with standard QM.}

\section{}

I must say that the ideas presented about spin in this Appendix may not constitute the core part of the paper but nevertheless I think they could be considered as serious proposals pointing at a promising research direction.

We demonstrated in Appendix J that superposition of energy states can emerge by set-theoretic arguments without pre-assuming the Schroedinger equation and its linear structure. This argument denies physical superposition and leads us to a significant Hilbert space feature. However, confirming the idea that physical superposition is impossible does not by itself solve the reality problem. In standard QM, it is often mentioned that the state of the particle is indefinite before measurement and that the particle can neither be in all states simultaneously nor be in only one state. But what we will show is that the denial of such argument leads to the orthogonality condition for discrete energy states. Thus, suppose we are dealing with a two-energy state system whose state is given by the general form $\Psi(x,t_c)=c_n\Psi_n(x,t_c)+c_m\Psi_m(x,t_c)$ (the wave functions $\Psi_n(x,t_c), $ $\Psi_m(x,t_c)$ are always definite but there can be infinite pairs of $c_n,$ $c_m$).

By assuming that the particle's ontic state can not be indefinite and thus different from the superposed states, we are led to:
$$\mathbb{P}(B_n\rightarrow \{x \in \mathbb{R}, t_c \}\cup B_m\rightarrow \{x \in \mathbb{R}, t_c \})=1$$
and having in mind that $$B\rightarrow \{x \in \mathbb{R}, t_c \}=S_{nn}\cup S_{mm} \cup S_{nm} \cup S_{mn}$$ where
\begin{itemize}
    \item $S_{nn}=B_n\rightarrow \{x \in \mathbb{R}, t_c \}$
    \item $S_{mm}=B_m\rightarrow \{x \in \mathbb{R}, t_c \}$
    \item $S_{nm}=T_n\rightarrow \{x \in \mathbb{R}, t_c \}\cap A_m\rightarrow \{x \in \mathbb{R}, t_c \}$
    \item $S_{mn}=T_m\rightarrow \{x \in \mathbb{R}, t_c \}\cap A_n\rightarrow \{x \in \mathbb{R}, t_c \}$
\end{itemize}
 as well as that $B_n\rightarrow \{x \in \mathbb{R}, t_c \} \cap B_m\rightarrow \{x \in \mathbb{R}, t_c \}=\varnothing$, we are led to the equation 
 $$Re\left(c_nc_m^*\int_\mathbb{R}\Psi_n(x,t_c)\Psi_m^*(x,t_c)dx \right)=0$$
 which holds $\forall c_n$, $c_m \in \mathbb{C}$ such that $|c_n|^2+|c_m|^2=1$. Since we have freedom of choice of the coefficients we can first choose $c_n=\frac{1}{\sqrt{2}}$ and $c_m=\frac{i}{\sqrt{2}}$, which leads to $Im\left(\int_\mathbb{R}\Psi_n(x,t_c)\Psi_m^*(x,t_c)dx\right)=0$.  Secondly we can choose $c_n=\frac{1}{\sqrt{2}}$ and $c_m=\frac{1}{\sqrt{2}}$, which naturally leads to $Re\left(\int_\mathbb{R}\Psi_n(x,t_c)\Psi_m^*(x,t_c)dx\right)=0$. Those two results urge us to conclude that
 $\int_\mathbb{R}\Psi_n(x,t_c)\Psi_m^*(x,t_c)dx =0$
 
 Thus by rejecting two propositions of standard QM, we are led to Hilbert space structure for energy states. In other words, by accepting that particle is always in one state despite the interference effects gets us to Hilbert space picture. Of course, we could start the other way around and that is to start from Hilbert space properties (superposition and orthogonality equations) and derive that particle's ontic state in one of the two superposed states.

Now, since orthogonality and superposition are properties that ensure the definiteness of the particle's state for energy case, someone can reach to the conclusion that this definiteness can also be true for all observables, which I think is an asset comparing to Bohm's theory where only positions are real before measurement.leading to an asymmetric treatment of quantum observables. What we will do now is to extend the idea to electron spin states. The analysis will be much more difficult now, because our theory is a theory of diffusion in physical space and spin description requires abstract state vectors detached from position. In that sense, what is going to be presented is not a derivation of Hilbert space for spin but a mapping between Hilbert space tools/operations and extended probability theory.
We'll begin by mentioning for the product sets of the forward and backward particles that:
$$T \rightarrow \{x \in \mathbb{R},t_c;s_i\} \subset  T \rightarrow \{x \in \mathbb{R},t_c;\uparrow or \downarrow\} \equiv T \rightarrow \{x \in \mathbb{R},t_c\}XT \rightarrow\{ \uparrow or\downarrow\}$$
where $i=1,2$, $s_1=\uparrow$ and $s_2=\downarrow$, and similarly for the backward sets
$$A \rightarrow \{x \in \mathbb{R},t_c;s_i\} \subset  A \rightarrow \{x \in \mathbb{R},t_c;\uparrow or \downarrow\} \equiv A\rightarrow \{x \in \mathbb{R},t_c\}XA \rightarrow\{ \uparrow or\downarrow\}$$
But since we showed that $T \rightarrow \{x \in \mathbb{R},t_c\} $, $A \rightarrow \{x \in \mathbb{R},t_c\}$ non-real sets (they represent non-real events) that are essential for the description of reversible diffusion, we can easily conclude that $T \rightarrow \{x \in \mathbb{R},t_c;s_i\}$ and $A \rightarrow \{x \in \mathbb{R},t_c;s_i\}$ are non-real sets too since they contain the latter ones as parts of those.

Next, we write the equations  
$$T \rightarrow \{x \in \mathbb{R},t_c;\uparrow or \downarrow\}=T \rightarrow \{x \in \mathbb{R},t_c;\uparrow \}\cup T \rightarrow \{x \in \mathbb{R},t_c;\downarrow \}$$ and
$$A \rightarrow \{x \in \mathbb{R},t_c;\uparrow or \downarrow\}=A \rightarrow \{x \in \mathbb{R},t_c;\uparrow \}\cup A \rightarrow \{x \in \mathbb{R},t_c;\downarrow \}$$
Now, we will consider this 2-level spin system we talked about before for which the probability of having spin up is real and equal to $|c_1|^2$, while for spin down it is real and equal to $|c_2|^2$. We will attempt to explain linearity and derive the orthonormality equations by just using pure probability theory.

For this to be done we must first make a mapping between probability space and Hilbert vector space. According to that mapping, sets will correspond to vectors and the probability of intersection between two sets corresponds to the inner product between vectors. The latter proposal comes from the idea that since sets are equivalent to vectors, the measure of the overlap between sets (which is the probability of their intersection) will be equivalent to the measure of the overlap between vectors (which is their inner product). And what about vector addition? How is it translated in the probability set description? And the answer is that it is the union between sets (which is the "sum" operation between them) Therefore, we have the following correspondences/isomorphisms:
\begin{enumerate}
    \item $T \rightarrow \{x \in \mathbb{R},t_c;\uparrow or \downarrow\} \cong\omega_0$
    \item $A \rightarrow \{x \in \mathbb{R},t_c;\uparrow or \downarrow\} \cong \omega_0'$ 
    \item $T \rightarrow \{x \in \mathbb{R},t_c;\uparrow \} \cong \omega_1$ 
    \item $T \rightarrow \{x \in \mathbb{R},t_c;\downarrow \} \cong \omega_2$ 
    \item $A \rightarrow \{x \in \mathbb{R},t_c;\uparrow \} \cong \omega_1'$
    \item $A \rightarrow \{x \in \mathbb{R},t_c;\downarrow \} \cong\omega_2'$
    \item set union ($\cup$)$\cong$ vector sum ($+$)
    \item $\mathbb{P}(T \rightarrow \{x \in \mathbb{R},t_c;s_i\} \cap A \rightarrow \{x \in \mathbb{R},t_c;s_j\} )\cong \omega_i*\omega_j'$ where operation "$*$" represents the inner product. Also, $i,j=0,1,2$, $s_0=\uparrow or \downarrow$, $s_1=\uparrow$ and $s_2=\downarrow$
    
\end{enumerate}

In addition, since our sets are non-real sets, the vectors that correspond to them will be complex vectors which given the equations for $T \rightarrow \{x \in \mathbb{R},t_c;\uparrow or \downarrow\}$ and $A \rightarrow \{x \in \mathbb{R},t_c;\uparrow or \downarrow\}$ we wrote down before, as well as due to correspondences 1-7, we have that 
$$\omega_0=\omega_1+\omega_2$$ and 
$$\omega_0'=\omega_1'+\omega_2'$$
In other words, superposition emerges from the fact that forward and backward stochastic particles can have either spin up \textit{or} spin down, not both, and this is also true for the quantum particle.

Now, we should take into advantage of the fact that the quantum particle has spin up with probability $|c_1|^2$ which is equivalent to saying that it has spin up at whichever point in spacetime with probability $|c_1|^2$. Mathematically, we can write 
$$\mathbb{P}(B \rightarrow \{x \in \mathbb{R},t_c;\uparrow \})=|c_1|^2$$
which is the same as saying that
$$\mathbb{P}(T \rightarrow \{x \in \mathbb{R},t_c;\uparrow \}\cap A \rightarrow \{x \in \mathbb{R},t_c;\uparrow \} )=|c_1|^2$$
According to correspondence 8, this is equivalent to
$$\omega_1*\omega_1'=|c_1|^2$$
Similarly, we can derive 
$$\omega_2*\omega_2'=|c_2|^2$$

Given those two equations and since $T \rightarrow \{x \in \mathbb{R},t_c;s_i\} \rightleftarrows A \rightarrow \{x \in \mathbb{R},t_c;s_i\}$, where $\rightleftarrows$ represents the "entanglement" between sets (see Appendices A, B and E) we can say that $\omega_1'=\omega_1^*$ and $\omega_2'=\omega_2^*$ (see better justification below in this Appendix). And now, given those two findings and by using the argument that $B \rightarrow \{x \in \mathbb{R},t_c;\uparrow \}\cap B \rightarrow \{x \in \mathbb{R},t_c;\downarrow \} =\varnothing$ and that $\mathbb{P}(B \rightarrow \{x \in \mathbb{R},t_c;\uparrow \}\cup B \rightarrow \{x \in \mathbb{R},t_c;\downarrow \} )=1$, as in the energy states, we obtain
$$\omega_2*\omega_1'=0$$
and 
$$\omega_1*\omega_2'=0$$

Of course, this result emerges if we first define the following: $\omega_1=c_1\phi_1$, $\omega_2=c_2\phi_2$, $\omega_1'=c_1^*\phi^*_1$ and $\omega_2'=c_2^*\phi^*_2$, where $\phi^*_1*\phi_1=\phi^*_2*\phi_2=1$ so that both $\omega_i*\omega_i'=|c_i|^2$ and $\omega_i'=\omega_i^*$ ($i=1,2$) can be true. And now, it's easy to use the fact that the particle has a definite ontic state as well as that $\omega_i'=\omega_i^*$ to derive very easily (as before) the equation
$$Re(c_1c_2^*\phi_{1}*\phi^*_{2})=0$$
which by using the same method as in energy states (freedom of choice of complex coefficients) leads us to the orthogonality conditions.

Moreover, it is obvious that since
$$\mathbb{P}(B \rightarrow \{x \in \mathbb{R},t_c;\uparrow or \downarrow\})=1$$
which is equivalent to
$$\mathbb{P}(T \rightarrow \{x \in \mathbb{R},t_c;\uparrow or \downarrow\}\cap A \rightarrow \{x \in \mathbb{R},t_c;\uparrow or \downarrow\})=1$$
we derive that
$$\omega_0*\omega_0'=1$$
So, to summarize we have
\begin{itemize}
    \item $\omega_1*\omega_1'=|c_1|^2$ and $\omega_2*\omega_2'=|c_2|^2$
    \item $\omega_1*\omega_2'=\omega_2*\omega_1'=0$
    \item $\omega_0*\omega_0'=1$
    
\end{itemize}
From those relations and having into consideration the events/sets in which those vectors correspond, we reach to the conclusion that
\begin{itemize}
    \item $\omega_1\equiv c_1|  \uparrow \rangle$ and $\omega_2\equiv c_2|  \downarrow \rangle$
    \item $\omega_1'\equiv c_1^* \langle \uparrow |$ and $\omega_2'\equiv c_2^* \langle \uparrow |$
    \item $\omega_0\equiv|  \psi \rangle$ and  $\omega_0'\equiv \langle \psi |$
    
\end{itemize}
where each vector must obey to the equations $\langle \uparrow  |  \uparrow \rangle=\langle \downarrow  |  \downarrow \rangle=1$, $\langle \uparrow  |  \downarrow \rangle=\langle \downarrow  |  \uparrow \rangle=0$ and $\langle \psi |  \psi \rangle=1$. And of course, having already derived previously (from correspondences 1-7) the equations $\omega_0=\omega_1+\omega_2$ and $\omega_0'=\omega_1'+\omega_2'$, we can derive that $|  \psi \rangle=c_1|  \uparrow \rangle+c_2|  \downarrow \rangle$ and $\langle \psi  |=c_1^*\langle \uparrow  |+c_2^*\langle \downarrow  |$.

In conclusion, linearity emerges from the union of sets (which represents the "or" for the potentialities), complex structure from non-reality of sets (essential for reversible diffusion), orthonormality equations from the correspondence between inner product and probability of intersection of sets and from the idea that the particle occupies only one state. Moreover, the Born rule doesn't need a projection (or collapse) postulate but is taken from pure probability theory that describes probabilities independent of the act of measurement. And just a small note:\textit{Instead of measure of overlap, for a more conceptual understanding, we could say that both inner product and set intersection refer to measures of "commonness" or similarity between the two vectors/sets. On the one hand, inner product says how much two vectors resemble one another when it comes to direction and probability of intersection of sets says how much two sets resemble one another when it comes to their representation on the Venn diagram. Both resemblances have to do with geometrical representation}

But, I will say it again. What I provided is not a derivation of Hilbert space for spin but a mapping between (specific) operations in probability set theory and operations in Hilbert space. Also, I must add that my analysis can be extended to whichever basis we choose beyond spin. This means that all observables are definite before measurement. However, nowhere in this model non-contextuality is assumed and that means that Kochen-Spekker theorem \cite{kochen2011problem} does not apply to us. In other words, all observables, including position, are "contextual beables".

It must be emphasized that there are alternative explanations for why $\omega_1'=\omega_1^*$ beyond the "entangled" set explanation.
For instance, someone can think of it this way: since the stochastic process is reversible (forward and backward processes must be treated symmetrically), it is logical to say that if $\omega_1'=f(\omega_1)$, then it would not be reasonable to have $\omega_1=g(\omega_1') \neq f(\omega_1')$ (there must be a bidirectional relation between the two vectors). So, in order to have the equation $\omega_1=f(\omega_1')=f(f(\omega_1))$ as well as $\omega_1'*\omega_1=f(\omega_1)*\omega_1 \in \mathbb{R}$, we must have $f(z)=z^*$ where $z$ is a complex vector.

Alternatively, we can say that if $T\rightarrow \{x\in \mathbb{R},t_c ;\uparrow\} \rightarrow \omega_1$ then by reversing time we get $A\rightarrow \{x\in \mathbb{R},t_c,\uparrow \} \rightarrow \omega_1'=f(\omega_1)$, By reversing time again  we are led to the expression $T\rightarrow \{x\in \mathbb{R},t_c;\uparrow \} \rightarrow f(\omega_1')=f(f(\omega_1))$. However, we already know from the hypothesis that $T\rightarrow \{x\in \mathbb{R},t_c;\uparrow \} \rightarrow \omega_1$ which consequently leads to $f(f(\omega_1))=\omega_1$, which in combination with $\omega_1'*\omega_1=f(\omega_1)*\omega_1 \in \mathbb{R}$, leads us to conclude that $f(z)=z^*$ and thus that $\omega_1'=\omega_1^*$.

But why should we make the mapping between sets and vectors? Why do we need vectors? Why can't we just talk about the probabilities of $T\rightarrow \{x\in \mathbb{R},t_c ;s_i\} $ and $A\rightarrow \{x\in \mathbb{R},t_c ;s_i\}$ (where $i=1,2$)? Because if we did that, we would get into contradictions.  Suppose those sets were well-defined and specific numbers and we tried to assign probabilities to them, these would be given by the equations $\mathbb{P}(T\rightarrow \{x\in \mathbb{R},t_c ;s_i\})=c_i\int_{\mathbb{R}}\rho^T(x,t_c|s_i)dx$ and $\mathbb{P}(A\rightarrow \{x\in \mathbb{R},t_c ;s_j\})=c_j^*\int_{\mathbb{R}}\rho^A(x,t_c|s_j)dx$.
For the particle to be in a definite ontic spin state which is one of the two states $s_i$ and $s_j$, we must have $$\mathbb{P}(T\rightarrow \{x\in \mathbb{R},t_c ;s_i\}\cap A\rightarrow \{x\in \mathbb{R},t_c ;s_j\})=c_ic_j^*\int_{\mathbb{R}}\rho^T(x,t_c|s_i)\rho^A(x,t_c|s_j)dx=0$$ for $i \neq j$. For this equation to hold, we must definitely have $\rho^T(x,t_c|s_i) \neq (\rho^A(x,t_c|s_j))^*$ or equivalently $\rho^T(x,t_c|s_i) \neq \rho^T(x,t_c|s_j)$. However, this is impossible because the position wave function (and position distribution) of an electron with spin up is the same with the electron that has spin down. More briefly, position distribution is completely irrelevant to the spin of the particle. As a consequence, an analysis using probabilities for those events can not be valid and thus switching to vectors through our mapping is completely unavoidable, especially if those discrete states are not energy states (energy is related to diffusion and thus to position). It must be noted though that the solution to the ill-defined probabilities is given in Appendix V (subsection d).

It must be noted that the union between two sets can be considered equivalent to their "sum" only when the sets are disjoint. Then and only then, can we use the correspondence 7. In our case, we have that
$$B\rightarrow \{x\in \mathbb{R},t_c;\uparrow and \downarrow \}=\varnothing \Rightarrow$$
$$\Rightarrow T\rightarrow \{x\in \mathbb{R},t_c;\uparrow and \downarrow \}\cap A\rightarrow \{x\in \mathbb{R},t_c;\uparrow and \downarrow \}=\varnothing $$
which implies that $T\rightarrow \{x\in \mathbb{R},t_c;\uparrow  \}\cap T\rightarrow \{x\in \mathbb{R},t_c;\downarrow  \}=\varnothing$ or $A\rightarrow \{x\in \mathbb{R},t_c;\uparrow  \}\cap A\rightarrow \{x\in \mathbb{R},t_c;\downarrow  \}=\varnothing$ (we will assume the first one) since T and A sets describe distinct random experiments. Due to this mutual exclusivity of T sets we are allowed to use correspondence 7 and thus derive vector superposition in a safe way. And of course, by complex conjugating we can also get that $\omega_0^*=\omega_1^*+\omega_2^*$. A more detailed derivation of the mutually exclusive relation using propositional logic is done in the next Appendix.

I must add that in this Appendix, it is worthwhile examining alternative ideas. For instance, we restricted our analysis at the ensemble level to explain how statistical reversibility is responsible for Born rule through the intersection process of the ensemble trajectories (for more detailed mathematical analysis that is very similar to the one developed below, see Appendix S). However, we will now see how our reversibility discussion can still be valid at the individual path level.

Undoubtedly, quantum systems are described by a Heisenberg uncertainty principle and by an evolution that is reversible in time. Therefore, it is logical to assume in every case that quantum particles undergo a reversible diffusion process. This axiom does not change. However, the intersection process from which the Born rule emerges may not emerge from the statistical reversibility, but from individual path reversibility which exists anyway. Recall that even in classical diffusion, the microscopic laws that describe the individual motion of the particle are reversible despite the statistical/macroscopic irreversibility. The only difference here is that we want in the quantum case the preservation of statistical reversibility (see Appendix K). Below, we will demonstrate how the language of set intersection can emerge from an analysis of SDEs.

Before that, we will refer to the classical pure diffusion which was discussed in Appendix C. The SDE (at the mesoscopic level) that describes this Brownian particle\textit{ emerges from time-symmetric microscopic Newtonian/Hamiltonian dynamics} \cite{zwanzig1973nonlinear} (despite the fact that the SDE is irreversible due to coarse-graining) and it is the Langevin equation \cite{pavliotis2014langevin}:
$$\frac{dx^{(cl)}(t)}{dt}=\sqrt{2D^{(cl)}}\gamma(t)$$
where $\gamma(t)$ is a Gaussian white noise which has mean $0$ and is uncorrelated in time:
\begin{itemize}
    \item $<\gamma(t)>=0$
    \item $<\gamma(t_1)\gamma(t_2)>=\delta(t_1-t_2)$
\end{itemize}

Now, we'll talk about the stochastic process that describes the forward diffusion given by the forward Schroedinger's equation. The Schroedinger-Langevin equation is the following:
$$\frac{dx(t)}{dt}=\sqrt{2D}\gamma(t)=\sqrt{\frac{i\hbar}{m_\rho}}\gamma(t)$$
where $D=\frac{i\hbar}{2m_\rho}$ the Schroedinger diffusion coefficient and the noise has the same properties as in the classical case (the only difference in Schroedinger diffusion is the diffusion coefficient that is imaginary). Of course, this equation can take the following equivalent form:
$$\frac{dx(t)}{dt}=\sqrt{\frac{\hbar}{m_\rho}}\gamma_{sys}(t)$$
with complex noise $\gamma_{sys}(t)=\sqrt{i}\gamma(t)$ and diffusion constant equal to $D_{sys}=\frac{\hbar}{2m_\rho}$.

Before moving on, we need to mention that according to our set theory, the particle that obeys to Schroedinger's diffusion dynamics experiences two "parallel mathematical realities" at the same time, one in which it follows reversible trajectories (real microscopic trajectories) and one in which it follows irreversible trajectories (non-real microscopic trajectories). From our set-theoretic description, we have $T \rightarrow  \{ x \in F  ,t_c \}=B \rightarrow  \{ x \in F  ,t_c \} \cup\left(T \rightarrow  \{ x \in F  ,t_c \} \cap \left(A \rightarrow  \{ x \in  F ,t_c \}\right)^C\right) $, which means that: \textit{\{the forward moving particle is located at space-time point $(x,t_c)\}$ $=$ \{the time-symmetric forward moving particle is located at space-time point $(x,t_c)\}$$\lor$\{ the time-asymmetric forward moving particle is located at space-time point $(x,t_c)\}$.}

 If we take into account that $\sqrt{i}=\frac{1+i}{\sqrt{2}}$, we can easily recognize that the equation can take the form:
$$\frac{dx(t)}{dt}=\frac{dx_{rev}(t)}{dt}+\frac{dx_{irr}(t)}{dt}$$
where $x_{rev}(t)$ is the stochastic process that participates in the real Langevin equation:
$$\frac{dx_{rev}(t)}{dt}=\sqrt{\frac{\hbar}{m_\rho}}\gamma_{Re}(t)$$ with real noise $\gamma_{Re}(t)=\frac{\gamma(t)}{\sqrt{2}}$ and $D_1=\frac{\hbar}{2m_\rho}$ and $x_{irr}(t)$ is a different (non-real) stochastic process that participates in the imaginary Langevin equation:
$$\frac{dx_{irr}(t)}{dt}=\sqrt{\frac{\hbar}{m_\rho}}\gamma_{Im}(t)$$ 
with imaginary noise $\gamma_{Im}(t)=i\frac{\gamma(t)}{\sqrt{2}}$ and $D_2=\frac{\hbar}{2m_\rho}$. The equation for $\frac{dx(t)}{dt}$ is logically consistent given that Schroedinger trajectories, as we've demonstrated in our set-theoretic analysis must be given by the equation $x(t)=x_{rev}(t)+x_{irr}(t)$, since the particle is in a "superposition" of two stochastic processes (one is real and the other non-real). What we will show now is that indeed $x_{irr}(t)$ is emergent from coarse-graining over non-real processes that are in fact irreversible. It's odd to mention that $x_{rev}(t)$ indeed corresponds to  stochastic trajectories that always emerge from reversible Newtonian dynamics, as we said in the classical diffusion case.

In order to continue, it is wise to express each $\frac{dx}{dt}$ in terms of instantaneous velocities. An effective way of description is by saying that $\frac{dx}{dt}$ may not be an instantaneous velocity but can be approximately considered as the result of "averaging" the actual behavior (averaging the actual instantaneous velocities) of the particle at very small time steps $\Delta t=\epsilon<<1$, which also don't approach zero. In that sense, it is wise to suggest that effectively:
$$\frac{dx_{rev}(t)}{dt}=\bar{u}_{rev, \epsilon}(t)=\sum_{i=1}^{n}\bar{u}^{(i)}_{rev,\epsilon}(t)\chi_{G_i}(t)$$
Similarly
$$\frac{dx_{irr}(t)}{dt}=\bar{u}_{irr, \epsilon}(t)=\sum_{i=1}^{n}\bar{u}^{(i)}_{irr,\epsilon}(t)\chi_{G_i}(t)$$
where $t \in G=\{t_1, t_2\dots t_{n-1},t_n\}$, $n>>1$ and $ t_{i+1}-t_i=\Delta t=\epsilon$. Also, $\chi$ is the characteristic function and $G_i=\{t_i\}$. Previously, we had indirectly derived the relation $$\frac{dx_{irr}(t)}{dt}=i\frac{dx_{rev}(t)}{dt}$$
that must hold $\forall t\in G$. According to the previous effective descriptions, we are obviously led to $$\bar{u}_{irr, \epsilon}(t)=i\bar{u}_{rev, \epsilon}(t)$$
$\forall t\in G$. In addition, we've already mentioned that real Langevin dynamics with Gaussian white noise (despite the fact that the noise is $0.71$ times the strength of $\gamma(t)$) emerges from time-reversible Newtonian dynamics which means that $\forall t\in[t_1-\epsilon/2,t_n+\epsilon/2]$: 
$$u_{rev,in}(-t)=-u_{rev,in}(t)$$ where $u_{rev,in}(t)$ is the instantaneous Newtonian velocity. This implies that $\forall t \in G$:
$$\bar{u}_{rev,\epsilon}(-t)=-\bar{u}_{rev,\epsilon}(t)$$ 

Now, if we apply time-reversal operation in the equation that connects the real and non-real time-averaged velocities and we accept the fact that $\bar{u}_{irr,\epsilon}(-t)=-\bar{u}_{irr,\epsilon}(t) $, which is a necessary condition for reversibility of trajectories, we arrive at an equation that contradicts the previous equation of non-real motion (the equation for the imaginary time-averaged velocity). Or alternatively, by applying time-reversal, we observe that $\bar{u}_{rev,\epsilon}(-t)$ is not a solution to the same equation of motion. In other words, $\bar{u}_{irr,\epsilon}(-t)=-i \bar{u}_{rev,\epsilon}(-t)\neq i \bar{u}_{rev,\epsilon}(-t)$ and that happens because of the transformation $i\rightarrow -i$ under time reversal.
From all this information, someone can immediately recognize that the non-real trajectories that are described by the imaginary velocity $u_{irr,in}(t)$ are indeed irreversible.

And just to have a small intuition on what we've done: assume $S[x_{Schr,in}(t)]=S[x_{rev,in}(t)]\cup S[x_{irr,in}(t)]$ is the set of all microscopic forward trajectories and $S[x_{Schr,in}(t')]=S[x_{rev,in}(t')]\cup S[x_{irr,in}(t')]$ is the set of all microscopic backward trajectories then $S[x_{real,in}]\equiv S[x_{rev,in}(t)] \equiv S[x_{rev,in}(t')]=S[x_{Schr,in}(t)]\cap S[x_{Schr,in}(t')]$. Of course, the forward and backward irreversible trajectories are completely incompatible with each other and with the reversible trajectories meaning that the measure of their overlap is zero: $$\mu\left(S[x_{irr,in}(t')]\cap S[x_{irr,in}(t)]\right)=0$$ 
$$\mu\left(S[x_{rev,in}(t')]\cap S[x_{irr,in}(t)]\right)=0 $$
$$\mu \left(S[x_{irr,in}(t')]\cap S[x_{rev,in}(t)]\right)=0$$

The above equations are also compatible with the fact that in our set-theoretic framework (Appendix A), the non-real sets are mutually exclusive with each other as well as with the real sets, which is extremely easy to prove by just using the set identity $\Gamma \cap \Gamma^c=\varnothing$.

In other words, by taking the intersection of forward and backward actual trajectories (we're talking about the two pairs of reversible and irreversible trajectories from which our Langevin equations emerge through coarse-graining), someone obtains the reversible trajectories which are the real quantum trajectories and this intersection mathematically leads to the intersection of sets that describe those trajectories from which Born rule emerges (see Appendices K, G and Chapter III). Also, we have to mention again that since $\gamma_{Re}(t)$ is real and proportional to $\gamma(t)$, it will have the same statistical properties with it (see bullets above), despite the fact that its strength is $0.71$ times the strength of the noise $\gamma(t)$. This means that Markovianity for the quantum particle (real diffusion) is maintained. The resulting stochastic process will be given by the Langevin equation:
$$\frac{dx^{(q)}(t)}{dt}=\sqrt{2D^{(q)}}\gamma_{Re}(t)=\sqrt{\frac{\hbar}{m_p}}\gamma_{Re}(t) \equiv \frac{dx_{rev}(t)}{dt} $$
 where the quantum diffusion coefficient is $\frac{\hbar}{2m_\rho}$, which is in agreement with Furth's results (see Appendix C).

  Finally, just some small notes: in our model, we could instead use the center of mass in our discussion. So, we could instead consider the total stochastic process equal to $x(t)=x_{com}(t)=\frac{x'_{rev}(t)+x'_{irr}(t)}{2}=\frac{x_1(t)+x_2(t)}{2}$ instead of a pure sum. We should always remember that the particle exists in a "spatial mathematical superposition" following two different trajectories and thus the total stochastic process of the system of particles should be considered equivalent to the stochastic process of the center of mass. If we have two independent Wiener process (let's use Wiener processes for simplicity) for two different particles of mass $m_p$ for which we have that $dx_1(t)=\sigma_1 dw_1(t)$ and $dx_2(t)=\sigma_2 dw_2(t)$, we can easily show that $dx_{com}(t)=\sigma dw(t)=\frac{1}{2}(\sigma_1dw_1(t)+\sigma_2 dw_2(t))$ where $\sigma=\frac{1}{2}\sqrt{\sigma_1^2+\sigma_2^2}$ (see also Appendix Q), In our quantum case, it is proven that the aforementioned equations emerge from Schroedinger's diffusion analysis with $\sigma=\sqrt{2D_{sys}}$ and $\sigma_{1,2}=\sqrt{2D_{1,2}}=\sigma \sqrt{2}$, where $D_{1,2}=2D_{sys}=\frac{\hbar}{m_p}$ (the 2 appears only in Schroedinger's diffusion constant denominator since the system has total mass $m_{sys}=2m_p$). In addition, someone may be fooled by the fact that $dw_2(t)=idw_1(t)$ and talk about dependence between the two Wiener processes due to the "proportionality" relation, but that is not true because $i$ is imaginary and is used to describe a Wiener process $w_2(t)$ that is not related to $w_1(t)$ at all and in fact describes irreversible pseudo-trajectories as demonstrated in the previous discussion. Thus, these trajectories are not just the same trajectories but just with different speeds, as in the case where the proportionality constant is a real number. Also, in order to talk about time-averaged velocities that we discussed before, we should better work with the discrete-time model which can be derived from the continuous-time overdamped Langevin equation and according to \cite{pavliotis2014langevin} has the general form:
$$x_{n+1}-x_n=\sqrt{2D\Delta t}\gamma_n$$
where \textit{timestep size}$=\Delta t<<1$, but nevertheless finite. In this case, we can more safely say that 
$$ \bar{u}(t_n\rightarrow t_{n+1})=u_n=\frac{x_{n+1}-x_n}{\Delta t}=\sqrt{\frac{2D}{\Delta t}}\gamma_n$$
and make all the discussion we made before on proving the irreversibility of non-real motion. Nevertheless, we used the previous formalism just for practical convenience.

And now, from our discrete-time quantum SDE we can more easily see how classical behavior emerges for large enough masses, since $m_p \rightarrow \infty$ leads to zero diffusion constant and thus to zero time-averaged stochastic velocity. In this way, stochasticity is lost and classical behavior emerges naturally.

 Of course, all this analysis using non-real sets and non-real trajectories is done in order to maintain statistical reversibility (see Appendices B, K), which is a property that classical diffusion fails to have. Thus, someone may say that all this abstract formalism is a description of something much more simple. And establishing the existence of this sub-quantum field through our stochastic formalism, helps us understand why quantum nature appears wave-like from a physical perspective. More specifically, if our sub-quantum field is a massless entity that carries energy and momentum (and not matter) then we are automatically led into the definition of the wave, which is defined as the disturbance in which energy, momentum but not matter are transferred. Of course, that field is not a classical wave, but it probably has wave-like properties due to its massless structure and it undergoes fluctuations all the time leading to both Brownian and wave-like behavior and that's why we have wave-like solutions to a diffusion equation that eventually produce interference effects, despite the fact that $\Psi \neq wave$. Despite all this "classical-like" analysis, I think the situation can be much different and this is due to non-locality that prevents such local views of reality (see Appendix R).

\section{}

The reason why the whole "area" of non-real sets is extended outside the real sample space is something that can be mathematically demonstrated. It's enough to prove that $\left(T \rightarrow  \{ x \in F  ,t_c \} \cap \left(A \rightarrow  \{ x \in  F ,t_c \}\right)^C\right) \cap \Omega^{(q)}=\varnothing$. Or, in propositional logic terms, we have to show that $\forall F\subseteq\mathbb{R}$:
$$\{x_1=x\in F;t_c\}\land \{\neg\{x_2= x \in F;t_c\}\}\land\{x_1=x \in \mathbb{R} ;t_c\} \land \{x_2=x \in \mathbb{R} ;t_c\}=false$$
(where $\neg$ symbol expresses negation) which is a shorthand for the propositional expression:
$$\left\{\lim_{\epsilon \to 0}\left\{\{x_1\in F;t_c\}\land \{\neg\{x_2 \in F;t_c\}\}\land\{x_1 \in \mathbb{R} ;t_c\} \land \{x_2 \in \mathbb{R} ;t_c\}\Big ||x_1-x_2| \leq \epsilon\right\}\right\}=false$$
We already know that
$$\{x_i=x \in \mathbb{R} ;t_c\}=\{x_i=x \in F ;t_c\}\lor \{x_i=x \in F^c ;t_c\} $$
where $i=1,2$, $F \cap F^c=\varnothing $ and $F \cup F^c=\mathbb{R} $. Now, given this and due to the fact that $\{x_i=x \in F ;t_c\}\land \{x_j=x \in F^c ;t_c\}=false$ for $i,j=1,2$ (since $\{x\in F\}\land \{x \in F^c\}=false$), as well as because of the logical identity that $\{\neg\{x_2= x \in F;t_c\}\}\land \{x_2= x \in F;t_c\}=false$, we finally obtain the result that we want and we prove that indeed non-real sets and real sample space are completely disjoint.

The use of propositional logic is also useful for the derivation of $T \rightarrow \{x \in \mathbb{R},t_c; \uparrow\}\cap T \rightarrow \{x \in \mathbb{R},t_c; \downarrow\} = \varnothing$ in Appendix L. Let $q.p. \equiv \textit{quantum particle}$, $f.p.=\textit{forward particle}$ and $b.p.\equiv \textit{backward particle}$. 
To proceed, we know that $B \rightarrow \{x \in \mathbb{R},t_c; s_1 \,and\, s_2 \}= \varnothing$ which in proposition terms means that $\{q.p. \rightarrow (x\in \mathbb{R},t_c); s_1 \, and \, s_2\}=false \Rightarrow \{f.p. \rightarrow (x\in \mathbb{R},t_c); s_1 \,and \, s_2\} \land \{b.p. \rightarrow (x\in \mathbb{R},t_c); s_1 \, and \, s_2\}=false \Rightarrow \{f.p. \rightarrow (x\in \mathbb{R},t_c)\} \land \{b.p. \rightarrow (x\in \mathbb{R},t_c) \}\land \{f.p. \rightarrow s_1 \, and \, s_2\}\land \{b.p. \rightarrow s_1 \,and\, s_2\}=false$. Since $\{f.p. \rightarrow (x\in \mathbb{R},t_c)\} \land \{b.p. \rightarrow (x\in \mathbb{R},t_c) \}=\{q.p. \rightarrow (x\in \mathbb{R},t_c) \}=true$, the latter expression leads us to $\{f.p. \rightarrow s_1 \,and\, s_2\}=false$ or $\{b.p. \rightarrow s_1 \,and\, s_2\}=false$. Assuming the first, it follows that $\{f.p. \rightarrow (x\in \mathbb{R},t_c); s_1 \,and\, s_2\}= \{f.p. \rightarrow (x\in \mathbb{R},t_c)\}\land \{f.p. \rightarrow s_1 \,and\, s_2\}=false$, which in set-theoretic language is translated into $T \rightarrow \{x \in \mathbb{R},t_c; \uparrow and \downarrow\}= \varnothing$ and thus to $T \rightarrow \{x \in \mathbb{R},t_c; \uparrow\}\cap T \rightarrow \{x \in \mathbb{R},t_c; \downarrow\}= \varnothing$

\section{}

For a detailed explanation of quantum entanglement, we will certainly do a future work where the topic will be explored thoroughly. For now, to understand better the stochastic origin of quantum entanglement, we will mention a classical example which will be able to provide a picture of why information about past interactions/forces can remain in a stochastic classical system even after the abolishment of those. 

Let's first consider a classical particle that is completely isolated. We decide to apply an external force to it and after some time we stop applying that force. Now, there are no forces exerted on this particle. Now, let's consider a classical particle surrounded by many other particles. All of those particles are initially at rest. At some time, I choose to apply a force on that one particle and it starts colliding with all the other particles. If I remove the external influence, there will still be forces exerted upon that one particle. Imagine that in the quantum case, this one particle is the quantum particle and all the other particles are the sub-quantum field. Also, the other forces that arise correspond (at the ensemble level) to the quantum force. And of course, as we said in Appendix I, if the previous external classical force in a quantum system is a multi-particle interaction force, the information about it is stored in the system via the quantum force. This quantum information of course will evolve in time but it is still there leading the particles to "remember" that they were at some time "connected" with one another (see more information about entanglement in Appendix P).

\section{}

We must remind the reader that the Born rule emerges from two independent forward and backward motions living in two distinct sample spaces. The sample space that describes both motions will be a product sample space. The probability distribution for this two-directional process will be a joint probability distribution that has both directions in time. Since our probability theory is in fact classical despite its non-conventional structure (see Appendix P), this suggests that the probabilities for such motion would be taken by a probability of the form $Prob_{T \cap A}=Prob(T\rightarrow\{x_1 \in F_1,t_c\}\cap A\rightarrow \{x_2\in F_2,t_c\})$. What would classical probability theory suggest is that we can be restricted to a subset of those trajectories if we take a condition of the type $C_{\epsilon}=\{|x_1-x_2|\leq \epsilon\}$. The probabilities that describe the trajectories that respect $C_\epsilon$ will be given by a general form of the type $Prob_{T \cap A}(\epsilon)=Prob(T\rightarrow\{x_1 \in F_1,t_c\}\cap A\rightarrow \{x_2\in F_2,t_c\}|C_\epsilon)$ where we can choose $F_1 \subset F_2$ such that $supF_2=supF_1+\epsilon$ and $infF_2=infF_1-\epsilon$. Now, if we want to refer to probabilities for joint processes in which forward and backward trajectories coincide we have to take the limit $\epsilon \rightarrow 0$ and it's natural that in such case $\lim_{\epsilon \to 0}F_2=F_1=F$ because of $\lim_{\epsilon \to 0}supF_2=supF_1$ and $\lim_{\epsilon \to 0}infF_2=infF_1$. And now, $Prob_{T \cap A}(\epsilon \rightarrow0)$ will describe probabilities for a time-symmetric particle (see Appendix L) that also maintains statistical reversibility (see Appendices B and K). Of course, the reversible trajectories are the only real trajectories that truly exist but they are an extremely tiny subset (non-empty though) of a huge set that also includes irreversible trajectories. However, those trajectories are non-real (or pseudo-trajectories) and the probabilities that describe them are complex.

Now what about the arrow of time? How does macroscopic irreversibility emerge from reversible microscopic dynamics when QM is proven to be reversible both at the microscopic and at the statistical level? I don't want to provide a complete solution to this problem but I will emphasize that an effort for the solution may indeed start from quantum mechanics and specifically from our approach. As previously noted, the intersection of forward and backward paths in our model is a mathematical idealization that is expressed by the limit $\epsilon \rightarrow 0$. However, $\epsilon$ can never be zero because in such case independence of forward and backward paths would be violated and we would not be able to produce a reversible diffusion theory which is identical to QM. That means that there are two possible interpretations. Either our math, with this limiting procedure, is a tool that can be used to describe dynamics that is one hundred percent reversible or it describes a true fundamental limitation about the exact reversibility of trajectories. In other words, reversing time in the equations of motion for a specific forward quantum trajectory gives us the backward trajectory but slightly deviating from the original path. At the same time, Newtonian mechanics which is just a limit of quantum mechanics seems reversible but in fact it's not due to those small deviations. And yes, those small deviations don't cease to exist in Newtonian physics since background field fluctuations may not affect the motion in the same way they will affect an electron, but can still possibly produce deviations from Newtonian trajectories that are of the order of $10^{-7}nm$ for instance. However, when a large number of degrees of freedom is included in a classical (or even quantum) system, irreversibility becomes apparent. And that happens because reversing time at the whole system's evolution will not produce the exact reversed motion for the individual particles. And that may be unobservable at the level of individual particles but when the number of participating particles gets larger and larger, those tiny deviations will contribute as a whole producing the large deviations that we observe macroscopically. We have to to note also that in the literature-I am not a proponent of that view-the most standard resolution of time's arrow is given by the Past Hypothesis \cite{albert2003time} according to which the universe started in a low-entropy state and it continuously evolves into a higher-entropy state dictating in this way the thermodynamical arrow.

\section{}
Proving that quantum systems are always in one ontic (contextual though) state before measurement record is essential (but not adequate) for the dissolution of what we call the measurement problem. Instead, we will also need the loss of interference between branches (after measurement) that allows us to effectively get only one definite outcome after measurement, which will correspond to one of the superposed states. We will now provide a demonstration of how information transfer to the apparatus about the actual state of the particle may indeed kill the interference terms. In other words, our approach is purely information-theoretic and not a mechanistic one, as were for example Zeh's \cite{zeh1970interpretation}  and Zurek's \cite{zurek2003decoherence} historically. It's worth to mention that, again, our tool is pure set theory with no Hilbert spaces at all.

Let us consider a quantum particle described by the set $B_p=B_p\rightarrow \{x_p\in F,t_c\}$ being in a superposition of two energy states. However, this superposition is only mathematical and reflects the potential states of the particle before measurement. As explained in Appendices F, J and L, the ontic state of the particle is definite and is only one of the two superposed states. After the interaction, the particle and apparatus become entangled (see Appendices I and N) and those two parts of the system behave as a whole and not as independent from each other. The entangled system is now described by the set $B_{sys}=B\rightarrow \{x_p\in F,t_c;x_{app} \in F',t_c\}$ and logically speaking, it will be in a mathematical superposition of two states. One in which the apparatus has already interacted with the quantum particle whose ontic state is the first energy state and one in which the apparatus has interacted with the quantum particle whose ontic state is the second energy state. Of course, the state of the apparatus after the entanglement will be affected in a way that depends on the ontic state of the quantum particle with which it interacts. It is therefore natural to say that:
$$B_{sys}=B\rightarrow \{x_p\in F,t_c;x_{app} \in F',t_c\}$$
$$=T\rightarrow \{x_p\in F,t_c;x_{app} \in F',t_c\} \cap A\rightarrow \{x_p\in F,t_c;x_{app} \in F',t_c\}$$
where
$$T\rightarrow \{x_p\in F,t_c;x_{app} \in F',t_c\}=T^{(1)}\rightarrow \{x_p\in F,t_c;x_{app} \in F',t_c\} \cup T^{(2)}\rightarrow \{x_p\in F,t_c;x_{app} \in F',t_c\}$$
and an analogous equation holds also for the A set. Of course, we have to mention that for $i=1,2$, we have $T^{(i)}\rightarrow \{x_p\in F,t_c;x_{app} \in F',t_c\} =T_p^{(i)}\rightarrow \{x_p\in F,t_c\}\cap T_{app}^{(i)}\rightarrow \{x_{app}\in F',t_c\}$
as well as $A^{(i)}\rightarrow \{x_p\in F,t_c;x_{app} \in F',t_c\} =A_p^{(i)}\rightarrow \{x_p\in F,t_c\}\cap A_{app}^{(i)}\rightarrow \{x_{app}\in F',t_c\}$.

Assuming that the apparatus is a perfect measuring device, it will not just change state after measurement based on the ontic state of the observed particle, but it will also receive the information that this particle is on that state. So, the T set for an ideal measuring apparatus can be written as $T_{app}^{(i)}\rightarrow \{x_{app}\in F',t_c\}=T_{app}^{(i,\xi_i)}\rightarrow \{x_{app}\in F',t_c\}$ where proposition
$\xi_i=\{\textit{apparatus measures } B_p=B_p^{(i)}\}\equiv \{\textit{apparatus measures } T_p=T_p^{(i)} , A_p=A_p^{(i)}\}$ and the same thing can be done for B and A sets. Taking into account all mathematical relations we have got, it turns out that the set that corresponds to the entangled system particle-apparatus is:
$$B\rightarrow \{x_p\in F,t_c;x_{app} \in F',t_c\}=S_{p,app}^{(1,1)}\cup S_{p,app}^{(2,2)} \cup S_{p,app}^{(1,2)} \cup S_{p,app}^{(2,1)}$$ (a more rigorous shorthand derivation is provided later in the same Appendix) where:
\begin{itemize}
    \item $S_{p,app}^{(1,1)}=B_p^{(1)}\rightarrow \{x_p\in F,t_c\} \cap  B_{app}^{(1,\xi_1)}\rightarrow \{x_{app}\in F',t_c\}$
    \item $S_{p,app}^{(2,2)}=B_p^{(2)}\rightarrow \{x_p\in F,t_c\} \cap  B_{app}^{(2,\xi_2)}\rightarrow \{x_{app}\in F',t_c\}$
    \item $S_{p,app}^{(1,2)}=S_{p,app}^{(T_1)} \cap S_{p,app}^{(A_2)}$
    \item $S_{p,app}^{(2,1)}=S_{p,app}^{(T_2)} \cap S_{p,app}^{(A_1)}$
    \item $S_{p,app}^{(T_1)}=T_p^{(1)}\rightarrow \{x_p\in F,t_c\} \cap  T_{app}^{(1,\xi_1)}\rightarrow \{x_{app}\in F',t_c\}$
    \item $S_{p,app}^{(A_2)}=A_p^{(2)}\rightarrow \{x_p\in F,t_c\} \cap  A_{app}^{(2,\xi_2)}\rightarrow \{x_{app}\in F',t_c\}$
    \item $S_{p,app}^{(T_2)}=T_p^{(2)}\rightarrow \{x_p\in F,t_c\} \cap  T_{app}^{(2,\xi_2)}\rightarrow \{x_{app}\in F',t_c\}$
    \item $S_{p,app}^{(A_1)}=A_p^{(1)}\rightarrow \{x_p\in F,t_c\} \cap  A_{app}^{(1,\xi_1)}\rightarrow \{x_{app}\in F',t_c\}$

\end{itemize}

We already know that $B_p^{(1)}\rightarrow \{x_p\in F,t_c\}\cap B_p^{(2)}\rightarrow \{x_p\in F,t_c\}=\varnothing $ due to the absence of physical superposition for the quantum particle even before measurement (see Appendices F and J where linear superposition is derived from this classical assumption, which is a natural assumption since our theory is classical). From this mutually exclusive relation, we can derive that $T_p^{(1)}\rightarrow \{x_p\in F,t_c\}\cap T_p^{(2)}\rightarrow \{x_p\in F,t_c\}=\varnothing$ or $A_p^{(1)}\rightarrow \{x_p\in F,t_c\}\cap A_p^{(2)}\rightarrow \{x_p\in F,t_c\}=\varnothing$ using propositional logic (see Appendix M analysis for spin case). Those two results, based on the definition of proposition $\xi_i$ lead us respectively to $T_p^{(i)}\rightarrow \{x_p\in F,t_c\} \cap  A_{app}^{(j,\xi_j)}\rightarrow \{x_{app}\in F',t_c\}=\varnothing$ and $A_p^{(i)}\rightarrow \{x_p\in F,t_c\} \cap  T_{app}^{(j,\xi_j)}\rightarrow \{x_{app}\in F',t_c\}=\varnothing$ for $i,j=1,2$ and $i \neq j$. In any case, we will have $S_{p,app}^{(1,2)}=S_{p,app}^{(2,1)}=\varnothing$. The set that describes the entangled system will thus be equal to
$B_{sys}=\left(B_p^{(1)}\rightarrow \{x_p\in F,t_c\} \cap  B_{app}^{(1,\xi_1)}\rightarrow \{x_{app}\in F',t_c\}\right)\cup \left(B_p^{(2)}\rightarrow \{x_p\in F,t_c\} \cap  B_{app}^{(2,\xi_2)}\rightarrow \{x_{app}\in F',t_c\}\right)$. By taking the probability at each side of this set expression and knowing that it holds $\forall F \subseteq \mathbb{R}^{p}$ and $\forall F' \subseteq \mathbb{R}^{app}$. we are finally led to:
$$\rho_{sys}(x_p,t_c;x_{app},t_c)=|c_1|^2|\psi_p^{(1)}(x_p,t_c)|^2 |\psi_{app}^{(1)}(x_{app},t_c)|^2+|c_2|^2|\psi_p^{(2)}(x_p,t_c)|^2 |\psi_{app}^{(2)}(x_{app},t_c)|^2$$
where $|c_1|^2$ and $|c_2|^2$ are the probabilities which correspond to the quantum particle being in 1st and 2nd state respectively. This result agrees with standard QM.

Furthermore, since 
$$B^{(1)}\rightarrow \{x_p\in F,t_c;x_{app} \in F',t_c\}\cap B^{(2)}\rightarrow \{x_p\in F,t_c;x_{app} \in F',t_c\}=\varnothing$$
due to the fact that $B_p^{(1)}\rightarrow \{x_p\in F,t_c\} \cap B_p^{(2)}\rightarrow \{x_p\in F,t_c\}=\varnothing$ (which is also the reason why we obtain the above form of $\rho_{sys}$), it is easy to derive the quantum state for the entangled system in the same way as we did in Appendices F and J. This quantum state will be equal to
$$\Psi_{sys}(x_p,t_c;x_{app},t_c)=c_1\psi_p^{(1)}(x_p,t_c)\psi_{app}^{(1)}(x_{app},t_c)+c_2\psi_p^{(2)}(x_p,t_c)\psi_{app}^{(2)}(x_{app},t_c)$$
and is true for any case of external environment, even if this environment does not behave as a good measuring device (i.e. it is a system that is not meant to measure the energy at all). Below, we will explain more rigorously the loss of factorization due to interaction.

We can easily observe from the set expression of $B_{sys}$ that we eventually derived, and the mutual exclusivity relation between $B^{(1)}$ and $B^{(2)}$ that was derived before (and the expression for $\rho_{sys}$) that the entangled system after measurement will be in one of the two states 1 or 2 obeying to the rules of conventional (non-extended) real probability theory. But, in order to be sure that the state will indeed be one of the two states and not some indefinite state, we must show (like in Appendix L) that $$ Prob\left( B^{(1)}\rightarrow \{x_p\in \mathbb{R}^p,t_c;x_{app} \in \mathbb{R}^{app},t_c\}\cup B^{(2)}\rightarrow \{x_p\in \mathbb{R}^p,t_c;x_{app} \in \mathbb{R}^{app},t_c\} \right)=1$$
which is indeed true since it gives us the relation $|c_1|^2+|c_2|^2=1$ that was already derived in pre-measurement situation. In this way, measurement problem is completely dissolved.

Now, someone may wonder that if pre-measurement total state is different from after-measurement state, how does it make sense to say that in both cases an individual particle is always in one ontic state as was demonstrated from our set description? In other words, why ontic definiteness holds in both situations that are completely distinct? The difference has to do with the fact that the particle in the first case is always in one state but its behavior is constantly affected by the other state which acts as a "ghost state". In other words, the potential of the particle to follow one of the two states at its initial preparation/measurement context (see Khrennikov again) constantly "haunts" the particle even after it "enters" into one specific state (we see that "non-local memory" phenomenon appear also in quantum entanglement). After measurement, the influence by the other state is just lost (decoherence) and the particle is still in that state but now it behaves as if the other state/potentiality never existed at all. Now, someone may be confused and say that superposition still remains and thus despite the fact that the particle is in only one state, since we've failed to produce definite outcomes, we've actually not solved the measurement problem. But we do not need to produce definite outcomes since our theory is a classical objective/frequentist probability theory and thus this superposition is an expression we will get from an ensemble of infinite particles. That resembles the statistical interpretation of Einstein \cite{ballentine1970statistical}, but in our case, the reason why quantum theory describes ensembles is justified by presenting it as a classical objective/frequentist probability theory in which even the wave function has a clear probabilistic character, not just statistical (statistical interpretation  just postulates the ensemble approach without justifying the probabilistic origin of quantum formalism) or epistemic. Thus, we know that after measurement, the individual particles will be in just one of the superposed states as demonstrated through our set-theoretic analysis and completely unaffected by the other state resulting in the so-called "effective collapse" for an individual particle's state. And this phenomenon after many experiments will eventually produce the $\rho_{sys}$ we derived. 

Recall that PBR theorem \cite{pusey2012reality} rules out only $\psi-$epistemic hidden-variable models and not $\psi-$nomological models that treat the wave function as an objective property of the world (in our approach the wave function is a non-subjective complex probability distribution). Also another problem of epistemic approaches is that they describe knowledge, but the question "knowledge about what?" remains unanswered. On the other hand, as we've shown, probabilities derived by the wave function and its complex conjugate describe specific sets that refer to specific propositions (see for instance Appendices K, L).

Now, to discuss the non-conventional nature of quantum probability theory that produces extra interference terms and according to Feynman violates classical logic and classical probability theory \cite{feynman1951concept}, we will look at Khrennikov's contextual probability theory \cite{khrennikov106073quantum}. According to it, the sum of probabilities that refer to different contexts/experimental conditions (left slit open or right slit open) was never meant to describe the case where we have both potentialities coexisting in a single experiment. This sum is meant to just describe the probabilities for a joint experiment which is described by a product sample space that consists of two sub-experiments: the one in which one slit is open and the other in which the other slit is open. Therefore, in the case that we know what will happen when one of the slits is open, we can not say what will happen when both slits are open even by using classical probability theory. Therefore, classical probability theory is not violated and probabilities can still describe a hidden reality and not be intrinsic as standard QM suggests. 

We must mention that what we're presenting here is not a non-classical probability theory that violates classical logic but rather a non-conventional probability framework using a hyper-sample space that includes also pseudo-events. In this way, what we know from classical probability theory doesn't lose its meaning here (we thus strongly agree with Khrennikov's classical view of quantum probability using the contextual approach as discussed above). 

In our approach specifically, interference can be seen as a result of Born rule (see Appendix G) and superposition (it can be taken as given by the linearity of the Schroedinger equation whose existence was derived in Appendices B and L, or derived from the logical no-(state) superposition assumption for diffusion processes made in Appendices F and J without assuming Schroedinger equation). And the process with which we make those derivations is totally based on classical probability theory. However, interference could emerge from a more general set expression like the one we wrote in Appendix L and that connects $B_{tot}$ with $S_{11}$, $S_{22}$, $S_{12}$, $S_{21}$. But why can't we just say that $B_{tot}=S_{11}\cup S_{22}$? why postulate the existence of the other two sets that produce the interference terms? 

We will work with non-entangled systems just for simplicity and we will assume that a quantum particle can be in one of two energy states 1 and 2. According to Chapter III and Appendix G, the event that describes a quantum particle that obeys to a reversible diffusion process will be the following:
$$B \rightarrow \{x \in F,t_c\}=\lim_{\epsilon \to 0} \left\{T\rightarrow\{ x_1 \in F,t_c \} \cap A\rightarrow\{ x_2 \in F,t_c \}||x_1-x_2|\leq \epsilon  \right \}$$

And now, by making use of the equations $T \rightarrow \{x_1\in F,t_c\}=T_1 \rightarrow \{x_1 \in F,t_c\} \cup T_2 \rightarrow \{x_1 \in F,t_c\}$ and $A \rightarrow \{x_2\in F,t_c\}=A_1 \rightarrow \{x_2 \in F,t_c\} \cup A_2 \rightarrow \{x_2 \in F,t_c\}$ and by taking the probability in the left and right-hand side of the equation about the B set that we wrote before, we are easily led to interference. We did that to demonstrate that B set is a derivative set (subset of the "real sample space" which on its turn emerges as a subset of the true sample space which is the hyper-sample space) which is produced by the intersection of sets that come from completely distinct original/non-product sample spaces (let's call them $\Omega^T$ and $\Omega^A$, with the first one related to forward and the second to backward stochastic processes). Furthermore, B set doesn't just come from the intersection of T and A sets but also emerges under a significant condition/restriction $C_\epsilon=\{|x_1-x_2|\leq \epsilon\}$ where $\epsilon \rightarrow 0$. This means that T and A sets are the primary/fundamental sets and thus the property of the form $\Gamma=\Gamma_1\cup\Gamma_2$ will primarily be applied to T and A sets and as a result will not be valid for the B sets, giving rise in this way to two additional "interference sets" that describe non-real events/trajectories.

For similar reasons, for an entangled system of particles $p_1$ and $p_2$ (suppose we have the same exact case of entanglement as before but instead of an apparatus we have a single particle), in general, we have $$B_{sys}=B\rightarrow \{x\in F,t_c;y\in F', t_c\}  \neq B_{p_1}\rightarrow \{x_{}\in F,t_c\} \cap  B_{p_2}\rightarrow \{ y\in F',t_c\}$$   

To explain in more detail why this non-equality, that allows us to talk about quantum entanglement (since the particles can not be described separately) holds, we will write down the equation 
$$B_{sys}=\lim_{\epsilon \to 0}\{T\rightarrow \{x_{1}\in F,t_c;y_{1} \in F',t_c\} \cap A\rightarrow \{x_{2}\in F,t_c;y_{2} \in F',t_c\}|C_{p_1,\epsilon};C_{p_2, \epsilon}\}$$
where $C_{p_1,\epsilon} =\{|x_1-x_2|\leq \epsilon\}$ and $C_{p_2,\epsilon} =\{|y_1-y_2|\leq \epsilon\}$. Again, we apply the same logic as before. And that is that we can logically apply the equation $\Gamma=\Gamma_1 \cup \Gamma_2$ and $\Delta=\Delta_1 \cap \Delta_2$ but only between T sets and between A sets which are the primary sets. Thus, if we take the T set for the entangled system, then by using some shorthands, we analyze it in the following way: $T=T^{(1)}\cup T^{(2)}$ which leads to $T=(T_{p_1}^{(1)}\cap T_{p_2}^{(1)})\cup (T_{p_1}^{(2)}\cap T_{p_2}^{(2)})$ and the same shorthand analysis is done for the A set. In this way, we accomplish to produce $B_{sys}$ in a form that is completely similar to the one we wrote at the beginning of measurement discussion.

And to be fully consistent mathematically, the fact that we use the intersection formula for the B sets in the first two bullets of this Appendix does not contradict our claims. Why? Because the B sets that refer to the system as a whole can be written with respect to T and A sets and using the same methodology as before, we can obtain the intersection formula as presented in the first two bullets. And that can only happen when the B set describes an entangled system being at a specific eigenstate.

We have to mention again though that to derive the entangled state $\Psi_{sys}$ from above, we can just use the same logic as in the non-entangled one-particle case, but in this case our mutually exclusive B sets (see Appendix F, J) will refer to the system of particles 1 and 2 as a whole. Later, we have to also use the fact that T (or A) sets are the fundamental sets and thus we can safely apply the equation $T^{(i)} \rightarrow \{x_1 \in F,t_c;y_1 \in F',t_c\}=T_{p_1}^{(i)} \rightarrow \{x_{1} \in F,t_c\} \cap T_{p_2}^{(i)} \rightarrow \{y_{1} \in F',t_c\}$, where $i=1,2$, which eventually leads to the superposed state for the entangled system (like the $\Psi_{sys}$ we derived before but instead of apparatus we have particle $p_2$). Then, after taking the Born rule for the entangled system, we can derive our joint probability distribution and we can confirm that it indeed arises from a set expression that is identical in form to $B\rightarrow \{x_p\in F,t_c;x_{app} \in F',t_c\}$ we had mentioned at the start of measurement discussion.

And to be completely precise on our mathematical formalism that we used in Appendix J, we will just say (based on our shorthand description) that the mutual exclusivity relation between B sets $B_n\cap B_m=\varnothing$, from which superposition was derived, is not an invalid description. And we say that, because it is in fact emergent from the the generally accepted expression $B_{n,m}=\varnothing$ which gives us that $T_{n,m}\cap A_{n,m}=\varnothing \Rightarrow (T_{n} \cap T_m)\cap (A_{n} \cap A_m)=\varnothing \Rightarrow (T_{n} \cap A_n)\cap (T_{m} \cap A_m)=\varnothing \Rightarrow B_n \cap B_m= \varnothing$. The same thing can not be said about $B^{(\textit{n or m})} \equiv B_{n \lor m}$, for which we have that $B_{n \lor m} \neq B_n \cup B_m$ as we explained before, in our set-theoretic description of interference. That happens, because through the definition of the set $B_{n \lor m}$, we've taken into account the existence of one single context with both potentialities that lead to n and m states which of course does not always guarantee the additivity between probabilities that refer to different contexts (see Appendix V too), as discussed before in more detail. And that distinctive feature in QM, as we've shown, has its roots in classical set theory.

\section{}

As we mentioned in Appendix L, the Wiener processes $w_1(t)$ and $w_2(t)$ are completely independent and the number $i$ that connects them is of course not a proportionality coefficient between those two. Thus, to avoid any misleading argument that disputes the independence of those processes, let us say say that $w_1(t)\rightarrow w_{rev}(t)$ and $w_2(t)\rightarrow w_{irr}(t)$ with no relation of the type $w_{irr}(t)=Cw_{rev}(t)$ between those two processes, but by just saying that the irreversible Wiener process (the process that is not even theoretically reversible since it does not emerge from reversible Hamiltonian dynamics) is obviously non-real. 

We know from Wiener process theory and from Nelson's work that the covariance of two Wiener processes with the same diffusion constant $D$ is given by the general expression $Cov(dw_i(t),dw_j(t))=\delta_{ij}dt$ where $w_i(t)$ and $w_j(t)$ correspond to independent (real) Wiener processes when $i \neq j$. In our quantum case, we will use a generalized covariance that includes also Wiener processes that are non-real and which will be symbolized using an extra bar above it. In that case, we will have that $$\overline{Cov}(dw(t),dw(t))=dt$$

The above equation, due to the fact that $dx(t)=\frac{1}{2}(dx_{rev}(t)+dx_{irr}(t))\Rightarrow  dw(t)=\frac{1}{2 \sigma}(\sigma_1dw_{rev}(t)+\sigma_2 dw_{irr}(t))$, where $\sigma_1=\sigma_2=\sigma_{1,2}$, as well as because of the relation
$$\overline{Cov}(dw_i(t),dw_j(t))= \delta_{ij}dt$$
where $i,j=\textit{"rev"},\textit{"irr"}$, gives us that $\sigma_{1,2}=\sigma \sqrt{2}$ and $D_{1,2}=2D_{sys}=\frac{\hbar}{m_p}$. And given those results, we can easily derive that $w(t)=\frac{w_{rev}(t)+w_{irr}(t)}{\sqrt{2}}$. All those things agree with the Schroedinger's diffusion discussion, where Schroedinger's diffusion process gives rise to real and imaginary parts that contain the aforementioned derived quantities.

\section{}

We have seen that this theory makes some of the predictions of standard QM. However, could this theory make any new testable predictions? The answer is maybe. As we've mentioned previously, microscopic quantum motion is probably not perfectly reversible (see Appendix O). This could mean that there will be extremely tiny deviations from standard Born's rule. However, this idea should not be taken for sure yet, since someone could argue that this reversibility may in fact be exact and the limiting procedure is just a mathematical process we have to follow in order to express that exact reversibility. Anyhow, we provide the idea that possible deviations from the Born rule could be consistent with our theory in the case that someone takes seriously the fact that the Born rule is derived as a limit of a function $f(\epsilon)$ (see Appendix G) where $\epsilon \rightarrow 0$ and by not allowing the strict condition $\epsilon=0$ (given that $f(\epsilon)$ is not continuous at the point $\epsilon=0$ and this is clarified in Appendix O). At the end of the day, it's a matter of interpretation to decide which one of the two possibilities is correct, but a possible future experimental verification of extremely tiny deviations could be a good indicator for the validity of our theory. It would not guarantee it though because there is already a very small number of approaches that support deviations. The most famous of all is Valentini's approach \cite{valentini1991signal} in the context of dBB theory. However, this is not yet a standard view among Bohmians and differs from the more established views of Durr, Goldstein and Zanghi, who were mentioned in Chapter II also. 

\textit{Additionally, my theory claims two extra things: a)the diffusion constant is two times the Nelson's diffusion constant, based on the center of mass description of Schroedinger's diffusion, b) quantum trajectories are a tiny subset of classical Brownian trajectories. What that means is that possibly, if we isolate the ensemble of approximately reversible trajectories in a Brownian system, we may observe quantum-like behavior (i.e. double-slit interference phenomena or quantum probability rule satisfaction). And the most appropriate way of testing this is, I think, using Monte Carlo simulations, instead of real experiments.}

To sum up, confirming that contextual realism (see \cite{karakostas2012realism}) is in fact the ultimate feature of the quantum world expressed via non-local correlations, would possibly lead us implicitly to the unification of quantum mechanics and relativity. And I claim that since by simply looking at special relativity, we see something common with QM and this is that you can assign reality about the behavior of a system but this reality is not absolute and is completely relational to the observer and the observer in both cases is just another physical system. We have to accept reality, but we should always keep in mind that this reality is not classical and we can easily observe that if we try to talk about stochastic motion at the microscopic level. So, we can say that indeed the classical and quantum diffusion uncertainty principles (see Furth again) prove the exact statement of our SDE analysis in Appendix L. And this is that the more we zoom-in the particle's motion ($\Delta x<<1$), the more violent its motion will look like ($\Delta p>>1$) and this is why the concept of instantaneous velocity is not well-defined (see Appendix L).

However, we shall not forget that this sub-quantum field which is responsible for this Brownian motion is in fact a non-local field that behaves as a fluctuating and possibly continuous "whole" and not as a medium that consists of many particles that interact in a way that is local (see Appendix K) and that it allows for the surrounding particles for extremely tiny timescales to be free of interaction with the rest of the particles. Instead, everything behaves as one and the particle just responds to every violent fluctuation of the field. And as we said in Appendix K, this lack of discrete structure of the sub-quantum field is what prevents particle's information from being distributed into a vast number of surrounding particles whose number is estimated to be $N\sim 10^{20}-10^{40}$ particles (however, instantaneous kinetic energy fluctuations exist according to Chapter VI and Appendix K in QM if we think that $sup\{\Delta\hat{x}\Delta\hat{T}_K\}\neq 0$, which possibly reveals the inability of a well-defined definition of an instantaneous kinetic energy of the particle again due to violent fluctuations as in the momentum case). 

\textit{It's extremely important to note however, that the irreversibility that we observe in macroscopic phenomena (i.e. classical diffusion) may indeed be explained in the way we mentioned at the beginning of the Appendix without invoking information loss due to collisions but through identifying the irreversibility as an intrinsic feature even in microscopic motion due to this $\epsilon \neq 0$ (in this way our explanation offers an alternative approach to the standard anthropocentric/epistemic explanation). And that actually, happens due to the inability of the sub-quantum field to reproduce the exact same noise sequence in reversed order and direction due to zero probability measure of the event (despite the fact that the event is a non-empty event). In this way, extremely tiny deviations of backward noise from the original forward noise in its reversed form have a significantly low influence on quantum and mostly classical particles (large masses are extremely insensitive to the fluctuations) but they definitely affect a stochastic system with interacting molecules whose number is equal to $N\sim 10^{20}-10^{40}$ particles. Each molecule's reversed trajectory will exhibit extremely tiny deviations from the original one and someone could say that the result of that situation is similar to the quantum case where we have an approximately reversible system. But that is not true because this system is chaotic and thus if all those deviations apply to all particles (or even to some of them), the final state of the backward evolution will diverge a lot from the initial state of the forward one. As a result, reversing the whole evolution exactly is extremely and unbelievably unlikely. }

Given all those, we understand we can never separate the particle from the field (the sub-quantum field does not consist of discrete entities, one of which is the particle) and that non-separability between those two may possibly lead to a better foundational understanding of standard QFT's non-intuitive claims about fields' ontological superiority over localized entities (fields in QFT are different from our sub-quantum field though). But for now, it's too early to make any suggestions on how this theory can be compatible with QFT. For now, I think the theory is incomplete, but I definitely believe it's the right path to follow and that future work may provide a more generalized and rigorous version of it. Until then, we have to keep in mind that relations between the parts of the whole determine their behavior inside this whole, making relative information (not information obtained by human brain) a fundamental concept in QM (see also RQM which is also perspectival but does not provide any dynamics). And of course, the importance of it was realized in the measurement discussion where information acquisition about the particle's ontic state by the apparatus "kills" interference as demonstrated in our set-theoretic derivation. Now, regarding contextual realism for all observables, it is clearly adopted in our approach (see also Appendix V, last part). However, this does not mean that for some observables like position, non-contextual realism is excluded. But anyway, I think that this depends on anyone's philosophical preference and for me it's much more preferable and symmetric to establish contextual realism for all observables. 
 
\section{}

In this Section, we will explicitly show how the intersection process of the ensemble trajectories gives rise to the Born rule. The logic is completely the same as in Appendix L, but in this case, no SDEs will be used but only the Schroedinger forward and backward diffusions. Again, our analysis will justify not only why the Born rule must hold, but also why the Schroedinger equation has this specific form (diffusion form) by using once again the two-particle description, though this time not applied at the full microscopic level but at the local ensemble level.

In general, we need a reversible diffusion theory and thus the creation of two distinct probability spaces each of whom is related to forward and backward local ensemble trajectories respectively. This is done in Appendix B and allows us to construct a hyper-sample space in which sets for both reversible (real) ensemble trajectories and irreversible (non-real) ensemble trajectories are contained (the same method applies for the microscopic trajectories too in Appendix L). According to this method in Appendix B, we allow for $D\in \mathbb{C}$, which will give us a complex diffusion equation that is valid under time-reversal under the transformation $\Psi \rightarrow \Psi^*$ (according to Appendix L, the diffusion constant for Schroedinger diffusion is not imaginary though, but equal to $D_{sys}=\frac{\hbar}{2m_p} $ and the imaginary $i$ just "hits" the noise). Now, let's try to rewrite the forward complex diffusion (Schroedinger) equation in a slightly different form:
$$\frac{\partial \Psi(r,t)}{\partial t}=-\overrightarrow{\nabla} \cdot \left( \Psi(t,t)\left(-\frac{i \hbar}{2m_p}\frac{\overrightarrow{\nabla} \Psi(r,t)} {\Psi(r,t)}\right)\right)$$

From that, we can easily observe that the diffusion velocity (that's how Furth in his 1933 paper calls the current velocity in classical diffusion analysis and interpreted it as a local ensemble velocity) for the Schroedinger particle (which lives "two parallel mathematical realities" of real and non-real behavior as shown in Wiener process analysis of Appendix L) is equal to $\overrightarrow{\eta}^T_{sys}(r,t)=-\frac{i \hbar}{2m_p}\frac{\overrightarrow{\nabla} \Psi(r,t)} {\Psi(r,t)}=-\frac{i \hbar}{2m_p}\left(Re \left(\frac{\overrightarrow{\nabla} \Psi(r,t)} {\Psi(r,t)}\right)+iIm \left(\frac{\overrightarrow{\nabla} \Psi(r,t)} {\Psi(r,t)}\right)\right)$. Following the logic of the center-of-mass analysis of Appendix L, this expression can be written in the following form:
$$\overrightarrow{\eta}^T_{sys}(r,t)\equiv \overrightarrow{\eta}^T_{com}(r,t)=\frac{1}{2}(\overrightarrow{\eta}_{Re}^T(r,t)+i\overrightarrow{\eta}_{Im}^T(r,t))$$
where $\overrightarrow{\eta}_{Re}^T(r,t)=\frac{\hbar}{m_p}Im \left(\frac{\overrightarrow{\nabla} \Psi(r,t)} {\Psi(r,t)}\right)$ is the diffusion velocity (or local ensemble velocity) of the quantum particle that undergoes a reversible diffusion process and $\overrightarrow{\eta}_{Im}^T(r,t)=-\frac{\hbar}{m_p}Re \left(\frac{\overrightarrow{\nabla} \Psi(r,t)} {\Psi(r,t)}\right)=-\overrightarrow{u}_{osm}(r,t)$ is the diffusion velocity of the non-real particle undergoing irreversible diffusion (the reason why the first real ensemble motion is reversible and the second not will be analyzed in the second paragraph). The quantity $\overrightarrow{u}_{osm}(r,t)$ is Nelson's osmotic velocity as mentioned in Chapter II.

We can also easily derive that $\overrightarrow{\eta}_{Re}^T(r,t)=\overrightarrow{\eta}_{Bohm}(r,t)$ (see Chapter II)  and we should expect that, since according to our whole analysis (see also Appendix D) and according to weak measurement discoveries in 2011, the Bohm velocity has to be interpreted as a local ensemble velocity field (we remind that the local ensemble velocity is equivalent to the diffusion velocity according to Furth and to stochastic literature). And now, to complete what we said in the previous paragraph about reversibility, we must notice that reversing time in the diffusion velocities of both real and non-real particles gives us, according to the transformation $i \rightarrow -i$ that must be done simultaneously, that $\overrightarrow{\eta}_{Re}^A(r,t'=t_c)=- \overrightarrow{\eta}_{Re}^T(r,t=t_c)$ (quantum particle is subject to a real reversible diffusion) and $\overrightarrow{\eta}_{Im}^A(r,t'=t_c)\neq- \overrightarrow{\eta}_{Im}^T(r,t=t_c)$ (non-real particle is subject to a non-real irreversible diffusion). This is the exact reason why we can use the intersection of the whole family $S[r_{Schr}(t)]$ of all local ensemble forward paths with the family $S[r_{Schr}(t')]$ of all local ensemble backward paths to justify the Born rule and this was actually similarly done in Appendix L for the individual paths. So, now we can more clearly understand the actual origin of current and osmotic velocities that exist in stochastic mechanics literature. 

We will make two remarks about the previous: the first one is that our B, T and A quantum sets can very well describe the set of random microscopic trajectories of Appendix L, but can also apply to the ensemble/average trajectories that are non-empty, but are zero-measure subsets of this entire set of trajectories that are described by the hyper-sample space $\Omega_{h}^{(q)}$. We should be careful though and clarify that Langevin trajectories are ignorant about the position of the particle in spacetime (whether it's in in a location where the gradient of the probability distribution is positive or negative) and thus we say that the Langevin trajectory has no preferred direction right or left, in opposition to diffusion velocity that is a function of space and time and therefore carries that information.

The second one is that the non-real irreversible diffusion velocity is in fact identical to the one presented in classical irreversible diffusion by Furth in 1933 which has the form $\overrightarrow{u}_d=-D\frac{\overrightarrow{\nabla}\rho}{\rho}$ and it has opposite direction to the osmotic velocity field. 

And one small note. In Appendices I, N and P, we spoke about the concept of "memory" of past interactions which does not necessarily refer to multi-particle entanglement, but also to the interference experiment or even to the evolution of a single-particle wave packet. The common element for all those cases is that the particle "remembers" its past non-local correlation with the context (see Appendix P and Khrennikov) that was put into the system. However, this is not "memory" in the usual or non-Markovian sense. It is just a result of the deterministic flow of the quantum diffusion which is dictated by Bohm's velocity that has the phase structure in it. And we say deterministic, because our diffusion is fundamentally reversible and we can say that the particle's ensemble behavior now was absolutely predetermined at the time of context placement. However, that is about the ensemble evolution, which definitely affects the individual paths, but the individual paths themselves are memoryless due to the white noise (see Appendix L) that ensures Markovianity. 

And the quantity that plays a crucial role in the time-symmetric ensemble evolution of the particle is the quantum potential that does not play the role of a classical potential, but it represents a source of non-local active information (see Basil Hiley and Bohm \cite{bohm1984measurement}) that predetermines the ensemble evolution at the time of context placement. Now, why it should be non-local and not just local was discussed in Appendix I. However, I personally do not think that high-dimensionality of the quantum potential or the fact that it is inversely proportional to the probability density (and thus does not necessarily vanish as $r \rightarrow \infty$) are enough by themselves to justify non-locality (without of course using Bell's theorem as a justification). This issue will be thoroughly discussed in a future paper and we will show in detail why Bohm potential may indeed represent a non-local field. I want to emphasize though that the wave function should not be interpreted as a non-local field in 3D space, but as a complex probability field (or complex probability distribution) living in a high-dimensional space that describes the ensemble of all those pseudo-trajectories we discussed (or alternatively trajectories that do not exist in physical space) and generates all quantum phenomena that we know. Therefore, the status of the wave function remains strictly nomological rather than ontic or epistemic.

\section{}

 Before we move on, we need to mention that in a frequentist (or objective) probability theory like ours, normalization of the hyper-sample space is unavoidable, even when we have as subsets non-real sets with complex measures. Thus, when we start from two distinct sample spaces $\Omega^{T}$, $\Omega^{A}$ (from which we can get the real and artificially created sample space using the procedure described in Appendix O), we must keep in mind that those must be normalized. This suggests that $\Psi_{norm}(r_1,t_c)=\frac{\Psi(r_1,t_c)}{\int_{\mathbb{R}^3}\Psi(r_1',t_c)d^3r_1'}$ and that $\Psi^*_{norm}(r_2,t_c)=\frac{\Psi^*(r_2,t_c)}{\int_{\mathbb{R}^3}\Psi^*(r_2',t_c)d^3r_2'}$. And it is now noteworthy that the results that we derive in this case (i.e. Born rule, superposition) are again the same, but now a normalized complex probability distribution $\Psi_{norm}(r,t_c)$ is involved instead of a non-normalized one. In this case, we will also have that $Prob(\Omega_h^{(q)})=Prob(T \rightarrow  \{ r \in \mathbb{R}^3  ,t_c \} \cup A \rightarrow  \{ r \in \mathbb{R}^3  ,t_c \})=Prob(T \rightarrow  \{ r \in \mathbb{R}^3  ,t_c \})+Prob(A \rightarrow  \{ r \in \mathbb{R}^3  ,t_c \})-Prob(T \rightarrow  \{ r \in \mathbb{R}^3  ,t_c \} \cap A \rightarrow  \{ r \in \mathbb{R}^3  ,t_c \})=1 \Rightarrow \Omega_h^{(q)}=$ normalized. But how is it possible that $\mu(\Omega^{(q)})=\mu(\Omega^{T})=\mu(\Omega^A)=1$, while at the same time $\Omega^{(q)} \subset \Omega^{T}$ and $\Omega^{(q)} \subset \Omega^{A}$? The answer is given in Appendices A and H (it is based on the absence of monotonicity for complex measures). In this case, we will have that $\mu(T \rightarrow  \{ r \in \mathbb{R}^3  ,t_c \} \cap ( A \rightarrow  \{ r \in \mathbb{R}^3  ,t_c \})^C)=\mu(A \rightarrow  \{ r \in \mathbb{R}^3  ,t_c \} \cap ( T \rightarrow  \{ r \in \mathbb{R}^3  ,t_c \})^C)=0$, while at the same time those non-real sets are non-empty.
 
  From now on, in this Appendix, if we refer to $\Psi(r,t)$, we will be talking about $\Psi_{norm}(r,t)$. And having in mind that, we will attempt to explain what the operators may actually represent in our reversible diffusion framework by just performing a mapping between Hilbert space operator language and language of propositions (like in Appendices L, M). 
The ultimate goal here it to provide a derivation of the equation
$$<\hat{O}^B>(t_c)=\frac{\int_{\mathbb{R}^3}\Psi^*(r,t_c)\hat{O}\Psi(r,t)d^3r}{\int_{\mathbb{R}^3}\Psi^*(r,t)\Psi(r,t)d^3r}$$
where $\hat{O}$ represents a Hermitian operator that describes a specific observable. Now, the question is why the aforementioned expression for the average value of the quantum observable has the form it has in the Hilbert space language, but to answer that we will not assume Hilbert space at all.

In our diffusion description, to construct the physical quantities that describe the quantum particle, we must necessarily refer to the forward and backward moving particles first, as we did in the whole paper. Therefore, the first thing we will do is to write down the average value of the observable at time $t=t_c$ for the forward particle. Instead of pure probability set theory we will use more directly the trajectory description. The average value will thus be:
$$<\hat{O}^T> (t=t_c)= \frac{\sum_{n=1}^\infty N_{F_n}^T(t=t_c) \mathbb{E}_{F_n}^{TOT}[O^T(t=t_c)]}{N_{\mathbb{R}^3}^T(t=t_c)}$$
where $F_n=[r_n,r_{n+1}]$ is an infinitesimally small interval with measure $r_{n+1}-r_n=\Delta<<1$ and $\mathbb{E}_{F_n}^{TOT}[O^T(t=t_c)]$ is the local ensemble value of the observable $O$ for the f.p. (forward particle in Appendix L) at the spatial region $F_n$ at time $t=t_c$. The above relation reduces to saying that:
$$<\hat{O}^T>(t=t_c)=\sum_{n=1}^{\infty}\left(\frac{N_{F_n}^T(t=t_c)}{N_{\mathbb{R}^3}^T(t=t_c)}\right)\mathbb{E}_{F_n}^{TOT}[O^T(t=t_c)]$$
which is equivalent to:
$$<O^T>(t=t_c)=\sum_{n=1}^{\infty}prob\left(f.p. \rightarrow \{r_1 \in F_n,t_c\}\right)\mathbb{E}_{F_n}^{TOT}[O^T(t=t_c)]=\sum_{n=1}^{\infty} \mathbb{E}_{F_n}[O^T(t=t_c)]$$
where $\mathbb{E}_{F_n}[O^T(t=t_c)]$ is a measure of contribution of the ensemble value of $O^T$ at $F_n$ onto the overall average value of $O^T$ at $\mathbb{R}^3$. Alternatively, we can write the above result in this way:
$$\sum_{n=1}^{\infty} \mathbb{E}_{F_n}[O^T(t=t_c)]=\sum_{n=1}^{\infty} val_{F_n}[O^T(s_n,t=t_c)](r_{n+1}-r_n)=\int_{\mathbb{R}^3}val[O^T(r_1,t=t_c)]d^3r_1$$
where $s_n \in F_n$ and $val[O^T(r_1,t=t_c)]$ represents the $O-$density (i.e. energy density, momentum density etc.) of the forward moving particle at spacetime point $(r_1,t_c)$. Now, we will make the following statement:
$$val[O^T(r_1,t=t_c)]=\hat{O}_{r_1}\Psi(r_1,t_c)$$

In other words, the $O-$density of the forward-moving particle at spacetime point $(r_1,t_c)$ is considered to be equal to the operator $\hat{O}$ acted upon the wave function (which is the probability distribution for the forward-moving particle). We will now make use of that expression in order to derive the expression for the average value of this observable that corresponds to the quantum particle. And that expression will be presented of course in the Hilbert space language, similarly to the above expression.

We should now continue having in mind that, given that the number of trajectories passing through the interval $F_n$ at time $t=t_c$ is equal to $N_{F_n}^T(t=t_c)$, we can calculate that the number of reversible trajectories will be equal to  $$N_{F_n}^B(t_c)=prob(b.p. \rightarrow \{r_2 \in F_n,t'=t_c\})N_{F_n}^T(t=t_c)=\left(\frac{N_{F_n}^A(t'=t_c)}{N_{\mathbb{R}^3}^A(t'=t_c)}\right)N_{F_n}^T(t=t_c)$$
We can now write that:
$$<\hat{O}^B>(t_c)=\frac{\sum_{n=1}^\infty \left(\frac{N_{F_n}^A(t'=t_c)}{N_{\mathbb{R}^3}^A(t'=t_c)}\right)N_{F_n}^T(t=t_c) \mathbb{E}_{F_n}^{TOT}[O^T(t=t_c)]}{\sum_{n=1}^{\infty}\left(\frac{N_{F_n}^A(t'=t_c)}{N_{\mathbb{R}^3}^A(t'=t_c)}\right)N_{F_n}^T(t=t_c)} $$

Now, by dividing both numerator and denominator with $N_{\mathbb{R}^3}^T(t=t_c)$ and having in mind that $prob(f.p. \rightarrow \{r_1 \in F_n,t=t_c\})=\frac{N_{F_n}^T(t=t_c)}{N_{\mathbb{R}^3}^T(t=t_c)}$, that $prob\left(f.p. \rightarrow \{r_1 \in F_n,t=t_c\}\right)\mathbb{E}_{F_n}^{TOT}[O^T(t=t_c)]=\sum_{n=1}^{\infty} \mathbb{E}_{F_n}[O^T(t=t_c)]$, as well as that $\mathbb{E}_{F_n}[O^T(t=t_c)]= val_{F_n}[O^T(s_n,t=t_c)](r_{n+1}-r_n)$ (where $s_n \in F_n$), we finally get:
$$<\hat{O}^B>(t_c)=\frac{\sum_{n=1}^\infty (\Psi(s_n,t=t_c)(r_{n+1}-r_n))(val_{F_n}[O^T(s_n,t=t_c)](r_{n+1}-r_n))}{\sum_{n=1}^{\infty}\left(\Psi^*(s_n,t'=t_c)(r_{n+1}-r_n)\right)\left(\Psi(s_n,t=t_c)(r_{n+1}-r_n)\right)}$$
 which can be simplified further as follows:
$$<\hat{O}^B>(t_c)=\frac{\sum_{n=1}^\infty \Psi^*(s_n,t'=t_c)val_{F_n}[O^T(s_n,t=t_c)](r_{n+1}-r_n)}{\sum_{n=1}^{\infty}\Psi^*(s_n,t'=t_c)\Psi(s_n,t=t_c)(r_{n+1}-r_n)}$$
and now given that $r_{n+1}-r_n=\Delta<<1$, we eventually get that:
$$<\hat{O}^B>(t_c)=\frac{\int_{\mathbb{R}^3}\Psi^*(r,t'=t_c)val[O^T(r,t=t_c)]d^3r}{\int_{\mathbb{R}^3}\Psi^*(r,t'=t_c)\Psi(r,t=t_c)d^3r}$$

The above equation leads us to an expression equivalent to the standard expression for the average value of the observable of the quantum particle if we consider the equation we defined before and which is $val[O^T(r,t=t_c)]=\hat{O}_r\Psi(r,t_c)$. More specifically, we actually get that:
$$<\hat{O}^B>(t_c)=\frac{\int_{\mathbb{R}^3}\Psi^*(r,t'=t_c)\hat{O}\Psi(r,t=t_c)d^3r}{\int_{\mathbb{R}^3}\Psi^*(r,t'=t_c)\Psi(r,t=t_c)d^3r}$$
And just two small remarks: a) of course we have that $N_{F_n}^T(t=t_c)$, $N_{F_n}^A(t'=t_c) \in \mathbb{C}$ since those two quantities refer to (numbers of) trajectories, some of which are non-real/imaginary (see Appendices E, L, O), b) The equation we derived is also valid for the initial non-normalized wave function, since normalization constants in the expression above cancel out. 

Thus, the interpretation of $\hat{O} \Psi(r,t)$ as the $O-$density of the f.p. using our previous reasoning indeed leads us to the right expression of the observable average value for the quantum particle. And not only that, but that description is also compatible with the fact that $\hat{P}=-i\hbar \overrightarrow{\nabla}$ in standard QM since 
$$\hat{P} \Psi(r,t)=-i\hbar \overrightarrow{\nabla}\Psi(r,t)=\Psi(r,t) \overrightarrow{P}^T_{sys}(r,t) \equiv \textit{momentum density of f.p.}$$
The local ensemble momentum of the f.p. $\overrightarrow{P}^T_{sys}(r,t)$ is given by the formula $\overrightarrow{P}^T_{sys}(r,t)=m_{sys}\overrightarrow{\eta}^T_{sys}(r,t)=2m_p \overrightarrow{\eta}^T_{com}(r,t) $ in which $\overrightarrow{\eta}^T_{sys}(r,t)$ was derived in the previous Appendix as the local ensemble velocity field of the f.p. At the same time, given that we have $\int_{\mathbb{R}^3}\overrightarrow{\nabla}\rho(r,t)d^3r=0$, given the expression that we wrote in the previous Appendix about the f.p. local ensemble velocity and given the expression for the average value of a quantum particle observable, we can derive that: $$<\hat{P}^B>(t_c)=m_p<\overrightarrow{\eta}_{Bohm}(r,t_c)>$$ which is a logical result if we think that the Bohmian velocity is a local ensemble (or diffusion) velocity field.

In a similar way, we can also derive why the kinetic energy operator is equal to $\hat{T}_K=-\frac{\hbar^2}{2m_p}\overrightarrow{\nabla}^2$. Because if we calculate the $<\hat{T}_K>$, we will see that it contains an average kinetic energy term due to deterministic ensemble motion and an ensemble kinetic energy error correction term due to fluctuations (or the "zigzag" motion) related mathematically to the so-called quantum potential energy which can be both positive and negative. More specifically, a large part of quantum stochastic literature derives that:$$<\hat{T}_K>(t_c)=\int_{\mathbb{R}^3}\rho(r,t_c)\left(\frac{1}{2}m_p (\left(\overrightarrow{\eta}_{Bohm}(r,t_c)\right)^2+Q(r,t_c)\right)d^3r=<\hat{T}_K^{(flow)}>(t_c)+<\hat{T}_K^{(fluct)}>(t_c)$$
(which coincides with the average kinetic energy of Chapter II) with $$E_{K,Re}^T(r,t_c)=\frac{1}{2}m_p ((\overrightarrow{\eta}_{Bohm}(r,t_c))^2+Q(r,t_c)=\frac{(\overrightarrow\nabla S(r,t_c))^2}{2m_p}+Q(r,t_c)$$
being the local ensemble kinetic energy of the quantum particle and this can be understood if we look at the quantum Hamilton-Jacobi equation (see Chapter II) which is an expression for local ensemble energy conservation. We must say though that the above equation does not agree with Nelson's literature.

Now, let's return to Mita's paper \cite{mita2021schrodinger}, in which he used the polar transformation $\Psi=e^{R_M+iS_M}$ where $S_M=S/\hbar$ and $R_M=InR$ ($R$, $S$ refer to the variables of our approach for which we have $\Psi=R e^{iS/\hbar}$ and $S$ is the local ensemble action of the quantum particle given that $\overrightarrow{\eta}=\frac{1}{m_p}\overrightarrow{\nabla}S$). Mita did some long calculations and he derived his Eqs. (24), (26) and (30), from which we can conclude, using our framework (we won't do the calculations though because they are long) that for $V=0$:
$$-\frac{\hbar^2}{2m_p}\overrightarrow{\nabla}^2\Psi(r,t)=\left(\frac{(\overrightarrow\nabla S(r,t))^2}{2m_p}+Q(r,t)-\frac{i \hbar}{2}\frac{1}{\rho(r,t)} \overrightarrow{\nabla}\cdot \overrightarrow{J}(r,t)\right)\Psi(r,t)$$
And this expression right now, allows us to say that
$$\frac{\textit{K.E. density of f.p.}}{\textit{probability density of f.p.}}=\frac{\hat{T}_K \Psi(r,t)}{\Psi(r,t)}=\frac{1}{2}m_p ((\overrightarrow{\eta}_{Bohm}(r,t))^2+Q(r,t)+i\left(-\frac{ \hbar}{2}\frac{1}{\rho(r,t)} \overrightarrow{\nabla}\cdot \overrightarrow{J}(r,t)\right) \Rightarrow$$
$$\textit{local ensemble value of K.E. of f.p.=}E_{K,sys}^T(r,t)=E_{K,Re}^T(r,t)+i E_{K,Im}^T(r,t)$$ from which we see that we get pure sum as in the momentum case (momentum/kinetic energy of a system of two particles is equal to the sum of the momenta/kinetic energies of each particle while the velocity of the system is equal to the half of the vector sum of the two velocities due to the center-of-mass description). It is again easy to notice that the local ensemble kinetic energy of the quantum particle respects T-reversal symmetry, while the $E_{K,Im}^T(r,t)$ of the non-real particle does not (if we have T-symmetry then kinetic energy must remain invariant under T-reversal and, as we see, this holds only for the quantum particle).

Of course, it is also very easy to prove using again $\Psi=Re^{iS/\hbar}$ that: $$Re\left(\frac{i\hbar \partial_t \Psi(r,t)}{{\Psi(r,t)}}\right)=-\frac{\partial S(r,t)}{\partial t}=\textit{local ensemble value of total energy of the quantum particle}$$ and that when $V=0$, we get $\frac{i\hbar \partial_t \Psi(r,t)}{{\Psi(r,t)}}=\frac{\hat{T}_K \Psi(r,t)}{\Psi(r,t)}$ due to the Schroedinger equation whose form has been extensively justified (or "derived") from the reversible diffusion axiom in Appendices B, L and S. These two elements are enough for explaining why $\frac{i\hbar \partial_t \Psi(r,t)}{{\Psi(r,t)}}=\textit{local ensemble value of total energy of f.p.}=E_{TOT}^T(r,t)$ and thus for deriving that $\hat{H}=i \hbar \partial_t$. When a scalar potential $V(r)$ is involved, then it is obvious that local ensemble energy conservation for the f.p., will lead to $\frac{i\hbar \partial_t \Psi(r,t)}{{\Psi(r,t)}}=\frac{\hat{T}_K \Psi(r,t)}{\Psi(r,t)}+V(r)$ given that $(E_{TOT}^T(r,t))' \rightarrow E_{TOT}^T(r,t)+V(r) = E_{K}^T(r,t)+V(r)$, which leads to the Schroedinger equation when external potential is involved and which is:
$$i\hbar \frac{  \partial \Psi(r,t)}{\partial t}=-\frac{\hbar^2}{2m_p}\overrightarrow{\nabla}^2 \Psi(r,t)+V(r)\Psi(r,t)$$

It is also more than obvious that linearity of operators which is expressed through the equation $\hat{O} (\sum_{n=1}^\infty c_n \Psi_n(r,t=t_c))=\sum_{n=1}^\infty c_n(\hat{O} \Psi_n(r,t=t_c))$ is valid since the previous equation actually can be written as $\hat{O}\Psi(r,t=t_c)=\sum_{n=1}^\infty c_n(\hat{O} \Psi_n(r,t=t_c))$. And now, what this expression means is that $$val[O^T(r,t=t_c)]=\sum_{n=1}^\infty prob(f.p. \rightarrow \{\textit{n-th state}\})\left[val[O^T(r,t=t_c)]\right]_n$$
(where $\left[val[O^T(r,t=t_c)]\right]_n$ represents the $O-$density of the f.p. at the n-th state). And that result now makes absolute sense, confirming that linearity of operators is indeed a property that exists due to the logical and propositional structure behind QM and the Hilbert space language that it uses.

 \section{}

And now, we're moving onto a speculative part of the theory, for which more work needs to be done and at its current form should not be considered as finished. We are gonna talk a little bit about non-commuting observables but without smuggling in that $[\hat{x},\hat{p}]=i \hbar$ (we have derived it though indirectly in the previous Appendix, where we showed that $\hat{P}=-i \hbar \frac{\partial}{\partial x}$ and it would be very easy for us to show non-commutativity for observables of the form $\hat{O}_1=f(\hat{x})$ and $\hat{O}_2=g(\hat{P})$) and to take the product $\hat{O}_1\hat{O}_2$ seriously as a product between two observables. In such case,  we have to interpret $\hat{O}_1\hat{O}_2 \Psi(x,t_c)$ as $(O_1XO_2)-\textit{density of f.p. at}$ $(x,t_c)$ where $X$ represents a scalar product of two observables for which we have that the spacetime value of the one on the left is conditional to the spacetime value of the one on the right. We will begin by referring first to the ensemble values at infinitesimally small spatial intervals, similarly to the previous Appendix. Thus, we have that at time $t_c$: $(O_1XO_2)-\textit{value at}$ $F_n=(O_2-\textit{value at $F_n$})\cdot(O_1-\textit{value at $F_n$ given the $O_2-\textit{value at $F_n$}$})=V^{F_n}_2  V^{F_n}_{1|2}=W^{F_n}_{12}$, where $F_n$ is a random unknown spatial interval $[x_n,x_{n+1}]$ in which the particle is located at time $t_c$. Similarly, we have that $(O_2XO_1)-\textit{value at}$ $F_n=V^{F_n}_1  V^{F_n}_{2|1}=W^{F_n}_{21}$.

Suppose that conditioning on the local ensemble value of one observable of the f.p. does not give information about the local ensemble value of the other. In that case, $V^{F_n}_{1|2}$ and $V^{F_n}_{2|1}$ have indeterminate values (they exist but the information required for finding them is limited). As a result, $W^{F_n}_{12}$ and $W^{F_n}_{21}$ also have indeterminate values, and given that they are different expressions, they are non-equal with probability 1. Imagine two random limit expressions $A$ and $B$ that are completely distinct and their form is indeterminate. The probability of them having equal value is zero (despite the fact that it can happen), because each limit can give all possible real values.

The result of having $W^{F_n}_{12} \neq W^{F_n}_{21}$ $\forall F_n$ urges us to clearly suggest that transitioning to the continuum spectrum of position leads to $\frac{1}{\Psi(x,t_c)}\{(O_1XO_2)-\textit{density of f.p. at}$ $(x,t_c)\} \neq $ $\frac{1}{\Psi(x,t_c)}\{(O_2XO_1)-\textit{density of f.p. at $(x,t_c)$} \}$ and thus to $\{(O_1XO_2)-\textit{density of f.p. at}$ $(x,t_c)\} \neq \{(O_2XO_1)-\textit{density of f.p. at}$ $(x,t_c)\}\Rightarrow \hat{O}_1 \hat{O}_2 \Psi(x,t_c)\neq \hat{O}_2 \hat{O}_1 \Psi(x,t_c)$ which allows us to safely conclude that $\hat{O}_1 \hat{O}_2  \neq \hat{O}_2 \hat{O}_1$ for two observables/physical quantities for which the local ensemble value of the one can not determine the local ensemble value of the other. 

Such observables seem to be position and momentum and their "incompatibility" does not contradict classical mechanics, but in fact seems to completely agree with it. Classical mechanics does not say that that if position is known at a time $t_c$, then momentum is known too. There must be additional information and more specifically, in order to find $P_{cl}(x(t_c),t_c)$, we need to know also $x(t_c+\delta t)$, where $\delta t<<1$, since $P_{cl}(x(t_c),t_c)=m_p\left[\lim_{\delta t \rightarrow 0}\frac{x(t_c+\delta t)-x(t_c)}{\delta t}\right]$. If such information is not available, then momentum value at that point in spacetime is indeterminate. Similarly, in our "derivation" in this Appendix, we do not provide any extra information (i.e. mathematical form of wave function) and this leads to a similar result as in classical physics, namely that the local ensemble momentum of the f.p. at a given spacetime point $(x_c,t_c)$, which is given by the equation $P^T_{sys}(x_c,t_c)=-i \hbar \frac{\partial_x \Psi(x,t_c)|_{x=x_c}}{\Psi(x_c,t_c)}$ (see Appendix S and beginning of Appendix T), is determinate only when additional information exists. To state it simply, if the probability distribution of the f.p. is known, then knowing position at time $t_c$. leads to determinate local ensemble momentum, otherwise, if $\Psi$ is unknown, then it is indeterminate. 

And that also holds in reverse, namely if momentum in classical physics is known at time $t_c$, we still need $x(t_c+\delta t) $ to calculate $x(t_c)$. Similarly for the QM case, if $P_{sys}^T(x,t_c)$ is known (without knowing the position), then $x$ could be known if we solved $P_{sys}^T(x,t_c)=-i \hbar \frac{\partial_x \Psi(x,t_c)}{\Psi(x,t_c)}$ with respect to $x$, but this requires again extra information about the mathematical form of the wave function.

It's more than obvious that the same discussion also holds when the one observable/quantity is a function of position only and the other a function of momentum only, while also that it does not apply when both are functions of position or momentum and this is also compatible with classical physics. Knowing only the instantaneous momentum of the classical particle is enough on its own to determine the instantaneous kinetic energy since $E_K=\frac{P^2}{2m_p}$. In QM, we observe the "compatibility" of those observables since $[\hat{T}_K, \hat{P}]=0$. And to adjust this simple case into our previous discussion, we will have that 
$V^{F_n}_{2|1}$, $V^{F_n}_{1|2}$ are perfectly determinate. And that's because if ensemble momentum value of f.p. $P^T_{sys}(x,t_c)$ is known, then wave function can be computed by solving the differential equation $P^T_{sys}(x,t_c)=-i \hbar \frac{\partial_x \Psi(x,t_c)}{\Psi(x,t_c)}$ and will give solutions of the form $\Psi(x,t_c)=De^{\frac{i}{\hbar}G(x,t_c)}=D(x_0) e^{\frac{i}{\hbar}\int_{{x_0}}^{x}P^T_{sys}(x',t_c) dx'}$, where $\partial_x G(x,t_c)=P^T_{sys}(x,t_c)$ and $x_0$ a constant. Substituting this expression into the expression for the local ensemble kinetic energy of the f.p. that we derived in the previous Appendix, we derive a result that only depends only on $P^T_{sys}(x,t_c)$ and on the Bohm velocity which is also obvious that can be expressed with respect to $P^T_{sys}$ given that $\eta_{Bohm}(x,t_c)=\frac{Re(P^T_{sys}(x,t_c))}{m_p}$). In other words, the local ensemble energy of the f.p. can be completely determined given the local ensemble momentum of the f.p. since the first one can be written in a form that depends solely on the second one.

Of course, we can also do the opposite, which is that if we know the local ensemble kinetic energy of the f.p., we can find the local ensemble momentum. That requires solving the differential equation $E_{K,sys}^T(x,t_c)=- \frac{\hbar^2}{2m_p}\frac{\partial x^2 \Psi(x,t_c)}{\Psi(x,t_c)}$ and later substituting the expression of the wave function that we found into the expression for the local ensemble momentum.

This leads to perfect determinacy of $W^{F_n}_{12}$ and $W^{F_n}_{21}$ and to the equality relation between them: $W^{F_n}_{12}=W^{F_n}_{21}$. And this on its turn leads to $\hat{T}_K \hat{P}=\hat{P}\hat{T}_K$ in opposition to momentum-position pair, where indeterminacy of $V^{F_n}_{2|1}$ and $V^{F_n}_{1|2}$ leads to indeterminacy of $W^{F_n}_{12}$, $W^{F_n}_{21}$ and thus with probability 1 to their non-equality relation and as a consequence to $\hat{x} \hat{P} \Psi(x,t_c) \neq \hat{P} \hat{x} \Psi(x,t_c)$, which is generally true in standard QM. 

So, to see if two operators commute, we have to see if the local ensemble values $y_1$ and $y_2$ corresponding to those physical quantities can be written in a form $y_1=f(y_2)$ where $f$ is a known function and does not include any unknown parameters (i.e. wave function). So, to see why $[\hat{L}_x,\hat{L}_y]\neq 0$, for instance, we have to confirm that $L^T_{x,sys}\neq f(L^T_{y,sys})$, namely that the local ensemble quantities are not directly correlated mathematically. And by the way, since we're talking about angular momentum, we can do the same analysis for it and it will be very easy given that we know that $\overrightarrow{L}^T_{sys}=\overrightarrow{r}x\overrightarrow{P}^T_{sys}$ where $\overrightarrow{P}^T_{sys}$ is an already derived quantity. In other words, it will be extremely easy to prove that $$\overrightarrow{L}_{sys}^T(r,\theta,t)=\frac{\hat{L} \Psi(r,\theta,t)}{\Psi(r,\theta,t)}=\frac{\left(\hat{r}x\hat{P}\right) \Psi(r,\theta,t)}{\Psi(r,\theta,t)}=\overrightarrow{L}_{Re}^T(r,\theta,t)+i\overrightarrow{L}_{Im}^T(r,\theta,t) $$ (for a cylindrical coordinate quantum system for example) since $\frac{\hat{P}\Psi(r,\theta,t)}{\Psi(r,\theta,t)}=\overrightarrow{P}^T_{sys}(r,\theta,t)=\overrightarrow{P}^T_{Re}(r,\theta,t)+i\overrightarrow{P}^T_{Im}(r,\theta,t)$.

\section{}
\subsubsection{ANSWERS TO POSSIBLE OBJECTIONS }

Some people may object the idea of having non-differentiable trajectories in a physical theory since they are not physically realistic. More specifically, they are approximations that are done in order to construct models that will give us more general (but successful though) statistical results (i.e. Fokker-Planck equation that describes the probability density evolution of the Brownian particle). In fact, the overdamped limit that we have assumed in Appendix L (and that leads us to our SDEs) is an idealization in which the inertia is very small. This idealization helps us recover the Schroedinger equation but at the cost of introducing non-differentiable trajectories in which an infinitesimal displacement that happens at time $t+dt$ has absolutely no correlation with the displacement that happens at time $t$. 

The stochastic approach that we will demonstrate very briefly is closer to the more minimalistic (mathematically) stochastic approach by Bohm and Vigier, where the actual velocity of the particle was given as the sum between the deterministic Bohm velocity and a fluctuating velocity term $\overrightarrow{\xi}(t)$ with mean zero and with the finite noise correlation property $Cov[\overrightarrow{\xi}(t),\overrightarrow{\xi}(t+dt)] \neq 0$ which implies that the particle has memory of the previous kicks and that the trajectories will now be differentiable. This does not however destroy Markovianity, but it just introduces a deterministic character in the microscopic dynamics which is expected in realistic diffusion systems. Recall that in Newtonian dynamics, velocities are perfectly correlated in time, but still the dynamics is considered Markovian.

In order to proceed further, we recall that in driftless stochastic systems represented by the ordinary diffusion equation (at the non-realistic overdamped limit), we can say that the actual local velocity of the Brownian particle is equal to $\overrightarrow {\nu}_{cl}(r,t)=\overrightarrow{u}_{d,cl}(r,t)+\overrightarrow{\xi}_{cl}(t)=-D_{cl}\frac{\overrightarrow{\nabla}\rho_{cl}(r,t)}{\rho_{cl}(r,t)}+\overrightarrow{\xi}_{cl}(t)$ with $<\overrightarrow {\nu}_{cl}(r,t)>_L=\overrightarrow{u}_{d,cl}(r,t)$, where $L$ indicates that we're talking about \textit{local} mean value and not about a global average value in the whole $\mathbb{R}^3$ like in Appendix T. In other words, realistically speaking, the actual local velocity is given by the sum of the diffusion velocity and a fluctuating velocity field with mean zero ($<\overrightarrow{\xi}_{cl}(t)>_L=0$).

Similarly, in our case the instantaneous velocity of the forward-moving particle, given that we have found the current (or diffusion) velocity from Schroedinger diffusion equation (Appendix S), is given by the equation 
$$\overrightarrow{v}^T_{sys}(r,t)=\overrightarrow{\eta}^T_{sys}(r,t)+\overrightarrow{\xi}_{sys}^T(t)$$
 with $<\overrightarrow{\xi}_{sys}^T(t)>_L=0$. Of course, we also know that $\overrightarrow{\xi}_{sys}^T(t)=\frac{1}{2}\left(\overrightarrow{\xi}_{Re}^T(t)+i\overrightarrow{\xi}_{Im}^T(t)\right)$ since the stochastic velocity term corresponding to the superposed f.p. (namely to the system of two particles) is equal to the stochastic velocity of the center of mass of the two particles, hence the division by 2. The $i$ in front of $\overrightarrow{\xi}_{Im}^T(t)$ means that the stochastic velocity term of the non-real particle may be $\overrightarrow{\xi}_{Im}^T(t)$, but it is a non-real stochastic velocity term that describes irreversible pseudo-trajectories. Using those two equations and the equation $\overrightarrow{\eta}^T_{sys}(r,t)=\frac{1}{2}\left(\overrightarrow{\eta}_{Re}^T(r,t)+i\overrightarrow{\eta}_{Im}^T(r,t)\right)$ that was derived in Appendix S, we are led into:
$$\overrightarrow{v}^T_{sys}(r,t)=\frac{1}{2}\left(\left(\overrightarrow{\eta}^T_{Re}(r,t)+\overrightarrow{\xi}_{Re}^T(t)\right)+i\left(\overrightarrow{\eta}^T_{Im}(r,t)+\overrightarrow{\xi}_{Im}^T(t)\right)\right)$$
where $\overrightarrow{\eta}^T_{Re}(r,t)=\overrightarrow{\eta}_{Bohm}(r,t)$ and $\overrightarrow{\eta}^T_{Im}(r,t)=\overrightarrow{u}_{d}(r,t)$, That expression eventually leads us to
$$\overrightarrow{v}^T_{sys}(r,t)=\frac{1}{2}\left(\overrightarrow{v}^T_{Re}(r,t)+i\overrightarrow{v}^T_{Im}(r,t)\right)$$
in which $\overrightarrow{v}^T_{Re}(r,-t)=-\overrightarrow{v}^T_{Re}(r,t)$ and $\overrightarrow{v}^T_{Im}(r,-t)\neq-\overrightarrow{v}^T_{Im}(r,t)$ given that $\overrightarrow{\eta}_{Bohm}(r,-t)=-\overrightarrow{\eta}_{Bohm}(r,t)$, $\overrightarrow{u}_{d}(r,-t)\neq -\overrightarrow{u}_{d}(r,t)$, $\overrightarrow{\xi}_{Re}^T(-t)=-\overrightarrow{\xi}_{Re}^T(t)$ and $\overrightarrow{\xi}_{Im}^T(-t)\neq -\overrightarrow{\xi}_{Im}^T(t)$. 

What the above expression means is that the actual local velocity of the forward-moving particle can be written as the average of the actual local velocity of the reversible quantum particle and the actual local velocity of the irreversible non-real particle. And of course, the $i$ exists in front of $\overrightarrow{v}^T_{Im}(r,t)$ to tell us that $\overrightarrow{v}^T_{Im}(r,t)$ is indeed the instantaneous velocity that describes the non-real particle, but it is in fact a pseudo-velocity field that does not exist in reality, since it describes non-real trajectories. In other words, writing that $\overrightarrow{v}^T_{sys}(r,t)=\frac{1}{2}\left(\overrightarrow{v}^T_{Re}(r,t)+i\overrightarrow{v}^T_{Im}(r,t)\right)$ is equivalent to writing $\overrightarrow{v}^T_{sys}(r,t)=\frac{1}{2}\left(\overrightarrow{v}^T_{Re}(r,t)+\overrightarrow{v}^T_{Im}(r,t)\right)$ and denoting that $\overrightarrow{v}^T_{Im}(r,t)$ describes a non-real trajectory.

In that sense, the instantaneous quantum trajectories will be described by the equation
$$\overrightarrow{v}^B(r,t)=\overrightarrow{v}^T_{Re}(r,t)=\overrightarrow{\eta}_{Bohm}(r,t)+\overrightarrow{\xi}_{Re}^T(t)$$
and the set of those is given by the intersection of forward and backward actual paths described by the corresponding local velocity fields, and this leads us to the Born rule. We should note that $\overrightarrow{v}^T_{Im}(r,t)$ has a mathematical form completely analogous to that of classical (irreversible) diffusion and does not reverse sign under time-reversal. This suggests an absence of reversibility for the non-real processes at the statistical level (see justification in Appendix R via the use of the concept of chaos and not via coarse-graining), but even at the microscopic level too, again due to chaos. And of course, from the above equation, it is extremely natural to identify the diffusive motion of the quantum particle given that $\overrightarrow{\xi}_{Re}^T(t)$ is the real stochastic velocity term with the property $<\overrightarrow{\xi}^T_{Re}(t)>_L=0$ (inherited from the zero-mean value of $\overrightarrow{\xi}^T_{sys}(t)$)  and is added upon the quantum diffusion/current velocity that satisfies the quantum continuity equation
$$\frac{\partial \rho(r,t)} {\partial t}=-\overrightarrow{\nabla} \cdot \left(\rho(r,t) \overrightarrow{\eta}_{Bohm}(r,t)\right)$$

\textbf{It turns out that having a more physically realistic description of trajectories helps us have a more detailed picture of quantum motion given the quantum system that we have in front of us. And of course, the idea that Bohm velocity represents the local ensemble velocity field is not only mentioned in Bohm and Vigier's paper, but also in Tsekov and Heifetz's paper \cite{tsekov2017derivation}. In a different paper \cite{tsekov2021classical}, Tsekov also criticizes the SDE approach of Nelson given that the forward and backward drifts depend on the probability distribution itself which is a little bit problematic if you consider that the Wiener approach must refer to individual trajectories and therefore must not include ensemble quantities. And for me personally, the approach of this Appendix is much more accurate than the one in Appendix L. And of course, the same discussion about sets of trajectories that intersect producing the reversible trajectories applies here in the same way, having in mind the following statement (see Appendix L): $\{\textit{f.p. at $(x,t_c)$}\}=\{\textit{q.p. at $(x,t_c)$}\} \lor \{\textit{n-r.p at $(x,t_c)$}\}$, where $ $n-r.p.$\equiv\textit{"non-real particle"}$. }

\textbf{Another possible objection could be the following: how is it possible to obtain reversible stochastic trajectories from a complex diffusion equation that is meant to describe families of irreversible trajectories as mentioned before? And the answer to that is that, in fact, all trajectories are irreversible, including quantum trajectories too, due to the impossibility condition $|x_1-x_2|=\epsilon \neq 0$ (see Appendix O) imposed by probability theory (see Appendix R). But, quantum trajectories are chosen in a way so that the irreversibility is extremely small and unobservable ($\epsilon \rightarrow 0$). In that sense, the set of quantum trajectories $S[x_{rev}(t)]$ emerges as a tiny subset of the hyper-set $S[x_{Schr}(t)]$ of irreversible stochastic trajectories with the condition that the irreversibility approaches zero but never becomes zero. And those trajectories are considered as (approximately) reversible. However, as we've noted in Appendix R, the limiting procedure may in fact be a mathematical procedure from which we can obtain exact reversibility of quantum trajectories and may not indicate intrinsic microscopic irreversibility imposed by probability theory. In that case, time's arrow can be understood through the classic coarse-graining resolution, but I must admit that I am not a supporter of that idea (see Appendix R again, fifth paragraph).}

All this analysis though, does not necessarily undermine the SDE (Wiener process) analysis of Appendices L and Q. In the SDE approach, however, it is not right to say that $dx_{irr}(t)=i\sqrt{\frac{2\hbar}{m_p}}dw(t)$ is equivalent to saying that $dx_{irr}(t)=\sqrt{\frac{2\hbar}{m_p}}dw(t)$ but $x_{irr}(t)$ just refers to non-real trajectories, because, in such case, the non-real stochastic process would be pathwise identical to the real one which is reversible, and that of course does not make sense. So, in Appendix Q, we claimed that $w_{irr}(t)=iw(t)$ which corresponds to a completely different Wiener process from $w_{rev}(t)=w(t)$ and is both non-real and irreversible (see time-inversion analysis in the Langevin approach of Appendix L). 

Just a small note though: both real and non-real stochastic processes are given by an SDE of the type $dx_{gen}(t)=\sqrt{2D}dw_{gen}(t)$ indicating that they both are sub-processes of a classical driftless Wiener process. Both are irreversible, but the one of them (as we've said before) is approximately reversible and is given by the SDE $dx_{rev}(t)=\sqrt{\frac{2 \hbar}{m_p}}dw(t)$. Furthermore, it's worth to mention that if someone wants to avoid generalized covariances, that were used in Appendix Q and is fine with treating the Schroedinger diffusion constant as imaginary ($D_{sys}=\frac{i\hbar}{2m_p}$) instead of real (and equal to $\frac{\hbar}{2m_p}$), they can easily recognize that the following equation is indeed satisfied:
$$Cov(dx(t),dx(t))=2D_{sys}dt= \frac{i\hbar}{m_p}dt$$
with $dx(t)=\frac{1}{2}(dx_1(t)+dx_2(t))$, $dx_1(t)=\sqrt{\frac{2\hbar}{m_p}}dw(t)$ and $dx_2(t)=idx_1(t)$ and thus our center-of-mass description is still a valid one. Anyhow, I still prefer the approach with the differentiable trajectories, because it is physically more realistic and has more direct connection to the quantum formalism.

Another issue that someone might find problematic is that the description on a 2-particle diffusion process via a hyper-sample space is a little bit exotic. The answer to that is that, first of all, 2 (or even more) particles description can be done into a single event space in conventional probability theory. For instant, if we have a system of two particles 1 and 2, we can define a sample space $\Omega^\Gamma_{1,2}$ in which the sub-events $\Gamma\rightarrow \{x \in F,t\}=\Gamma_1\rightarrow \{x \in F,t\}\cup \Gamma_2\rightarrow \{x \in F,t\} \subseteq\Omega^\Gamma_{1,2}$ describe the events of finding particle 1 or 2 at spacetime point $(x,t)$ where $x \in F \subseteq \mathbb{R}$. A very similar situation holds in QM, but we also have pseudo-events involved in that case. In other words, we have the quantum hyper-sample space, which is the fundamental sample space that describes both real and non-real particles and of course the "real sample space" is emergent in that picture (see more analysis in the final subsection of this Appendix). Someone could go even further and suggest that since real sample space refers to specific time instants $t_c \in [0,t_0]$, that process of emergence holds at discrete time steps in that time interval, which could also allow for the stochastic trajectories to be differentiable. In another paper, we will show how Schroedinger equation and Born rule are derived from random walks through the forward and backward trajectory overlap.

\subsubsection{OPEN PROBLEMS}
As we've discussed, the open problems (for which there are ideas but I'm not ready to present them yet and the extent of this paper is extremely large) are: 1) further discussion on non-commutativity and non-locality, 2) derivation of the general/abstract form of Schroedinger equation. For the last one, very briefly, I believe that it represents a relation that equates measurable sets together and is derived from the position representation a)once you integrate the Schroedinger equation in its Hamiltonian form $\hat{H}\Psi(r,t)=i \hbar \frac{\partial}{ \partial t}\Psi(r,t)$ (which was derived) in the whole space region $\mathbb{R}^3$, b)once you generalize the probabilities of quantum events and the average value of the Hamiltonian of the forward-moving particle as measures of sets, and c)you account for the mapping between sets and vectors (the generalization of quantum events to sets and the mapping have already been shown in Appendix L). But no more discussion will be done since this is something that requires a calculus of sets which someone may need work to establish fully.

It is also obvious that the Heisenberg uncertainty principle can easily be proven given that we have derived the momentum operator and the formula for the average value of observables. We just have to follow the steps of standard QM (using the Cauchy-Schwarz inequality relations etc.). However, if we want to develop a purely stochastic (or dynamical) derivation, then that will be done in a future paper. Anyway, the physical explanation of quantum uncertainty principle has already been provided in Appendices D and R. In general, what I wanted to demonstrate in Appendix U is that the uncertainty principle may contain the non-commutative structure at its lower bound, but non-commutativity and uncertainty principle could possibly represent two different things. However, someone could choose a safer path and suggest that non-commutativity does not reflect the independence of the local ensemble value of two physical quantities (conditioning on the one is not enough by itself to find the other), but the inability of a well-defined mathematical definition of the one when the other is fixed (despite the fact that it physically exists) like in the quantum uncertainty principle case. In other words, that value indeterminacy that we've discussed in the first three paragraphs of the previous Appendix, and that leads to non-commutativity, may in fact hold due to the violent stochastic nature of quantum dynamics (see again Appendix R). 

To sum up, it would be immature to claim that this theory is ready to be the final theory for non-relativistic QM. And I say that not only because of the aforementioned open problems, but also because it is presented in a semi-rigorous, instead of a fully rigorous form (an introduction to a rigorous establishment of complex probability measures will be done in the next subsection). However, I strongly believe that it is the first, and in my perspective, the most appropriate step into the final non-relativistic theory of QM, given the alternatives that exist and the problems they face.

\subsubsection{A SMALL RIGOROUS INTRODUCTION TO COMPLEX PROBABILITY THEORY}

This part is not mandatory for the reader to read...In this paper, and especially in the Appendices, we attempted to give a flavor of why specific probability measures should be complex in the case that we want to describe a reversible diffusion theory, both at the microscopic and (as a result) at the ensemble level. However, as we've said, our presentation was semi-rigorous and not fully rigorous and the reason why I use some shorthands is because I want this manuscript to not exceed a certain magnitude and to give more emphasis to the physical aspect of quantum stochastic processes. I am not willing to provide the most rigorous form of the theory, but I just want to make a brief introduction to the rigorous mathematical logic behind the establishment of our non-real sets. To be more specific, this subsection will just give emphasis to the fact that we are supposed to work in a product sample space (real sample space) that emerges-if we want to be mathematically rigorous-from certain mathematical procedures starting from more primary sample spaces and probability sets. Of course, we've discussed that before (see Appendices A, G and certainly O, P), but now we will give an analysis of what exactly our T and A sets actually would mean for a mathematician. In that sense, this section could constitute a material that could perfectly replace Appendices A, B, where we first introduced non-real sets/pseudo-sets and their reason for existence. 

Let $\mu$ describe a generalized probability measure (we can also have complex measures) that satisfies the standard probability measure rules of countable additivity for disjoint sets and of normalizability for the total sample space and also has the property $\mu(\varnothing)=0$, but also lacks the property of positivity and monotonicity of classical probability measures (Appendix H, T). Also, let $\Omega^T_{in,t=t_c}$ be the initial (normalized) sample space referring to the forward-moving particle at the time instant $t=t_c$ and $\Omega^A_{in,t'=t_c}$ be the initial (normalized) sample space referring to the backward-moving particle at the time instant $t'=t_c$. We can write that $\forall F_1,F_2 \subseteq \mathbb{R}$ such that $F_1 \subset F_2$ for instance, it is true that $\Omega^T_{in,t=t_c}=T_{in}\rightarrow \{x_1\in F_1,t=t_c\} \cup (T_{in}\rightarrow \{x_1\in F_1,t=t_c\})^C$
and that $\Omega^A_{in,t'=t_c}=A_{in}\rightarrow \{x_2\in F_2,t'=t_c\} \cup (A_{in}\rightarrow \{x_2\in F_2,t'=t_c\})^C$. In order to describe the joint experiment where both forward and backward particles are included, we use the product sample space $\Omega_{h,in}=\Omega^T_{in,t=t_c}X\Omega^A_{in,t'=t_c}$. In that hyperset, from which the quantum hyper-sample space in which we were referring to in Appendices H and P, emerges under specific conditions, we can define the following set equations: the first one is $T_{in}\rightarrow \{x_1\in F_1,t=t_c\}X\Omega^A_{in,t'=t_c}=\left((T_{in}\rightarrow \{x_1\in F_1,t=t_c\}X\Omega^A_{in,t'=t_c})\cap (\Omega^T_{in,t=t_c}XA_{in}\rightarrow \{x_2\in F_2,t'=t_c\})\right)\cup\left((T_{in}\rightarrow \{x_1\in F_1,t=t_c\}X\Omega^A_{in,t'=t_c})\cap \left(\Omega^T_{in,t=t_c}X(A_{in}\rightarrow \{x_2\in F_2,t'=t_c\})^C\right)\right)$ which naturally emerges from the set equation $$T_{in}\rightarrow \{x_1\in F_1,t=t_c\}X\Omega^A_{in,t'=t_c}=$$ $$=T_{in}\rightarrow \{x_1\in F_1,t=t_c\}X\left(A_{in}\rightarrow \{x_2\in F_2,t'=t_c\} \cup (A_{in}\rightarrow \{x_2\in F_2,t'=t_c\})^C\right)$$ after we take into account that for independent events $A \subseteq \Omega_1$ and $B,D \subseteq \Omega_2$, we have that $AX(B \cup D)=(AXB)\cup(AXD)$ in the product space $\Omega=\Omega_1X\Omega_2$ and we also consider that $AXB=(AX \Omega_2)\cap(\Omega_1X B)$ and $AXD=(AX \Omega_2)\cap(\Omega_1X D)$. And now, if $D=B^C$ (or differently $B \cup D=\Omega_2$, with $B \cap D=\varnothing$), that gives us that $AXD=AXB^C$ (all those relations come from elementary set theory or from proposition equivalences).
Of course, the second equation is for the set $\Omega^T_{in,t=t_c}XA_{in}\rightarrow \{x_2\in F_2,t'=t_c\}$, whose mathematical expression will be obvious and emerges in a similar way as the previous one.

Now you can understand why I avoided using this language of modified product sets and used simpler math, and this is because the expressions are extremely lengthy. We can now take the condition $C_{\epsilon \rightarrow 0}=\left\{\lim_{\epsilon \rightarrow 0}\{|x_1-x_2| \leq \epsilon\}\right\}$, as we've shown in Appendix O, which will give rise to $\lim_{\epsilon \rightarrow 0}F_2=F_1=F$. Applying the latter into the expression we wrote for the modified product form of the $T_{in}$ set, we actually get our real and non-real sets:
$$\lim_{\epsilon \rightarrow 0}\left((T_{in}\rightarrow \{x_1\in F,t=t_c\}X\Omega^{A,F_E}_{in,t'=t_c})\cap (\Omega^T_{in,t=t_c}XA^{F_E}_{in}\rightarrow \{x_2\in F_{2 },t'=t_c\})\right)\Rightarrow \textit{r.set}$$
$$\lim_{\epsilon \rightarrow 0}\left((T_{in}\rightarrow \{x_1\in F,t=t_c\}X\Omega^{A,F_E}_{in,t'=t_c})\cap \left(\Omega^T_{in,t=t_c}X(A^{F_E}_{in}\rightarrow \{x_2\in F_2,t'=t_c\})^C\right)\right)\Rightarrow \textit{n-r.set}$$
where the spatial interval $F_E=[x_1-\epsilon,x_1+\epsilon]$ and the events $\Omega^{A,F_E}_{in,t'=t_c}=\Omega^{A}_{in,t'=t_c}|_{C_{\epsilon \rightarrow 0}}$ and $\Omega^T_{in,t=t_c}XA^{F_E}_{in}\rightarrow \{x_2\in F_2,t'=t_c\}=\Omega^T_{in,t=t_c}XA^{}_{in}\rightarrow \{x_2\in F_2,t'=t_c\}|_{C_{\epsilon \rightarrow 0}}$ are defined only in the product space where events have both $x_1$ and $x_2$ dependence.

In addition, we know very well that our non-real set/pseudo-set (as well as the real one) can never be empty since it emerges from the set $T_{in}\rightarrow \{x_1\in F,t=t_c\}X(A_{in}\rightarrow \{x_2\in F_2,t'=t_c\})^C$ (once we apply the condition $C_{\epsilon \rightarrow0}$ into the product set) which by definition can not be empty given that neither $T_{in}$ nor $A_{in}$ set is empty. In other words, non-real sets emerge from our mathematical inability to describe pseudo-events (irreversible trajectories) using empty sets. Theoretically, as we've mentioned in Appendices A and B, we could possibly have a set-theoretic description of reversible trajectories without including complex probability measures in the discussion by just denoting that $\Omega^A_{in,t'=t_c}\equiv \Omega^T_{in,t=t_c}$, but the problem is that a non-equilibrium stochastic version of such H-V theory leads to real irreversible diffusion-like equations, which break statistical reversibility and, for me, microscopic reversibility also (see chaos explanation above, that is responsible for both statistical and microscopic reversibility breaking, once you accept the existence of tiny microscopic asymmetry), which contradicts the initial hypothesis of individual path reversibility anyway.

It's worth to mention that if we apply the conditions $C_{\epsilon \rightarrow 0}$ and $F_1=\mathbb{R}$ to the product hyperset $\Omega_{h,in}$ in which we are working, we restrict ourselves to a specific kind of set which is equivalent to the quantum hyper-sample space, which is equal to:
$$\Omega_h^{(q)}=\lim_{\epsilon \rightarrow 0}\left((T_{in}\rightarrow \{x_1\in \mathbb{R},t=t_c\}X\Omega^{A,F_E}_{in,t'=t_c})\cup (\Omega^T_{in,t=t_c}XA^{F_E}_{in}\rightarrow \{x_2\in F_{2 },t'=t_c\})\right)$$
and it describes all possible real and non-real trajectories. More specifically, it contains 3 disjoint subsets, from which one of them is the emergent real sample space of reversible trajectories (it contains real probability events and it's normalized by construction as shown in Appendix G) and the other 2 refer to the entire family of pseudo-trajectories (forward and backward irreversible trajectories). More specifically:
$$\Omega_h^{(q)}=\Omega^T_{in,t=t_c}X\Omega^A_{in,t'=t_c}|_{C_{\epsilon \rightarrow 0};F=\mathbb{R}}=S_{r}\cup\ S_{n-r}^{(1)}\cup\ S_{n-r}^{(2)}$$
where:
\begin{itemize}
    \item $S_r=\Omega^{(q)}=\lim_{\epsilon \rightarrow 0}\left((T_{in}\rightarrow \{x_1\in \mathbb{R},t=t_c\}X\Omega^{A,F_E}_{in,t'=t_c})\cap (\Omega^T_{in,t=t_c}XA^{F_E}_{in}\rightarrow \{x_2\in F_{2 },t'=t_c\})\right)$
    \item $S_{n-r}^{(1)}=\lim_{\epsilon \rightarrow 0}\left((T_{in}\rightarrow \{x_1\in \mathbb{R},t=t_c\}X\Omega^{A,F_E}_{in,t'=t_c})\cap \left(\Omega^T_{in,t=t_c}X(A^{F_E}_{in}\rightarrow \{x_2\in F_{2 },t'=t_c\})^C\right)\right)$
    \item $S_{n-r}^{(2)}=\lim_{\epsilon \rightarrow 0}\left(\left((T_{in}\rightarrow \{x_1\in \mathbb{R},t=t_c\})^CX\Omega^{A,F_E}_{in,t'=t_c}\right)\cap (\Omega^T_{in,t=t_c}XA^{F_E}_{in}\rightarrow \{x_2\in F_{2 },t'=t_c\})\right)$
\end{itemize}

We should also not forget to mention that Schroedinger diffusion processes are of course ordinary irreversible diffusion processes (from which real diffusion emerges if we consider irreversibility as approaching zero). Thus, the total event space that describes it, is, as we said in the beginning, a normalized sample space. That means that for the forward case, we have $\mu(\Omega^T_{in, t=t_c})=1$ or alternatively $\mu(T_{in}\rightarrow \{x_1\in \mathbb{R},t=t_c\})=1$. This leads to $\mu(\Omega_{h,in})=\mu(\Omega^T_{in,t=t_c}X\Omega^A_{in,t'=t_c})=\mu(T_{in}\rightarrow \{x_1\in \mathbb{R},t=t_c\}X\Omega^A_{in,t'=t_c})=1$, that must hold $\forall F_2 \subseteq \mathbb{R}$ and thus also when inserting the condition that $F_2=F_E$, which means that $\mu(T_{in}\rightarrow \{x_1\in \mathbb{R},t=t_c\}X\Omega^{A,F_E}_{in,t'=t_c})=1$. And now, since
$T_{in}\rightarrow \{x_1\in \mathbb{R},t=t_c\}X\Omega^{A,F_E}_{in,t'=t_c}=S_{r}\cup\ S_{n-r}^{(1)}$
with $\mu(S_{r}^{})=1$ and $S_r \cap S_{n-r}^{(1)}=\varnothing$, we are led into $\mu(S_{n-r}^{(1)})=0$ (see this discussion in the beginning of Appendix T). 

This analysis that we did in this subsection gives the same exact predictions as before. The only difference is that here, we have considered that in order to be fully compatible with rigorous probability theory, we must first put forward and backward processes together in a common product sample space and then make the derivations. This approach does not cancel the previous approach, but it's just a more rigorous presentation that explains how this artificial independence (Appendix B) of forward and backward processes via those pseudo-paths can lead to a time-symmetric diffusion. And of course, if we look at our shorthand definitions of sets and derivations of the emergent sets and probabilities, we can clearly see a perfect match with the analysis done here.

What I think that is worth reminding is the role and importance of context in the set-theoretic structure of quantum mechanics. For instance, in Appendix P, we explained how the context of two potentialities coexisting in a quantum experiment leads to different set structure of the time-symmetric B set comparing to the case of having a joint experiment of two sub-experiments with different context each. Now the B set, which was denoted as $B_{1 \lor 2}$ now contains information about two potentialities coexisting in a single experiment and that is because the hyper-sample space itself, in which we're working, primarily has this information by its nature since it contains information about the apparatus-particle entangled system (the fact that sample space is tied to a specific context was actually implicitly mentioned by Kolmogorov, according to Khrennikov \cite{khrennikov106073quantum}). That contextual information is what led to loss of interference due to measurement and that same contextual information is encoded again into the B set through the proposition $\xi$ that eventually destroys interference sets due to the "no-superposition" set-theoretic argument (the argument and its mathematical consequences are  analyzed in Appendices F, J, M). 

Now, we will make an attempt to provide a proof that additivity of probabilities that describe events that emerge from different contexts $C_1,C_2$ is valid in a new experiment (by experiment we mean the conduct of many trials under a specific context) under the context $C_{1,2}$ that is constituted by both $C_{1}$ and $C_2$ and under a specific condition. This condition is that the experiment under $C_1$ and the experiment under $C_2$ are described by non-product sample spaces.

So, to set things more clearly:\textit{Let $C_i$ be the context that gives rise to a single potentiality i, where $i=1,2$ (in the case of having two distinct one-slit experiments, where each experiment has its own slit, we have that $i\equiv \textit{particle goes through i-th slit}$). Also, let $\Omega_i^B$ be the sample space of events taking place under context $C_i$ and let event $B_{i,F}$ such that $B_{i,F} \subseteq \Omega_i^B$. In our case, we have that $B_{1,F}$, $B_{2,F}$ both describe the same event but under different experimental context each. In addition, let $C_{1,2}$ be the context that is constructed by the coexistence of $C_1$, $C_2$ into a single experimental setup,\ let $\widetilde{\Omega}^B$ be the sample space of events under that context and let $\widetilde{B}_{1\lor 2,F}$ be such that $\widetilde{B}_{1\lor 2,F} \subseteq \widetilde{\Omega}^B$ with $1 \lor 2$ indicating the possible emergence either of potentiality 1 or of potentiality 2 under $C_{1,2}$. Suppose we also define the emergence of the two potentialities under the context $C_{1,2}$ to be mutually exclusive and suppose we make the assumption that $\nexists$ $\Omega^T$, $\Omega^A$ such that $\Omega^B=\Omega^TX\Omega^A$, we can very easily prove that:
$$\mu_{C_{1,2}}(\widetilde{B}_{1\lor 2,F})=p_1\mu_{C_1}(B_{1,F})+p_2\mu_{C_2}(B_{2,F})$$
with $p_1+p_2=1$ and with $p_1$, $p_2$ representing the probabilities of the emergence of potentialities 1 and 2 respectively under experimental context $C_{1,2}$}.

We start by mentioning that instead of starting the analysis from $\Omega_1^B$ and $\Omega_2^B$ and seeing if the context transition into $C_{1,2}$ can affect or not the probability measure expression which is defined in the new sample space $\widetilde{\Omega}^B$, we can begin our analysis directly from that new sample space. We start by our mutual exclusivity assumption, which is that $\widetilde{B}_{1,F} \cap \widetilde{B}_{2,F}=\varnothing$ and which due to the fact that $\widetilde{B}_{1\lor 2,F}=\widetilde{B}_{1,F} \cup \widetilde{B}_{2,F}$, gives rise to $\mu(\widetilde{B}_{1\lor 2,F})=\mu(\widetilde{B}_{1,F})+\mu(\widetilde{B}_{2,F})$. Now, we can use the conditional probability formula and obtain that $$\mu_{C_{1,2}}(\widetilde{B}_{1\lor 2,F})=\mu_{C_{1,2}}(\widetilde{B}_{F}|1)\mu_{C_{1,2}}(1)+\mu_{C_{1,2}}(\widetilde{B}_F|2)\mu_{C_{1,2}}(2)$$
By following the definitions of $p_1,p_2$ written before, the above expression can be written as follows:
$$\mu_{C_{1,2}}(\widetilde{B}_{1\lor 2,F})=p_1\mu_{C_{1,2}}(\widetilde{B}_{F}|1)+p_2\mu_{C_{1,2}}(\widetilde{B}_F|2)$$

In general, we're working in a sample space where both contexts $C_1,C_2$ coexist and this is something we can not avoid. What we can do though is to artificially make the potentiality 2 to approach zero (but never become zero, since in that case we would have that $C_{1,2}=C_1$, which is not well-defined given that potentiality 2 would not exist at all and thus we would return to the initial sample space $\Omega_1^B$) by allowing that $p_2\rightarrow 0$, which as a consequence, also gives rise to $p_1 \rightarrow 1$. So, given those, and since we should logically expect that $\lim_{p_2\rightarrow 0}\mu_{C_{1,2}}(\widetilde{B}_{1\lor 2,F})=\mu_{C_1}(B_{1,F})$, we eventually obtain that 
$\mu_{C_{1,2}}(\widetilde{B}_{F}|1)=\mu_{C_1}(B_{1,F})$. In the same way, we can also prove that $\mu_{C_{1,2}}(\widetilde{B}_{F}|2)=\mu_{C_2}(B_{2,F})$ and thus we are easily led into what we wanted to prove.

It should be noted that the quantum experimental analog of the limiting procedure that we applied to practically erase the one potentiality emerging from one of the contexts could be to make, for instance in the double-slit experiment, the one of the two slits extremely narrow or to move it in a place very far away from the electron gun, so that probability of the particle going through that slit approaches zero value. Mathematically speaking, the set $\widetilde{B}_{ 2,F}$ can be approximately considered as a point-like set (not empty though) which of course has approximately measure 0.

Having derived those results when $\Omega^B$ is considered a non-product sample space, we can easily see that this theorem can not be extended in cases where we have the product relation $\Omega^B=\Omega^TX\Omega^A$ under whatever context. And that is because in that case, we could write that under context $C_{1,2}$ we have that: $\forall\widetilde{B}_{1\lor 2,F} \subseteq \widetilde{\Omega}_B$, $\exists \widetilde{T}_{1\lor 2,F} \subseteq \widetilde{\Omega}_T,\widetilde{A}_{1\lor 2,F} \subseteq \widetilde{\Omega}_A$ such that $\widetilde{B}_{1\lor 2,F}=\widetilde{T}_{1\lor 2,F}X\widetilde{A}_{1\lor 2,F}$, which naturally leads to $\widetilde{B}_{1\lor 2,F}\neq \widetilde{B}_{1,F}\cup\widetilde{B}_{2,F}$ (see discussion in Appendix P in the context of quantum probability sets) and thus to "interference" of probabilities (these interference terms do not need to be complex and the only cases where complex numbers emerge is in quantum diffusion where set-theoretic extension unavoidably is used). 

What I wanted to demonstrate with all that is that, indeed the additivity of probabilities is not well-defined under context transition, once we don't know the nature of the events that we're dealing with. If those events are non-product though, then additivity is guaranteed for mutually exclusive potentialities. But it is not just that...The conditional probability measures that we have under context $C_{1,2}$ are identical to the probability measures that we have under $C_1,C_2$. In that sense, if we are dealing with quantum probabilistic systems, additivity is guaranteed for the wave function that emerges as a complex probability density from the non-product T set, but not allowed for the Born probability distribution that emerges from the product B set, and that is compatible with our contextual probability theorem. And given our previous analysis, we are allowed to use that $$\mu_{C_{1,2}}(\widetilde{T} \rightarrow\{x_1 \in F_1,t=t_c\}|i)=\mu_{C_{i}}({T_i} \rightarrow\{x_1 \in F_1,t=t_c\})=\int_{F_1} \Psi_{i,norm}(x_1,t=t_c)dx_1$$, as was explicitly done in Appendix P, but in a less strict mathematical way. We must note that $i$ represents the $i-th$ potentiality and $i=1,2$. Of course, $\Psi_{norm}$ represents the normalized probability density of the f.p. that lives within the normalized non-product hyper-set $\Omega_{in,t=t_c}^T$ and it was first mentioned in Appendix T.

The emphasis on contextuality must not only be restricted to probability spaces, but it could also be extended into the non-commutativity discussion of Appendix U. There, the context does not need necessarily to be interpreted as an experimental context but as a mathematical condition (i.e. "given" the local ensemble value of the one physical quantity of the f.p.) under which we can, or not, define the local ensemble value of the second physical quantity of the f.p.. It turned out that non-commutativity emerged as a result of conditional independence between the local ensemble values of two physical quantities. And as we've said before, uncertainty principle seems to be related with non-commutativity at first sight, but this relation may in fact exist due to the fact that the same quantities that are conditionally independent (i.e. $f(\hat{x}), g(\hat{P})$) seem to be the ones that due to the violent stochastic nature of quantum dynamics seem to obey to uncertainty relations. Or to express it the other way around, the uncertainty of kinetic energy and of momentum operator increase or decrease always simultaneously (they do not satisfy an uncertainty relation), which is also compatible with classical physics exactly because the instantaneous values and therefore the local ensemble values of each of them are fully conditionally dependent (the latter was demonstrated in Appendix U), which is also true in classical physics. But, once again, I will just say that non-commutativity could also be interpreted in the same way as uncertainty relations have.

\textit{For now, what we must keep in mind, based on our analysis, is that contextuality and non-locality seem to be two different sides of the same coin, and the coin is in fact the non-separability that exists in quantum diffusion processes. And as we explained in Appendix R, non-separability and the time-symmetric nature of quantum diffusion are largely related because both emerge once we deny the discrete structure of the sub-quantum medium, which would give rise to both causal relations between entangled particles (non-locality seems to be a fact of correlation and not of causation) and to time-asymmetric diffusion once we allow for the existence of infinitely many "quantum fluid" particles to collide inside a discrete medium. We should always also recall that the classical non-equilibrium diffusion models, in contrast to quantum diffusion, can be derived from Hamiltonian dynamics (under specific approximations like coarse-graining and infinite friction), and more specifically from causal relations between infinitely many particles in an oscillating bath. }

\subsubsection{AN ATTEMPT FOR THE DERIVATION OF THE ABSTRACT VECTOR FORM OF SCHROEDINGER EQUATION FOR NON-KINEMATICAL STATES}

In this subsection, we will attempt to derive the Schroedinger equation, in which the quantum state has no "information" about the kinematical state of the particle. Despite the fact that reversible diffusion is related to motion, this does not mean that the description that we will use about spin, for example, has any classical analogue in conventional probability theory, despite the fact that, still, classical probability theory will be used. 

We mentioned before that a description of non-kinematical observables (like quantum spin for example) in the context of our reversible diffusion theory would require a development of propositional calculus that would contain measurable sets and derivatives acting on them and also other exotic mathematical tools that are unknown so far. However, we can avoid such complications and we can instead consider a different approach.

Let us define a quantum super product space $\Omega_S^{(q)}=\Omega_h^{(q)}X \Omega_{n-k}$. The space $\Omega_{n-k}$ represents the sample space where all the possible "non-kinematical" sets and their subsets live (like for example the set that describes the "up" or "down" possible spin state of a particle interacting with a uniform magnetic field with which we want to work specifically and infinite other "non-kinematical" sets that describe "dead" or "alive"-type states (like Schroedinger cat case and other cases where the mutually exclusive states are discrete and finite, that in the standard QM language is mapped into vectors living in discrete finite dimensional Hilbert spaces) for example, which allows us to write that $\Omega_{sp}\cup(\cup_{n=1}^{\infty}S_{other}^{(n)})=\Omega_{n-k}$, of course with $\Omega_{sp}\cap S_{other}^{(n)}=\varnothing$ and $S_{other}^{(i)} \cap S_{other}^{(j)}=\varnothing$ for $i \neq j$. Just to be clear, a non-kinematical set is a set that describes states unrelated to quantities that are connected to motion (like position, energy, momentum, angular momentum etc.). Since we want to work on spin states, we are enforced to work in the product space $\Omega_h^{(q)}X\Omega_{sp}$ which is a tiny subset of the quantum super product space and refers to a specific sub-case of all the possible situations we could have. 

What we are achieving with this move is to be able to claim that our reversible diffusion theory which focuses on position representation already takes into consideration the possible existence of all possible systems. We can thus say, for example, that Schroedinger diffusion processes can also be described by the super product space, meaning that we can say that $\mu(T \rightarrow \{x_1 \in F,t\})=\mu(T \rightarrow \{x_1 \in F,t\}X\Omega_{n-k})=\int_{F}\Psi_{norm}(x_1,t)dx_1$ (it is obvious that this T-set lives in the quantum hyper-sample space $\Omega_h^{(q)}$ where $x_1$ and $x_2$ almost coincide and also $t_{forward}=t_{backward}=t$). In this way, the Schroedinger equation in the position representation is not emergent from a more fundamental equation (according to Dirac's formalism this is the general abstract vector form of the Schroedinger equation) from which the Schroedinger equation in the position representation emerges through multiplication with an $x$-bra from the left. Rather, it is an equation that has averaged out all additional information about the non-kinematical state of the particle and the treatment of 2-level spin systems (for instance) requires from us to work in the less general product sample space $\Omega_{h,sp}^{(q)}=\Omega_h^{(q)}X\Omega_{sp}$. 

Since $\Omega_{sp}$ is uncoupled to quantum diffusion (we assume the magnetic field is uniform), there should be no forward and backward particle description for spin sets. On the contrary, we have that $\Omega_{sp}=sp(\uparrow or \downarrow)=sp(\uparrow)\cup sp(\downarrow)$ for which we consider that $sp(\uparrow)$ and $sp(\downarrow)$ describe the quantum particle as having spin up and spin down respectively. For them, we have that $\mu(sp(\uparrow))=p_1=|c_1|^2 \in \mathbb{R}$ and $\mu(sp(\downarrow))=p_2=|c_2|^2 \in \mathbb{R}$ and $sp(\uparrow)\cap sp(\downarrow)=\varnothing$ which gives rise to $|c_1|^2+|c_2|^2=1=\mu(\Omega_{sp})$. Since forward and backward description is invalid for spin set discussion, the analysis here will slightly differ from Appendix L in the sense that, there, T and A sets were used also for spin. And as we will see, the approach developed in this subsection, has advantages compared to Appendix L, because here all probability measures can be well-defined without that causing issues with state orthogonality and thus with state physical definiteness before measurement (the ontic state must be one of the two superposed states). However, it is good to have all approaches in mind if we want to make the first step for the development of a complete non-relativistic quantum theory. In that sense, this is the only difference and the methodology we follow in this approach, as well as the justification of why the mapping between quantum probability sets and Hilbert space vectors is useful, are completely similar to Appendix L. 

In our case, presenting an interaction Hamiltonian is equivalent to modifying the quantum Hamilton-Jacobi equation by introducing an extra interaction term and by not changing the "quantum potential energy" which is the local ensemble kinetic energy due to fluctuations and the local ensemble kinetic energy due to deterministic flow which is the Bohmian flow that depends on the current velocity/gradient of the phase (see Appendix T) and that's because the motion in general is unaffected by the interaction. This means that the interaction will just induce an additional phase factor on the wave function that depends only on time. More specifically, following the notation of $s_1$ and $s_2$ from Appendix L, we can define the classical interaction Hamiltonian for a quantum particle with spin, which is $H_{q.p.}^{s_i}=-\overrightarrow{\mu}_{s_i}\cdot \overrightarrow{B}$, where $i=1,2$ and which gives rise to $\overrightarrow{F}_{q.p.}^{s_i}=-\overrightarrow{\nabla}H_{q.p.}^{s_i}=\overrightarrow\nabla(\overrightarrow{\mu}_{s_i} \cdot \overrightarrow{B})=0$ since $B$ and $\mu_{s_i}$ are spatially uniform (recall that $\overrightarrow{B}$ is the magnetic field and $\overrightarrow{\mu}_{s_i}$ is the intrinsic spin magnetic moment with $\overrightarrow{\mu}_{s_i} \propto \overrightarrow{S}_i$ \cite{sakurai1994modern}). The induced phase, will be given by the equation $-\frac{\partial (\delta S_i(r,t))}{\partial t}=H^{s_i}_{q.p}$, that finally gives rise to $\delta S_i(r,t)=-H_{q.p.}^{s_i}t$ and thus to $\Psi^{s_i}(r_1,t)=e^{-iH^{s_i}_{q.p.}t}\Psi_{norm}(r_1,t)$, where $\Psi_{norm}(r_1,t)$ is the initial interaction-free position representation wave function defined as $\Psi_{norm}(r_1,t)=\frac{\Psi(r_1,t)}{\int_{\mathbb{R}^3}\Psi(r_1',t)d^3r_1'}$. 

So, now we are introducing the T-set in the product space $\Omega^{(q)}_{h,sp}$ that describes the f.p. behavior in the case where the q.p. is a particle with spin interacting with a magnetic field. This is defined as $T' \rightarrow \{r_1 \in \mathbb{R}^3,t\}=T \rightarrow \{r_1 \in \mathbb{R}^3,t\}X\Omega_{sp}=T \rightarrow \{r_1 \in \mathbb{R}^3,t\}X(sp(\uparrow) \cup sp(\downarrow))=\left(T \rightarrow \{r_1 \in \mathbb{R}^3,t\}Xsp(\uparrow)\right)\cup \left(T \rightarrow \{r_1 \in \mathbb{R}^3,t\}Xsp(\downarrow)\right)$. Those subsets are in fact mutually exclusive since $\left(T \rightarrow \{r_1 \in \mathbb{R}^3,t\}Xsp(\uparrow)\right)\cap \left(T \rightarrow \{r_1 \in \mathbb{R}^3,t\}Xsp(\downarrow)\right)=T \rightarrow \{r_1 \in \mathbb{R}^3,t\}X(sp(\uparrow) \cap sp(\downarrow))=T \rightarrow \{r_1 \in \mathbb{R}^3,t\}X\varnothing=\varnothing$ (a similar result holds for the A-set too). Let's recall that the union between the T (or A) spin subsets is isomorphic to the pure sum between the kets $c_1 e^{-iH_{q.p}^{s_1}t}|  \uparrow \rangle\equiv c_1e^{-iH_{q.p}^{s_1}t}\phi_1(0)=c_1\phi_1(t)$ and $c_2 e^{-iH_{q.p}^{s_2}t}|  \downarrow \rangle \equiv c_2 e^{-iH^{s_2}_{q.p.}t} \phi_2(0)=c_2\phi_2(t)$ (we will see later how the exponential expressions that are related to the time evolution of vectors emerge) and let's note that the mutual exclusivity between the two sets is translated into zero overlap (orthogonality) between the kets, as expected. 

But beyond that, we can also see that the orthogonal bra-ket relations emerge quite easily. For instance, we have: $\mu\left((T \rightarrow \{r_1 \in \mathbb{R}^3,t\}Xsp(\uparrow)\cap(A \rightarrow \{r_2 \in \mathbb{R}^3,t\}Xsp(\downarrow))\right)=\mu\left((T \rightarrow \{r_1 \in \mathbb{R}^3,t\}\cap A\rightarrow \{r_2 \in \mathbb{R}^3,t\})X(sp(\uparrow)\cap sp(\downarrow))\right)=\mu((T \rightarrow \{r_1 \in \mathbb{R}^3,t\}\cap A\rightarrow \{r_2 \in \mathbb{R}^3,t\})X \varnothing)=\mu(\varnothing)=0$, which maps into $c_1^*c_2\phi_1^*(t)*\phi_2(t)=c_1^*c_2 e^{-i(H_{q.p.}^{s_2}-H_{q.p.}^{s_1})t}(\phi_1^*(0)*\phi_2(0))\equiv c_1^*c_2e^{-i(H_{q.p.}^{s_2}-H_{q.p.}^{s_1})t}\langle \uparrow |  \downarrow \rangle=0$ and similarly when the ket is a "spin-up" state and the bra is a "spin-down" state. And if in the above equation, $sp(\downarrow)$ is substituted by $sp(\uparrow)$, we obtain $\mu(B\rightarrow \{r \in \mathbb{R}^3,t\}Xsp(\uparrow))=|c_1|^2$ and that maps into $|c_1|^2\phi_1^*(t)*\phi_1(t)=|c_1|^2 \phi_1^*(0) * \phi_1(0)\equiv|c_1|^2\langle \uparrow |  \uparrow \rangle=|c_1|^2$ from which we get the normalization of the basis $\forall t \geq 0$, since $\phi_1^*(t)*\phi_1(t)=\phi_1^*(0)\phi_1(0)$ that holds at any such time $t$. Of course, we can easily derive the normalization not only for basis vectors but also for vectors representing the overall quantum state and which are the linear combinations of those basis vectors (in Appendix L, the superposed state was denoted with the ket $| \psi \rangle $). I also need to mention that the $\omega$-vectors that we used in Appendix L are the ones that are directly isomorphic to our quantum sets. For them, we have that $\omega_i(t)=c_i\phi_i(t)$ and $\omega(t)=\phi(t)$ and similarly for the $\omega'$-vectors.

\textit{Now, to be more intuitive in our representation, we can perform a more direct picture of this isomorphism. In that picture, the T and A-sets are "carriers" of vectors. The degree of such set overlap must correspond to the degree of overlap between the vectors that they "carry" (or their inner product measure). When two sets are non-overlapping at all, the vectors "carried" by them are perpendicular to each other and that is what we saw before, where $\omega_i*\omega'_j=\omega_i*\omega_j=\omega_i'*\omega_j'=0$ for $i \neq j$. Especially, the zero set overlaps that produce the relations $\omega_i*\omega_j=\omega_i'*\omega_j'=0$ for $i \neq j$ is responsible for the "forward" and "backward" state superposition (in the position representation of spin description) that we will present below and that produces the spinor structure of the wave function in the Hilbert space picture. And that is the thing here. We do not need spinor wave functions to produce those zero values. Instead, we have probability sets doing this work and spinor wave functions just emerge as convenient tools.}

Now, to derive the Schroedinger's equation, we must first see how this time evolution of the abstract state vectors emerges and why the set description that we have given for systems described by finite-dimensional Hilbert space vectors (like spin systems) is not incompatible with the probabilistic treatment and the operator description that we have extensively discussed so far in the position representation. This representation included both denumerable/continuous state spaces and discrete but not necessarily finite-dimensional Hilbert spaces (like Hilbert spaces whose elements are energy eigenstates).

{}


\begin{thebibliography}{} 

\bibitem{caves2007subjective} Subjective probability and quantum certainty, Caves, Carlton M and Fuchs, Christopher A and Schack, R{\"u}diger, Studies in HIstory and Philosophy of Science Part B: Studies in History and Philosophy of Modern Physics {\bf 38} 255--274 (2007)

\bibitem{rovelli1996relational} Relational quantum mechanics, Rovelli, Carlo, International journal of theoretical physics {\bf 35} 1637--1678 (1996)


\bibitem{peliti2023r} R. F{\"u}rth’s 1933 paper “On certain relations between classical statistics and quantum mechanics”[“{\"U}ber einige Beziehungen zwischen klassischer Statistik und Quantenmechanik”, Zeitschrift f{\"u}r Physik, 81 143--162], Peliti, Luca and Muratore-Ginanneschi, Paolo, The European Physical Journal H, {\bf 48} 4 (2023)


\bibitem{dewitt2015many} The many-worlds interpretation of quantum mechanics, Dewitt, Bryce Seligman and Graham, Neill {\bf 61} (2015)

\bibitem{carroll2014many} Many worlds, the born rule, and self-locating uncertainty, Carroll, Sean M and Sebens, Charles T 157--169 (2014)

\bibitem{wallace2010prove} How to prove the Born rule, Many worlds, Wallace, D (2010)

\bibitem{saunders2024finite} Finite frequentism explains quantum probability, Saunders, Simon (2024)

\bibitem{bohm1952suggested} A suggested interpretation of the quantum theory in terms of" hidden" variables. I, Bohm, David, Physical review, {\bf 85} 166 (1952)

\bibitem{durr2009bohmian} Bohmian mechanics, D{\"u}rr, Detlev and Goldstein, Sheldon and Tumulka, Roderich and Zangh{\'\i}, Nino, Compendium of quantum physics 47--55 (2009)


\bibitem{deotto1998bohmian} Bohmian mechanics revisited, Deotto, Enrico and Ghirardi, GianCarlo, Foundations of Physics, {\bf 28} 1--30 (1998)

\bibitem{tzemos2023unstable} Unstable points, ergodicity and Born’s rule in 2D Bohmian systems, Tzemos, Athanasios C and Contopoulos, George, Entropy {\bf 25} 1089 (2023)

\bibitem{tzemos2022born} Born’s rule in multiqubit bohmian systems, Tzemos, AC and Contopoulos, G, Chaos, Solitons \& Fractals {\bf 164} 112650 (2022)


\bibitem{nelson1966derivation} Derivation of the Schr{\"o}dinger equation from Newtonian mechanics, Nelson, Edward Physical review {\bf 150} 1079 (1966)

\bibitem{goldstein2013reality} Reality and the role of the wave function in quantum theory, Goldstein, Sheldon and Zangh{\`\i}, Nino and others, The wave function: Essays on the metaphysics of quantum mechanicsn {\bf 91} (2013)

\bibitem{kochen2011problem} The problem of hidden variables in quantum mechanics Kochen, Simon and Specker, Ernst P, 235--263 (2011)


\bibitem{beyer2021stochastic} On the stochastic mechanics foundation of quantum mechanics, Beyer, Michael and Paul, Wolfgang, Universe {\bf 7} 166 (2021)

\bibitem{wallstrom1989derivation} On the derivation of the Schr{\"o}dinger equation from stochastic mechanics, Wallstrom, Timothy C, Foundations of Physics Letters {\bf 2} 113--126 (1989)

\bibitem{davidson1979generalization} A generalization of the F{\'e}nyes—Nelson stochastic model of quantum mechanics, Davidson, Mark, Letters in Mathematical Physics {\bf 3} 271--277 (1979)

\bibitem{bacciagaluppi2005conceptual} A conceptual introduction to Nelson’s mechanics Bacciagaluppi, Guido 367--388 (2005)

\bibitem{mita2021schrodinger} Schr{\"o}dinger's equation as a diffusion equation, Mita, Katsunori, American Journal of Physics {\bf 89} 500--510 (2021)

\bibitem{bohm1989non} Non-locality and locality in the stochastic interpretation of quantum mechanics, Bohm, David and Hiley, Basil J, Physics Reports {\bf 172} 93--122 (1989)

\bibitem{folland1999real} Real analysis: modern techniques and their applications, Folland, Gerald B (1999)

\bibitem{bell1966problem} On the problem of hidden variables in quantum mechanics, Bell, John S, Reviews of Modern physics {\bf 38} 447 (1966)

\bibitem{pavliotis2014langevin} The Langevin equation, Pavliotis, Grigorios A, Stochastic Processes and Applications: Diffusion Processes, the Fokker-Planck and Langevin Equations 181--233 (2014)

\bibitem{zwanzig1973nonlinear} Nonlinear generalized Langevin equations, Zwanzig, Robert, Journal of Statistical Physics {\bf 9} 215--220 (1973)

\bibitem{gleason1975measures} Measures on the closed subspaces of a Hilbert space, Gleason, Andrew M, 123--133 (1975)

\bibitem{cramer1986transactional} The transactional interpretation of quantum mechanics, Cramer, John G, Reviews of Modern Physics {\bf 58} 647 (1986)





\bibitem{guerra1981structural} Structural aspects of stochastic mechanics and stochastic field theory, Guerra, Francesco, Physics Reports {\bf 77} 263--312 (1981)

\bibitem{redei2013quantum} Quantum logic in algebraic approach, R{\'e}dei, Mikl{\'o}s {\bf 91}
(2013)

\bibitem{ardakani2018time} Time Reversal Invariance in Quantum Mechanics Ardakani, Reza Moulavi, arXiv preprint arXiv:1802.10169 (2018)


\bibitem{callender2024insights} Insights into Quantum Time Reversal from the Classical Schr{\"o}dinger Equation, Callender, Craig (2024)

\bibitem{gao2022understanding} Understanding Time Reversal in Quantum Mechanics: A New Derivation Gao, Shao, Foundations of Physics {\bf 52} 114 (2022)

\bibitem{kocsis2011observing} Observing the average trajectories of single photons in a two-slit interferometer Kocsis, Sacha and Braverman, Boris and Ravets, Sylvain and Stevens, Martin J and Mirin, Richard P and Shalm, L Krister and Steinberg, Aephraim M, Science {\bf 332} 1170--1173 (2011)

\bibitem{antonakos2024aharonov} Aharonov--Bohm Effect as a Diffusion Phenomenon, Antonakos, Charalampos and Terzis, Andreas F, Foundations of Physics {\bf 54} 53 (2024)

\bibitem{balazs1979nonspreading} Nonspreading wave packets, Balazs, MVBNL and Berry, MV, Am. J. Phys. {\bf 47} 264--267 (1979)

\bibitem{feynman1951concept} The concept of probability in quantum mechanics, Feynman, Richard P {\bf 2} 533--542 (1951)

\bibitem{albert2003time} Time and chance, Albert, David Z (2003)

\bibitem{zeh1970interpretation} On the interpretation of measurement in quantum theory, Zeh, H Dieter, Foundations of Physics {\bf 1} 69--76 (1970)

\bibitem{zurek2003decoherence} Decoherence, einselection, and the quantum origins of the classical, Zurek, Wojciech Hubert, Reviews of modern physics {\bf 75} 715 (2003)

\bibitem{khrennikov106073quantum} Quantum probabilities’ as context depending probabilities, Khrennikov, Andrei, E-print: arXiv: quant-ph/0106073 {\bf 224} (2001)

\bibitem{ballentine1970statistical} The statistical interpretation of quantum mechanics, Ballentine, Leslie E, Reviews of modern physics {\bf 42} 358 (1970)

\bibitem{pusey2012reality} On the reality of the quantum state, Pusey, Matthew F and Barrett, Jonathan and Rudolph, Terry, Nature Physics {\bf 8} 475--478 (2012)

\bibitem{valentini1991signal} Signal-locality, uncertainty, and the subquantum H-theorem. I, Valentini, Antony, Physics Letters A {\bf 156}  5--11 (1991)

\bibitem{karakostas2012realism} Realism and objectivism in quantum mechanics, Karakostas, Vassilios, Journal for general philosophy of science {\bf 43} 45--65 (2012)

\bibitem{bohm1984measurement} Measurement understood through the quantum potential approach, Bohm, David and Hiley, Basil J, Foundations of Physics {\bf 14} 255--274 (1984)

\bibitem{tsekov2017derivation} Derivation of the local-mean stochastic quantum force, Tsekov, Roumen and Heifetz, Eyal and Cohen, Eliahu, Fluctuation and Noise Letters {\bf 16} 1750028 (2017)

\bibitem{tsekov2021classical} Classical and Quantum Brownian Motion Tsekov, Roumen, arXiv preprint arXiv:2105.05646 (2021)


\bibitem{sakurai1994modern} Modern quantum mechanics, Sakurai, Jun John and San Fu Tuan {\bf 3} (1994)

\end{thebibliography}
\end{document}